\documentclass[]{aastex701}
\usepackage{graphicx}

\usepackage{longtable, booktabs, threeparttablex}
\usepackage{chemformula}
\usepackage[english]{babel}
\usepackage[autostyle]{csquotes}
\usepackage{graphicx}
\usepackage{multirow}
\usepackage{xcolor}
\usepackage[version=4]{mhchem}
\usepackage{amsmath}
\usepackage{chemformula}
\usepackage{dirtytalk}
\usepackage{tikz}
\usepackage{textcomp}
\usepackage{savesym}
\savesymbol{tablenum}
\usepackage{siunitx}
\restoresymbol{SIX}{tablenum}
\usepackage{natbib}
\usepackage{csquotes}
\usepackage{float}
\usepackage{chngpage}
\usepackage{caption}
\usepackage{silence}
\begin{document}

\title{Interstellar Complex Organic Molecules and Molecular Outflows in NGC 1333 IRAS 4B and 4B$'$ Observed Using NOEMA}

\author[0000-0002-3976-322X]{Collette C. Sarver}
\email{ccsarver@wisc.edu}
\affiliation{Department of Chemistry, University of Wisconsin-Madison \\ 1101 University Ave, Madison, Wisconsin 53706, USA}

\author[0009-0007-3950-335X]{Morgan M. Giese}
\email{mgiese@uwo.ca}
\affiliation{Department of Physics \& Astronomy, The University of Western Ontario, London, Canada, N6A 3K7}
\affiliation{Institute for Earth and Space Exploration, The University of Western Ontario, London, Canada, N6A 3K7}

\author[0000-0002-0500-4700]{Dariusz C. Lis}
\affiliation{Jet Propulsion Laboratory, California Institute of Technology \\ 4800 Oak Grove Drive, Pasadena, CA, 91109, USA}
\email{dariusz.c.lis@jpl.nasa.gov}

\author[0000-0001-6015-3429]{Susanna L. Widicus Weaver}
\affiliation{Department of Chemistry, University of Wisconsin-Madison \\ 1101 University Ave, Madison, Wisconsin 53706, USA}
\affiliation{Department of Astronomy, University of Wisconsin-Madison \\ 475 N Charter St, Madison, Wisconsin 53706, USA}
\email{slww@chem.wisc.edu}

\begin{abstract}
We present interferometric observations with the Northern Extended Millimeter Array (NOEMA) of NGC 1333 IRAS 4B and 4B$'$, two young stellar objects (YSOs) within the Perseus molecular cloud.  YSOs, especially Class 0 protostars, are rich in gas-phase interstellar complex organic molecules (iCOMs), but relatively little is known about the spatial distribution of these important prebiotic molecules. Various molecules were imaged within IRAS 4B and the surrounding area in the 145 GHz spectral range, revealing hundreds of molecular transitions within the warm inner envelope of IRAS 4B. This work provides derived physical parameters (rotational temperature, column density, velocity shift with respect to local standard of rest, and spectral line width full width at half-maximum) for 11 molecules toward IRAS 4B including methanol (\ch{CH3OH} and isotopologues), methyl formate (\ch{HCOOCH3}), dimethyl ether (\ch{CH3OCH3}), acetaldehyde (\ch{CH3CHO}), ethanol (\ch{C2H5OH}), glycolaldehyde (\ch{CH2OHCHO}), acetone (\ch{CH3COCH3}), and isocyanic acid (\ch{HNCO}). Out of these molecules, 10 are iCOMs (excluding \ch{HNCO}) that trace the inner envelope of IRAS 4B and mark it as a rich source of organic material. There is a strong spatial correlation between the continuum emission of IRAS 4B and iCOM molecular emission and among various iCOMs with each other. Physical parameter and integrated intensity maps show that IRAS 4B is a binary protostar system that drives a jet in a north-south orientation and an outflow in a northwest-southeast orientation. IRAS 4B$'$ also drives a west-east outflow but shows no evidence of molecular emission cospatial with its continuum emission.  
\end{abstract}

\keywords{Astrochemistry (75) --- Cosmochemistry (331) --- Star-forming regions (1565) --- Young stellar objects (1834) --- Radio Interferometry (1346) --- Submillimeter astronomy (1647) --- Complex organic molecules (2256) --- Interstellar Molecules (849) --- Chemical abundances (224)}

\section{Introduction} \label{sec:intro}
Spectral line surveys of molecular rotational transitions provide a reliable way of detecting molecules within various astronomical environments. Broadband line surveys that use radio interferometers like the Atacama Large Millimeter Array (ALMA) and Northern Extended Millimeter Array (NOEMA) can reveal the spatial distribution and kinematics of molecules and provide insights into the chemical and physical conditions in these environments  \citep{jorgensen_prosac_2007}. Interstellar complex organic molecules (iCOMs), defined as molecules with six or more atoms that contain carbon along with potentially hydrogen, oxygen, and nitrogen \citep{herbst_complex_2009}, are of particular interest in star-forming regions. While not being chemically complex in the terrestrial sense, these iCOMs are prebiotic molecules that are the precursors to life and are detected at every stage of star formation from dense molecular clouds to protoplanetary disks \citep{herbst_complex_2009}. Most iCOM formation happens on the surfaces of icy dust grains, where complex chemistry can occur through energetic and thermal processing \citep{greenberg_1972,hagen_1979,allamandola_1988,Gerakines1996,Oberg2009,Henderson2015,AbouMrad2017,Yocum2019,Yocum2021,sarver}.  Gas-phase reactions can occur to form additional iCOMs once temperatures in the hot core phase of star formation are sufficient to desorb species from the ice surfaces \citep{herbst_complex_2009}.

The masses of Class 0 sources are dominated by the envelope, which surrounds the protostar and provides the material feeding the accretion process onto the protoplanetary disk and forming star. The outer envelope, which scales on order of thousands of astronomical units (au) in size, is formed of quiescent gas with a direct chemical inheritance from the parent molecular cloud and can be traced with different ions such as \ch{N2D+} and \ch{DCO+} \citep{Caselli-Ceccarelli-2012}. The warm inner envelope of the protostar reaches temperatures above 100 K which allows for iCOMs trapped in the ices to sublimate and become detectable via rotational spectroscopy at radio and millimeter wavelengths. To extract excess angular momentum, Class 0 sources are also associated with high velocity molecular jets and low velocity molecular outflows, where dust and gas are propelled away from the system along an axis perpendicular to the disk, along the magnetic field lines \citep{what-traces-what-2021}. Jets are highly collimated, fast ($>$20 km s$^{-1}$) gas moving away from the protostar, typically traced by O-bearing molecules such as carbon monoxide (\ch{CO}), silicon monoxide (\ch{SiO}), sulfur monoxide (\ch{SO}), and formaldehyde (\ch{H2CO}) \citep{what-traces-what-2021}, as well as atomic oxygen [OI] \citep{gusdorf-2017-OI}. Low velocity molecular outflows can also be traced by other molecules like hydrogen cyanide (\ch{HCN}). Ice sputtering can be traced with low energy transitions of methanol (\ch{CH3OH}) and \ch{H2CO}, which occurs through the interaction between the outflows and the ice mantles, where molecules are released from ice grains due to shocks \citep{what-traces-what-2021}. By combining detections of multiple molecules, a complete picture of the physical and chemical conditions of a star-forming region can be constructed.

NGC 1333 is a star-forming region within the Perseus molecular cloud at a distance of $235\pm18$ parsecs \citep{Hirota-2008}. This low-mass star-forming region includes a multitude of young stellar objects, including IRAS 4B (also known as 4BW, 4BI, and Per-emb-13), a Class 0 source, with a companion separated by approximately 11$''$ labeled here as 4B$'$ (other names include 4BE and 4BII) \citep{di_francesco_2001,sandell-2001,jorgensen_prosac_2007,choi-2011,maury_2019,prodige-2026}. Although formed from the same parent molecular cloud, 4B and 4B$'$ are quite different. 4B is known to drive molecular outflows in generally the north-south orientation as seen in multiple molecular tracers such as CO, \ch{CH3OH}, deuterated methanol (\ch{CH2DOH}), \ch{H2CO}, dideuteriomethanone (\ch{D2CO}), SO, \ch{SiO}, \ch{HCN}, deuterated hydrogen cyanide (\ch{DCN}), carbon monosulfide (\ch{CS}), methyl cyanide (\ch{CH3CN}), acetaldehyde (\ch{CH3CHO}), and isocyanic acid (\ch{HNCO}) \citep{blake_1995,jorgensen_prosac_2007,choi-2011,sakai-ch3oh-2012,prodige-2026}. The northern outflow lobe of 4B is redshifted while the southern one is blueshifted, with comparable brightness. \ch{H2} knots have been observed in the southern outflow \citep{choi-2011}. There is also evidence of a second bipolar outflow or arc from 4B, observed in \ch{CH3OH} and \ch{H2CO} with a northwest-southeast orientation \citep{di_francesco_2001,sakai-ch3oh-2012,prodige-2026}. The inner envelope of IRAS 4B has a few iCOM detections including \ch{CH3OH} and its isotopologues (\ch{^{13}CH3OH}, \ch{CH3^{18}OH}, \ch{CH2DOH}, and \ch{CH3OD}), methyl formate (\ch{HCOOCH3}), \ch{CH3CN}, a methyl cyanide isotopologue (\ch{^{13}CH3CN}), dimethyl ether (\ch{CH3OCH3}), propionitrile (\ch{C2H5CN}), and glycolaldehyde (\ch{CH2OHCHO}) \citep{jorgensen-2005,maret-2005,sakai_detection_2006,Parise-2006,sakai-ch3oh-2012,de-simone-2017,busch_prodige_2025}. Less is known about its companion, 4B$'$, with very few molecular detections claimed: only \ch{HCN} and \ch{^{13}CO} \citep{choi-2001}. 4B$'$ is not detected in the infrared, but only in the millimeter and submillimeter bands \citep{choi-2011}. Only recently has 4B$'$ been observed to drive molecular outflows in CO, \ch{SiO}, \ch{CH3OH}, and \ch{H2CO} emission \citep{prodige-2026}, but these outflow have yet to be characterized. Additionally, multiple \ch{H2O} masers are associated with the region surrounding IRAS 4 (which includes the sources 4A, 4B, 4B$'$, and 4C) with many \ch{H2O} masers clustering toward the inner envelope of 4B \citep{Rodriguez-2002,masers-2008}. 

The majority of previous observations of IRAS 4 focus on the brighter source 4A, with only a few on 4B, and fewer still differentiating the neighboring continuum source 4B$'$. Previous observations of IRAS 4B include a molecular line survey using the Caltech Submillimeter Observatory (CSO) that was part of a deep broadband survey of 30 star-forming regions \citep{cso}. The goal of this survey was to identify iCOMs within these regions and explore the link between physical environment and molecular complexity. Due to the lack of spatial resolution of the single-dish observations using the CSO, follow-up observations with interferometric capabilities were therefore necessary to complete this objective. Single-dish observations typically lack the angular resolution to differentiate the spatial distribution of iCOMs within the warm inner envelopes of protostars and outflows traveling away from the source. Differentiation of molecular emission complexity between outflows and envelopes is made possible with interferometry and can yield information on whether gas-phase or ice-phase production routes are dominate \citep{what-traces-what-2021}.

A recent study by \cite{prodige-2026} observed the outflows and their morphology in IRAS 4B and 4B$'$ using Northern Extended Millimeter Array (NOEMA) in band 3 (226.5 GHz) as a part of a PROtostars \& DIsks: Global Evolution large program (PRODIGE; PIs: P. Caselli and Th. Henning). That study confirmed iCOM detections of \ch{CH3OH}, \ch{CH2DOH}, \ch{CH3CN}, and \ch{CH3CHO} in the north and south outflows of 4B. Their results include integrated intensity ratio maps showing the spatial distribution of these molecules with respect to \ch{CH3OH}. Their analysis focuses on molecular abundances derived at two positions toward the northern and southern outflow lobes, as well as the effects of shocks on chemical composition. Additionally, they noted the outflow driven from 4B$'$ for the first time, but did not discuss the outflow any further. Here we present observations of IRAS 4B and 4B$'$ using NOEMA in band 2, which allows for additional iCOM molecular detection characterization and the analysis of iCOM distribution within the envelope and outflows. These results include an in-depth chemical analysis including derived parameter maps, molecular correlation matrices, and the first characterization of the outflow driven by IRAS 4B$'$. This manuscript is structured as follows: Section \ref{sec:obs} describes the observations, data reduction, and imaging of IRAS 4B and 4B$'$. Section \ref{sec:results-and-discussion} discusses the results, which includes line identification and analysis (\ref{subsec:line-id-analysis}), derived physical parameter maps (\ref{subsec:parameter-maps}), and molecular emission in the high-resolution spectral windows (\ref{subsec:highres}). The discussion in Subsection \ref{subsec:discussion} presents implications of the data. Conclusions are summarized in Section \ref{sec:conclusions}.
\setcounter{footnote}{0}

\section{Observations} \label{sec:obs}
IRAS 4B was observed with IRAM/NOEMA\footnote{IRAM is supported by INSU/CNRS (France), MPG (Germany), and IGN (Spain).} during two nights on October 19th 2022 for 3.9 hours and November 6th 2022 for 3.5 hours (on-source time). Observations were carried out in the 2 mm band in C configuration (12C-E023 for Oct 10 2022 and 12C-N007 for Nov 06 2022) with the local oscillator frequency set at 145.250 GHz. NOEMA's C configuration has baselines that range from 24.0 m to 368.0 m. The pointing center of the observation was at  $\alpha$(J2000) = $03^h29^m12^s.50$,  $\delta$(J2000) = 31\textdegree$13'08''.0$.

The upper and lower sidebands (denoted USB and LSB) had a spectral resolution of 2000.0 kHz and an effective spectral coverage of 127374 to 135500 MHz and 142862 to 150986 MHz, respectively. An additional 28 high-resolution windows were strategically placed to cover transitions of iCOMs of interest: \ch{CH3OH}, \ch{H2CO}, \ch{CH3CN}, \ch{CH3CHO}, and \ch{CH3OCH3}, among others. These high-resolution windows had a spectral resolution of 62.5 kHz and were split evenly between the LSB and USB. Table \ref{tab:table-summary-hr-spw} lists both sidebands as well as the 28 high-resolutions windows analyzed here and their respective frequency ranges. The resulting synthesized beam for the LSB was $1.46\times1.62''$ (PA$=27.72$\textdegree, from north toward east) and for the USB was $1.26\times1.46''$ (PA$=23.04$\textdegree, from north toward east).

IRAS 4B observations were obtained using the PolyFiX correlator. Phase and amplitude calibration was done using observations of 0333+321. The flux calibration used 2010+723 and MWC349. Finally, the bandpass calibration was performed using 3C84 and a spectral index of -0.55 from the SPIDX catalog. Receiver temperatures ranged from 20--70 K throughout the two nights of observing. 

Each high and low-resolution spectral window was reduced and cleaned in a similar fashion as described by \cite{will_w3_2023} and \cite{Morgan-2025}. First the GILDAS\footnote{\url{http://www.iram.fr/IRAMFR/GILDAS}} package CLIC and IRAMS's pipeline was used to generate UV tables. These UV tables were self-calibrated with another GILDAS package MAPPING, and then cleaned with the same software. A natural robust weighing mode was used for imaging along with the hogbom deconvolution method. After the final round of cleaning, MAPPING's command of `go noise' was used to determine the root-mean squared (rms) noise value of each cube. Continuum subtraction was done with the software package STATCONT \citep{STATCONT}, which determined the continuum at each pixel within the datacube using rms values determined in the previous step. Table \ref{tab:table-summary-hr-spw} lists the associated rms in mJy beam$^{-1}$ for each of the sidebands and high-resolution spectral windows.

\begin{deluxetable}{ccc}[ht]
\tablecaption{\label{tab:table-summary-hr-spw}Summary of sidebands and high-resolution spectral windows.}
\tablehead{
    \colhead{\textbf{Spectral Window}} & \colhead{\textbf{Frequency Range}} & \colhead{\textbf{rms}} \\
    \colhead{} & \colhead{(MHz)} & \colhead{(mJy beam$^{-1}$)}
    }
    \startdata
        LSB & 127374.0 -- 135500.0 & 1.2\\
        1 & 127820.2 -- 127883.7 & 7.5\\
        2 & 127948.2 -- 128075.7 & 8.7\\
        3 & 128588.2 -- 128843.7 & 5.5\\
        4 & 129100.2 -- 129163.7 & 5.3\\
        5 & 129356.2 -- 129420.7 & 5.5\\
        6 & 130188.2 -- 130315.7 & 5.7\\
        7 & 132236.2 -- 132299.7 & 5.6\\
        8 & 132556.2 -- 132939.7 & 4.9\\
        9 & 133132.2 -- 133323.7 & 4.8\\
        10 & 133580.2 -- 133643.7 & 5.3\\
        11 & 134156.2 -- 134283.7 & 5.2\\
        12 & 134860.2 -- 134923.7 & 5.7\\
        13 & 134988.2 -- 135051.7 & 6.0\\
        14 & 135244.2 -- 135307.7 & 6.6\\
        USB& 142862.0 -- 150986.0 & 1.1\\
        15 & 143116.1 -- 143180.0 & 6.3\\
        16 & 143436.1 -- 143564.0 & 5.7\\
        17 & 143692.1 -- 143756.0 & 5.7\\
        18 & 144524.1 -- 144652.0 & 5.7\\
        19 & 144716.1 -- 144908.0 & 5.9\\
        20 & 145036.1 -- 145164.0 & 6.1\\
        21 & 145548.1 -- 145612.0 & 5.6\\
        22 & 145932.1 -- 145996.0 & 5.6\\
        23 & 146316.1 -- 146380.0 & 5.3\\
        24 & 146508.1 -- 146636.0 & 5.5\\
        25 & 146956.1 -- 147212.0 & 5.8\\
        26 & 148044.1 -- 148172.0 & 5.8\\
        27 & 149516.1 -- 149580.0 & 6.7\\
        28 & 150092.1 -- 150668.0 & 7.4\\
        \enddata
        \tablecomments{Noise in rms is listed for each continuum corrected datacube used for line analysis. The velocity bin used to estimate the rms was the velocity resolution for each band.}
\vspace{-30pt}      
\end{deluxetable}

\begin{figure}[ht]
    \centering
    \includegraphics[width=0.8\linewidth]{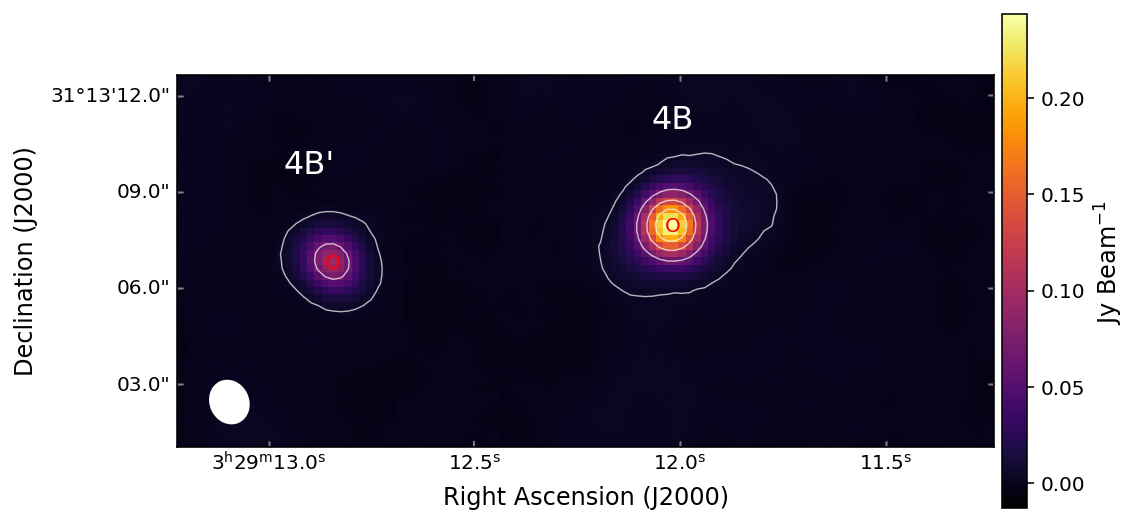}
    \caption{Continuum image of NGC 1333 IRAS 4B (right) and IRAS 4B$'$ (left) at 147 GHz. Contours are 5 evenly spaced levels (1 level $=0.0599$ Jy beam$^{-1}$) from $3\times \sigma_{cont}$ ($\sigma_{cont} = 0.0013$ Jy beam$^{-1}$) to the maximum continuum flux at 0.2436 Jy beam$^{-1}$ (maximum continuum flux in 4B$'$ is 0.0958 Jy beam$^{-1}$). Red circles are peak continuum pixels selected for spectral line analysis. The synthesized beam is shown in the lower left corner and has a physical diameter of 438 AU at 1.5$''$.}
    \label{fig:continuum-image}
\end{figure}

Figure \ref{fig:continuum-image} shows the continuum emission of IRAS 4B (right) and IRAS 4B$'$ (left) at 147 GHz, labeled in accordance with previous studies, at a separation of about 11$''$  consistent with previous observations that spatially resolve the two sources \citep{di_francesco_2001,jorgensen_prosac_2007,choi-2011,de_simone_seeds_2020,prodige-2026}. The peak continuum flux is centered on IRAS 4B with a value of 0.2436 Jy beam$^{-1}$, while the fainter 4B$'$ has a peak continuum flux of 0.0958 Jy beam$^{-1}$. \cite{jorgensen_prosac_2007} studied the continuum of IRAS 4B at 230 GHz, which showed similar results as presented here, with 4B having much stronger emission. The continuum emission from 4B is approximately 2.5 times that of 4B$'$ found in this work, as compared to the ratio of 2.7 found by \citet{jorgensen_prosac_2007}. As the flux ratios between 230 and 147 GHz in IRAS 4B and 4B$'$ are similar, there is no evidence for free free contamination in 4B$'$. Finally, the source sizes in this work are larger than the synthesized beam size, indicating that emission is spatially resolved and no beam coupling correction factor is needed when deriving molecular column densities.

\section{Results \& Discussion} \label{sec:results-and-discussion}
\subsection{Results: Line Identification and Analysis}\label{subsec:line-id-analysis}
Line identification and analysis were conducted in a nearly identical fashion as \cite{will_w3_2023} and \cite{Morgan-2025}, with the only difference being a change in source and hence molecules included in the fit. The Global Optimization and Broadband Analysis Software for Interstellar Chemistry (GOBASIC) was used to fit molecular emission in each of the low-resolution sidebands \citep{rad_gobasic_2016}. GOBASIC assumes local thermodynamic equilibrium (LTE) conditions, optically thin molecular transitions, and fits Gaussian line shapes to simultaneously fit multiple molecules and components to account for line blending instead of fitting each spectral feature individually in broadband spectra. GOBASIC generates the model spectrum using a rotational partition function and a modified population diagram equation from \citet{Goldsmith1999}. The model is then compared to the observational data by a least-squares optimization where the residuals of the four dimensional parameter vector are minimized. These four parameters are determined for each molecule and molecular component: rotational temperature (K), column density (molecules cm$^{-2}$), spectral line width full width at half-maximum (FWHM in km s$^{-1}$), and velocity shift of the emission lines with respect to with respect to local standard of rest (km s$^{-1}$). Uncertainty of the optimized parameters are determined by calculating the curvature of the parameter matrix summed on all observational data points with the assumption that the fitting function at all points has same distribution of error (see Appendix A of \citet{rad_gobasic_2016}).

GOBASIC imports spectral line catalogs mainly from databases such as the Cologne Database for Molecular Spectroscopy (CDMS) \citep{cdms-muller-2001,cdms-mullner-2005} and the Jet Propulsion Laboratory (JPL) Spectral Line Catalog \citep{jpl-PICKETT-1998} to conduct a broadband spectral line analysis by comparing the observed spectra to multiple molecules cataloged in these databases. These catalogs provide the partition function and transition frequencies for pure rotational or rovibrational lines of molecules, many of which are found in interstellar or circumstellar environments. This information is vital for astronomers to accurately detect molecules and model the physical and chemical conditions in these environments. If multiple catalog versions exist for the same molecule, the catalog with the most recently updated values is used. GOBASIC uses the provided catalog information, truncates the data to the given frequency range, and simulates a predicted spectrum to compare with observations from user-defined initial guesses of calculated parameters. The interpolated partition function for the predicted spectra uses the form in Equation \ref{equation:partition-function} as seen in \cite{cso}: 

\begin{equation}
    \label{equation:partition-function}
    Q(T)=\alpha T ^{\beta} [\gamma + exp(\frac{\epsilon}{T})]
\end{equation}

 Updated or new interpolated partition functions for molecules not reported in previous works (\citealt{cso,Morgan-2025}) are listed in Table \ref{tab:par-coefficients}. The interpolated partition function coefficients $\alpha$, $\beta$, $\gamma$, and $\epsilon$ from Equation \ref{equation:partition-function} are determined by fitting this equation to the partition function values given in the catalog documentation. The percent error comparing the interpolated partition function used and the tabulated catalog partition function values is described by Equation \ref{equation:percent-error} and is listed in Table \ref{tab:par-coefficients}. \ch{CH3OH} ($v_t$=0--2), $cis$-\ch{CH2OHCHO} ($v_t$=0), and ethanol ($gauche/anti$ combined \ch{C2H5OH}) all have updated catalog partition functions and associated interpolated coefficients from previous works, so their updated values used in this work are listed in Table \ref{tab:par-coefficients}. 

\begin{equation}
    \label{equation:percent-error}
    \text{Percent Error \%} =  |\frac{Q(T)_{interpolated} - Q(T)_{catalog}}{Q(T)_{catalog}}| \times 100
\end{equation}

\begin{deluxetable}{cccccccc}[h]
    \tablecaption{Interpolated partition function coefficients used in GOBASIC spectral analysis. \label{tab:par-coefficients}}
    \tablehead{\colhead{\textbf{Molecule}} & \colhead{\textbf{$\alpha$}} & \colhead{\textbf{$\beta$}} & \colhead{\textbf{$\gamma$}} & \colhead{\textbf{$\epsilon$}} & \colhead{\textbf{Percent Error (\%)}} & \colhead{\textbf{Database}} & \colhead{\textbf{Date of Entry}}}
    \startdata
        \ch{CH3OD} & 1.5835 & 1.6738 & 0.4241 & -565.7370 & [0.0085, 15.58] & CDMS & Aug. 2024 \\
        \ch{CH3OH} & 2.7025 & 1.6690 & 0.7400 & -392.990 & [0.40, 11.72] & CDMS & May 2016 \\
        \ch{CH2OHCHO} & 6.9077 & 1.5 & 0 & 0 & [0.0077, 2.69] & CDMS & June 2021 \\
        \ch{C2H5OH} &6.7212&1.49877&0.496432&-60.0823&[0.00085, 2.88]&CDMS& Nov 2016\\
    \enddata
    \tablecomments{The range in percent error from tabulated catalog partition function values and interpolated partition function values is listed as percent error [min, max] in units of percent.}
\end{deluxetable}

\begin{figure}[th]
    \centering
    \includegraphics[width=0.89\linewidth]{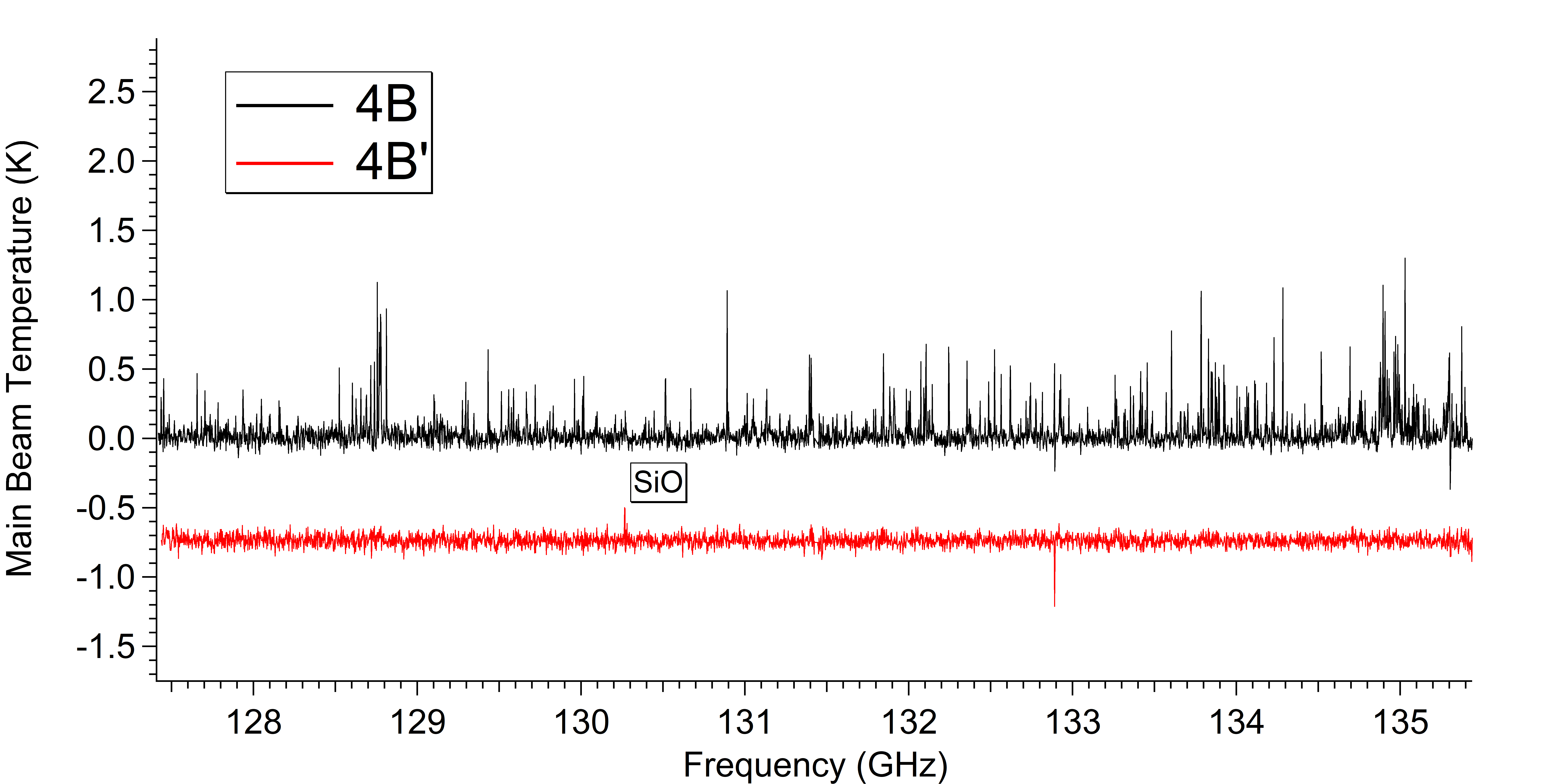}
    \includegraphics[width=0.89\linewidth]{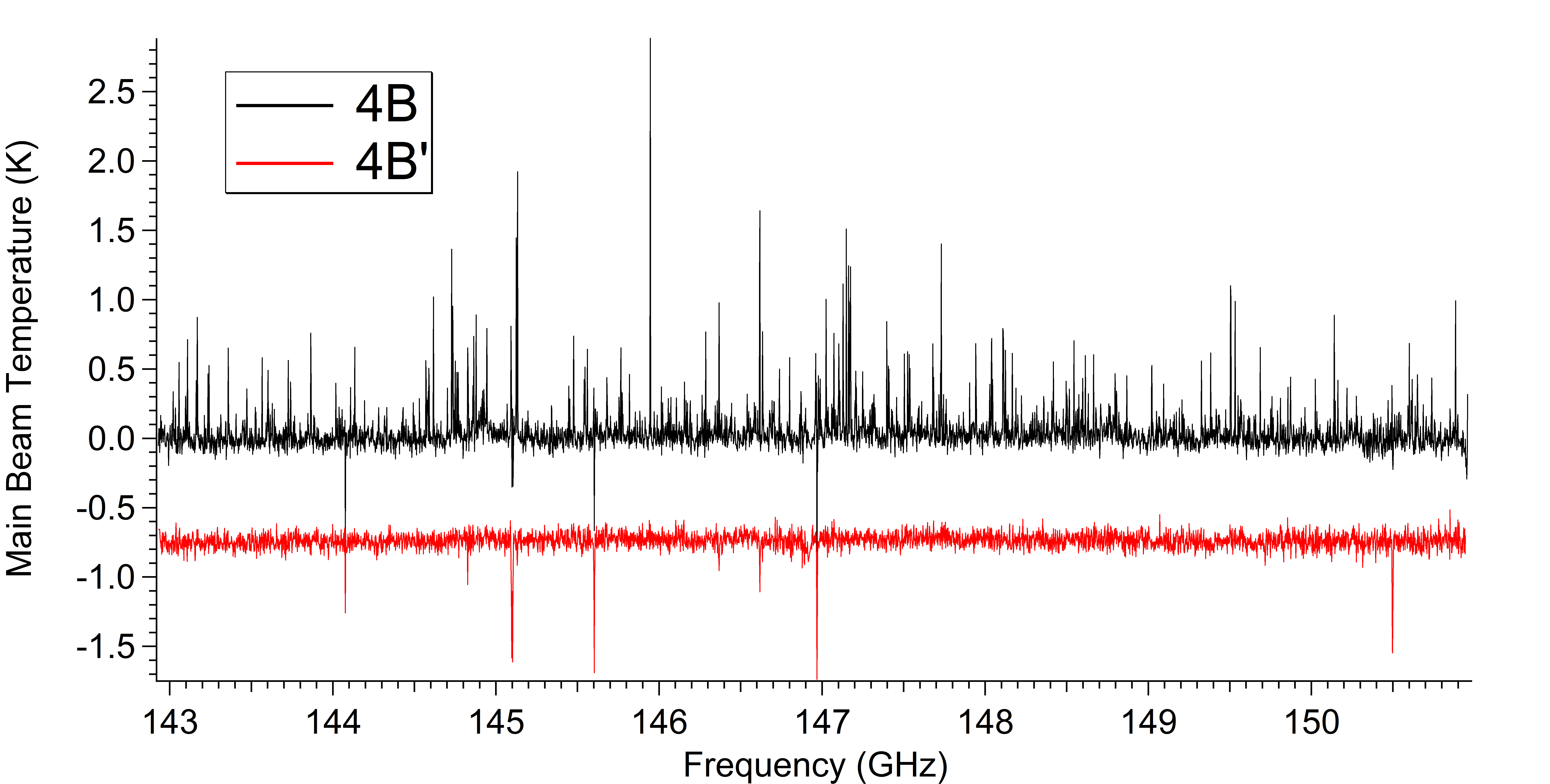}
    \caption{Extracted spectra of IRAS 4B (black) and 4B$'$ (red) for the lower sideband (top) and the upper sideband (bottom). Spectra are offset for clarity, and the one emission line of SiO in 4B$'$ is labeled.}
    \label{fig:extracted-lsb-usb-rc-lc}
\end{figure}
\vspace{-24pt}
To begin the line analysis, the spectra from peak continuum pixels toward both IRAS 4B and 4B$'$ were extracted from each of the sidebands. The positions of these extracted pixels are listed in Table \ref{tab:ra-dec-continuum-core} and are indicated in Figure \ref{fig:continuum-image} by two red circles. The extracted spectra, see Figure \ref{fig:extracted-lsb-usb-rc-lc}, were baseline corrected with a first-order polynomial using the python library pybaselines\footnote{\url{https://github.com/derb12/pybaselines}} using the imodpoly algorithm \citep{pybaselines}. Hundreds of molecular transitions are seen in IRAS 4B (black spectrum), while only a single emission line of SiO at 130268.7 MHz is present in the spectrum toward the core of IRAS 4B$'$ (red spectrum). Based on the flux density ratio between IRAS 4B and 4B$'$ and the rms values listed in Table \ref{tab:table-summary-hr-spw}, molecular emission lines were expected to be observed in the spectrum towards IRAS 4B$'$. The fact these emission lines were not present in the extracted spectrum suggests a low molecular abundance towards IRAS 4B$'$. Additionally, some absorption lines were observed toward both cores and are discussed below. Thus the spectral analysis presented here focuses on IRAS 4B. GOBASIC is capable of fitting the emission of all molecules present in a spectrum simultaneously. Therefore, the spectra of the LSB and USB were combined and any absorption features and the intermediate frequency between the sidebands were ignored for the purpose of spectral analysis. Matches to transitions for molecules previously observed in IRAS 4B with the CSO in the 223--251 GHz range were tried first \citep{cso}, then detections of other molecules included in other studies such as \citealt{blake_1995,di_francesco_2001,sakai_detection_2006,jorgensen_prosac_2007,sakai-ch3oh-2012,koumpia_2016,busch_prodige_2025} and \citet{prodige-2026} were investigated. Lines that were still unidentified were then cross referenced with known lines from the CDMS and JPL spectral databases. Every reasonable molecule that had previously been detected in either the ISM or a circumstellar environment with a rest frequency that fell within a window of $\pm$3 MHz of a transition observed here was analyzed using GOBASIC. Each molecular fit was constrained to be within a reasonable set of physical parameters for a hot corino. Resultant fits were individually evaluated based on visual matches to the observed line widths and relative intensities to determine if the prediction represented the data well. If so, the best fit determined via least squares analysis was used in subsequent analyses.

\begin{deluxetable}{ccc}[th]
    \tablecaption{Right ascension and declination (J2000) of the peak continuum flux pixel of IRAS 4B and 4B$'$ used for spectral extraction and analysis.\label{tab:ra-dec-continuum-core}}
    \tablehead{\colhead{\textbf{Continuum Core}}  & \colhead{\textbf{Right Ascension (J2000)}} & \colhead{\textbf{Declination (J2000)}}}
    \startdata
    IRAS 4B & 03$^h$29$^m$12$^s$.02 & 31:13$'$:08$''$.12 \\
    IRAS 4B$'$ & 03$^h$29$^m$12$^s$.85& 31:13$'$:06$''$.97 \\
    \enddata
\end{deluxetable}

\begin{figure}[h]
    \centering
    \includegraphics[width=0.98\linewidth]{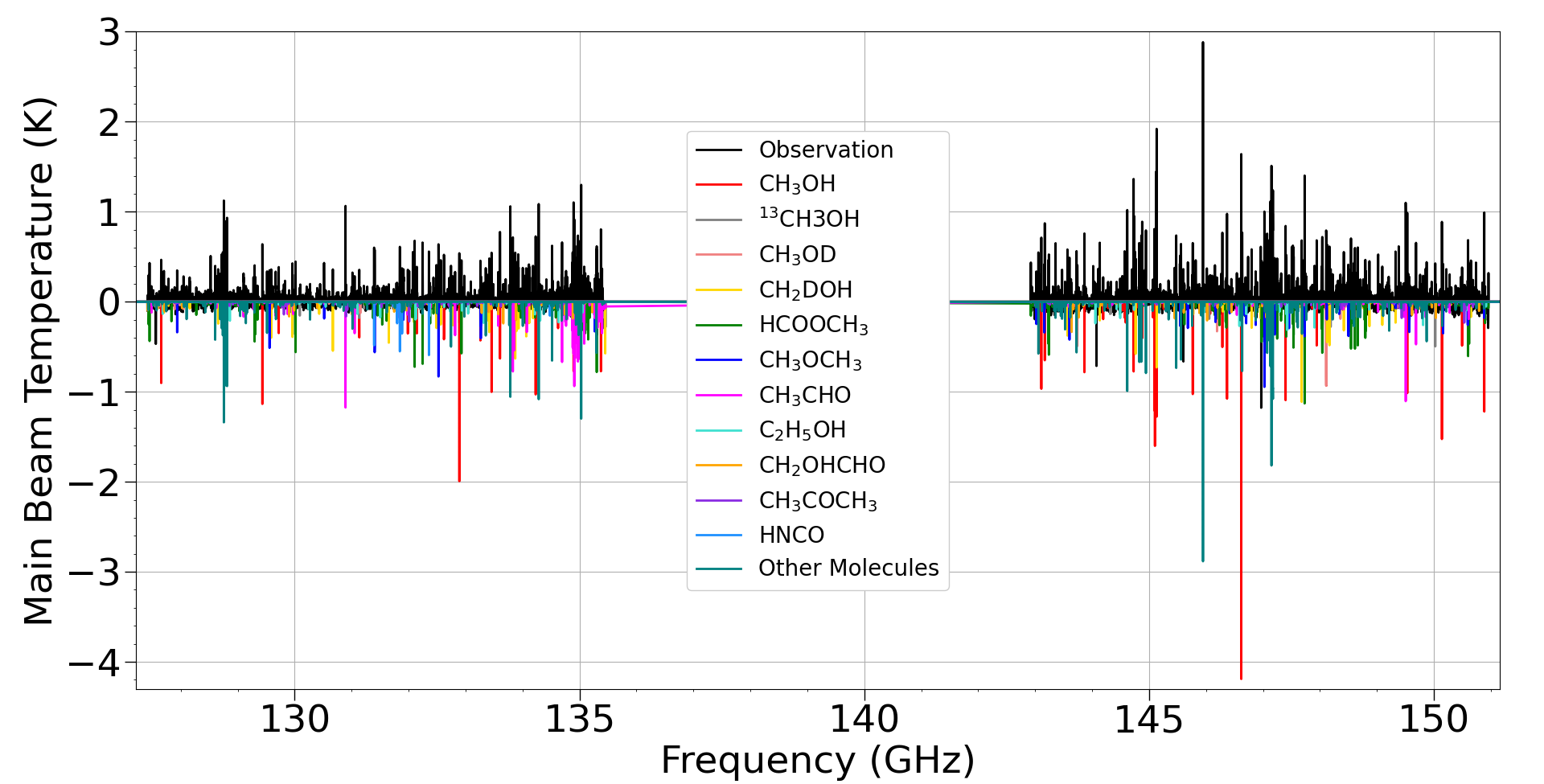}
    \includegraphics[width=0.47\linewidth]{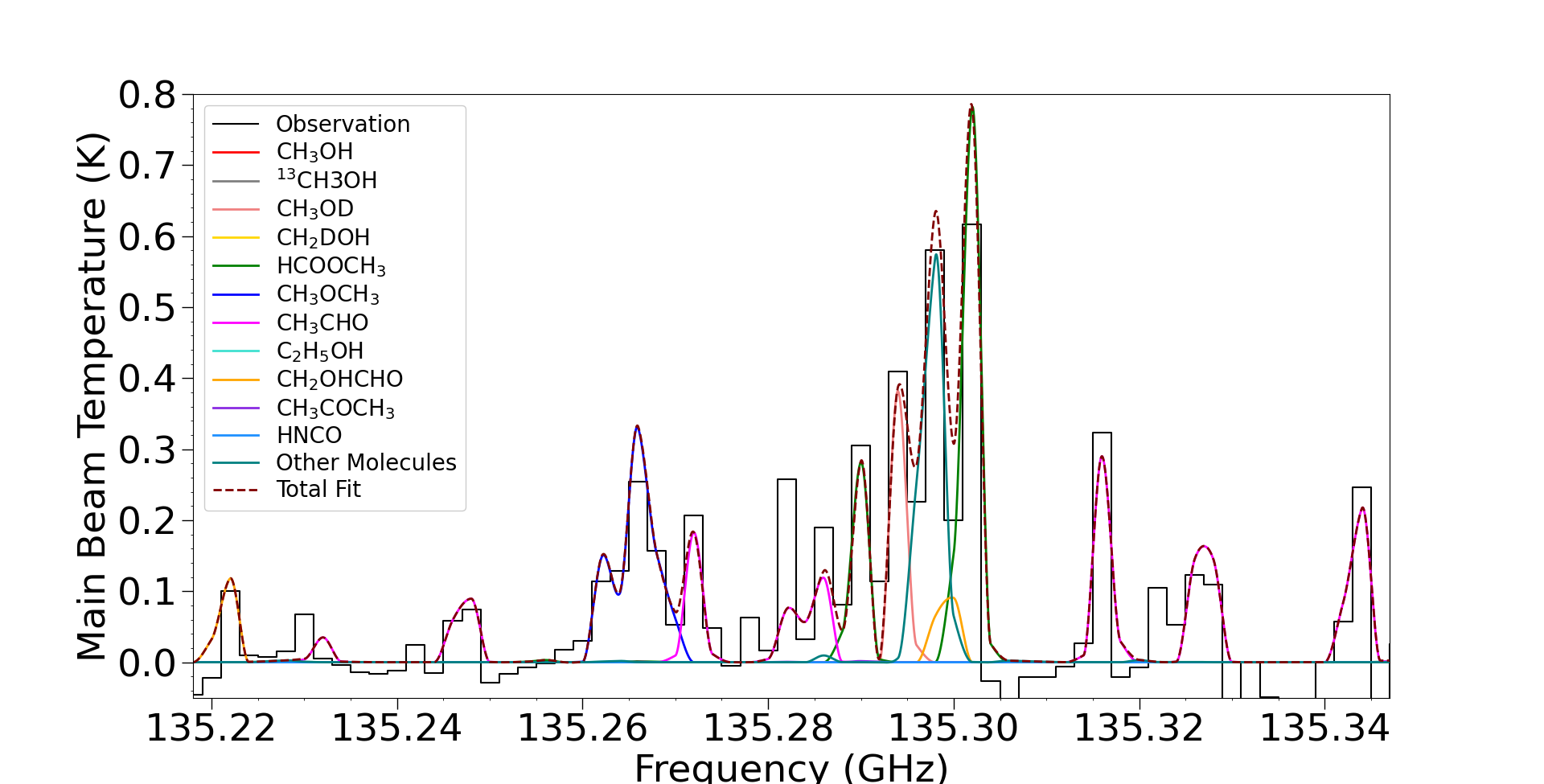}
    \includegraphics[width=0.47\linewidth]{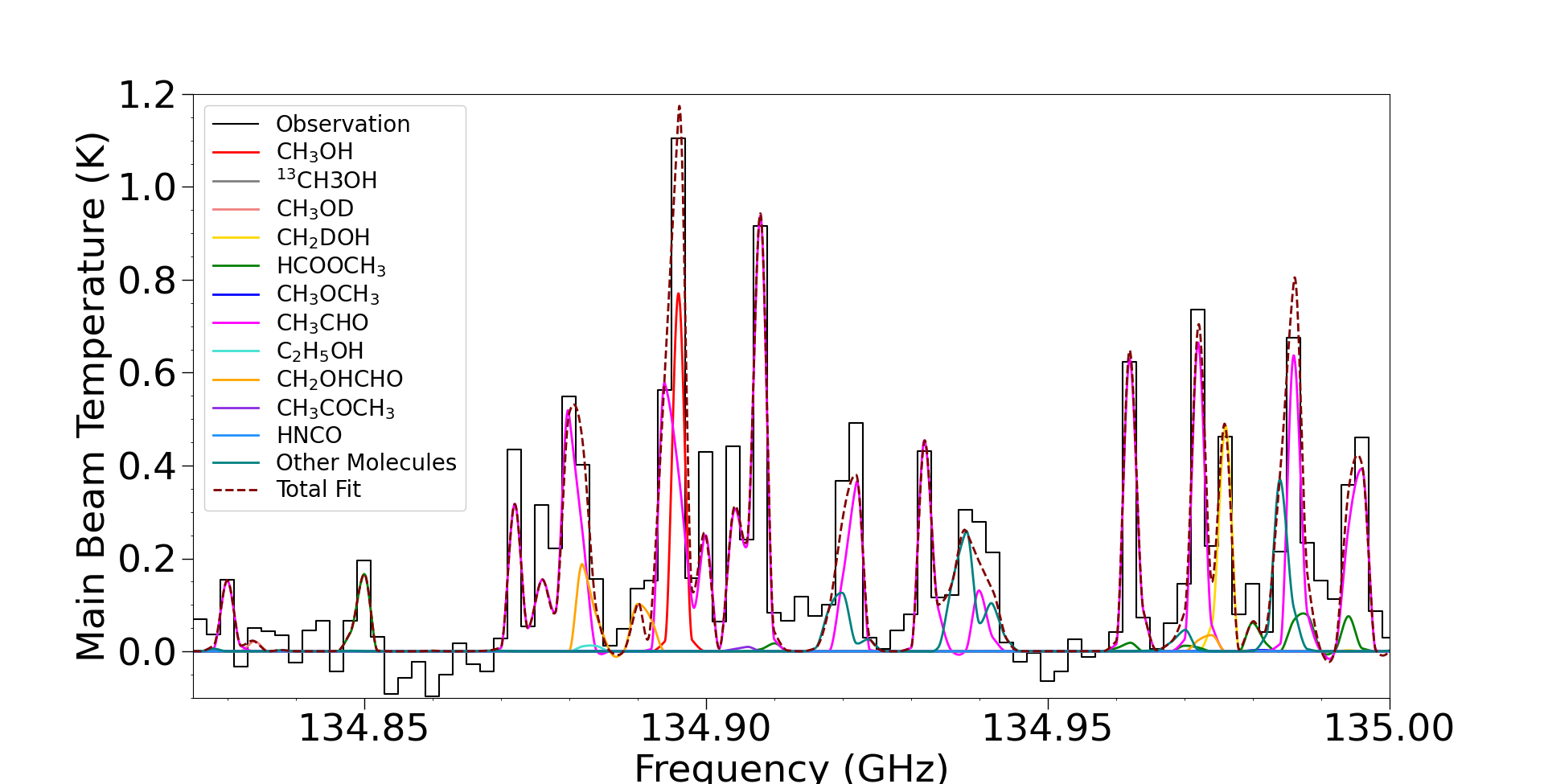}
    \includegraphics[width=0.47\linewidth]{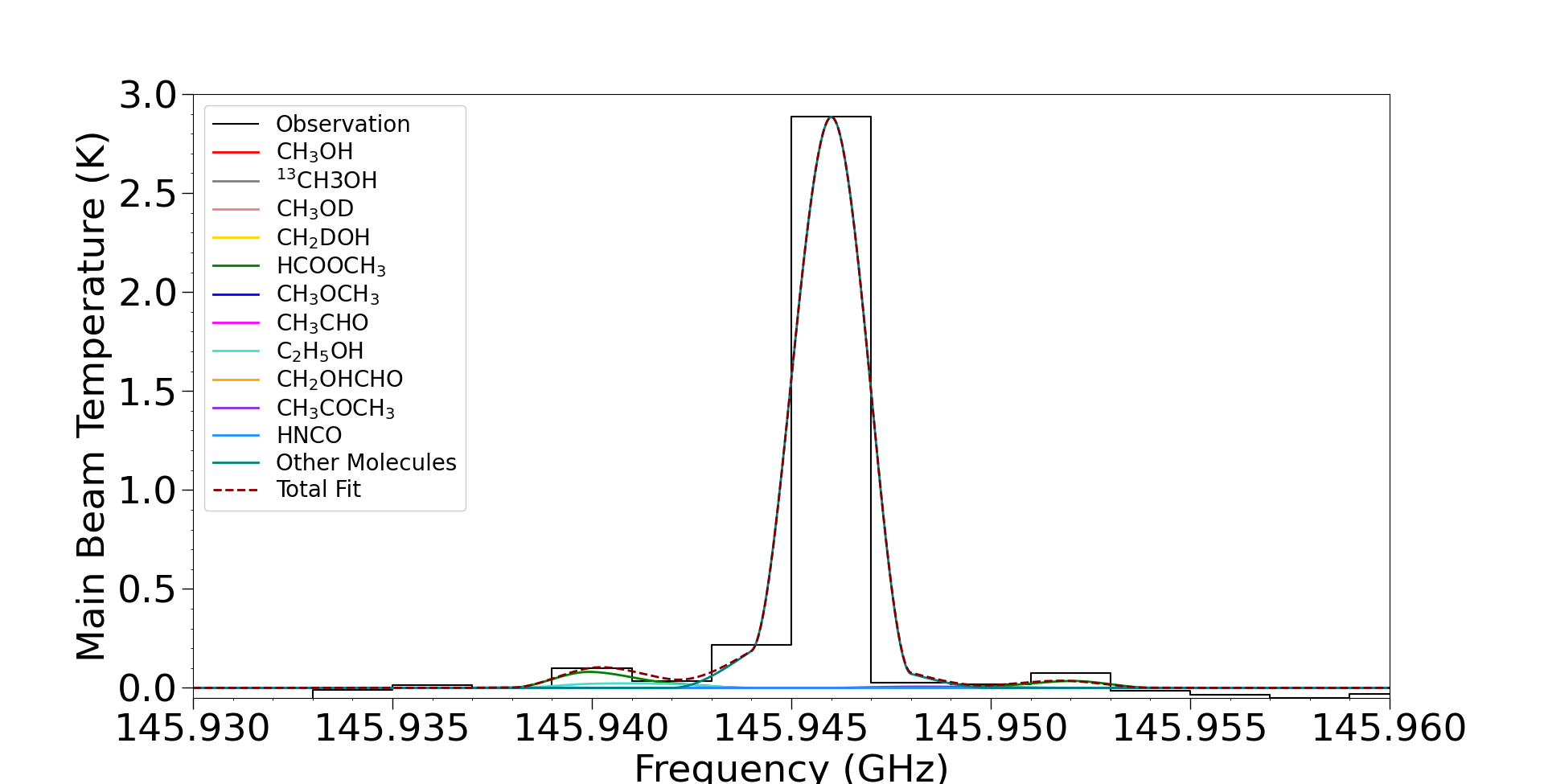}
    \includegraphics[width=0.47\linewidth]{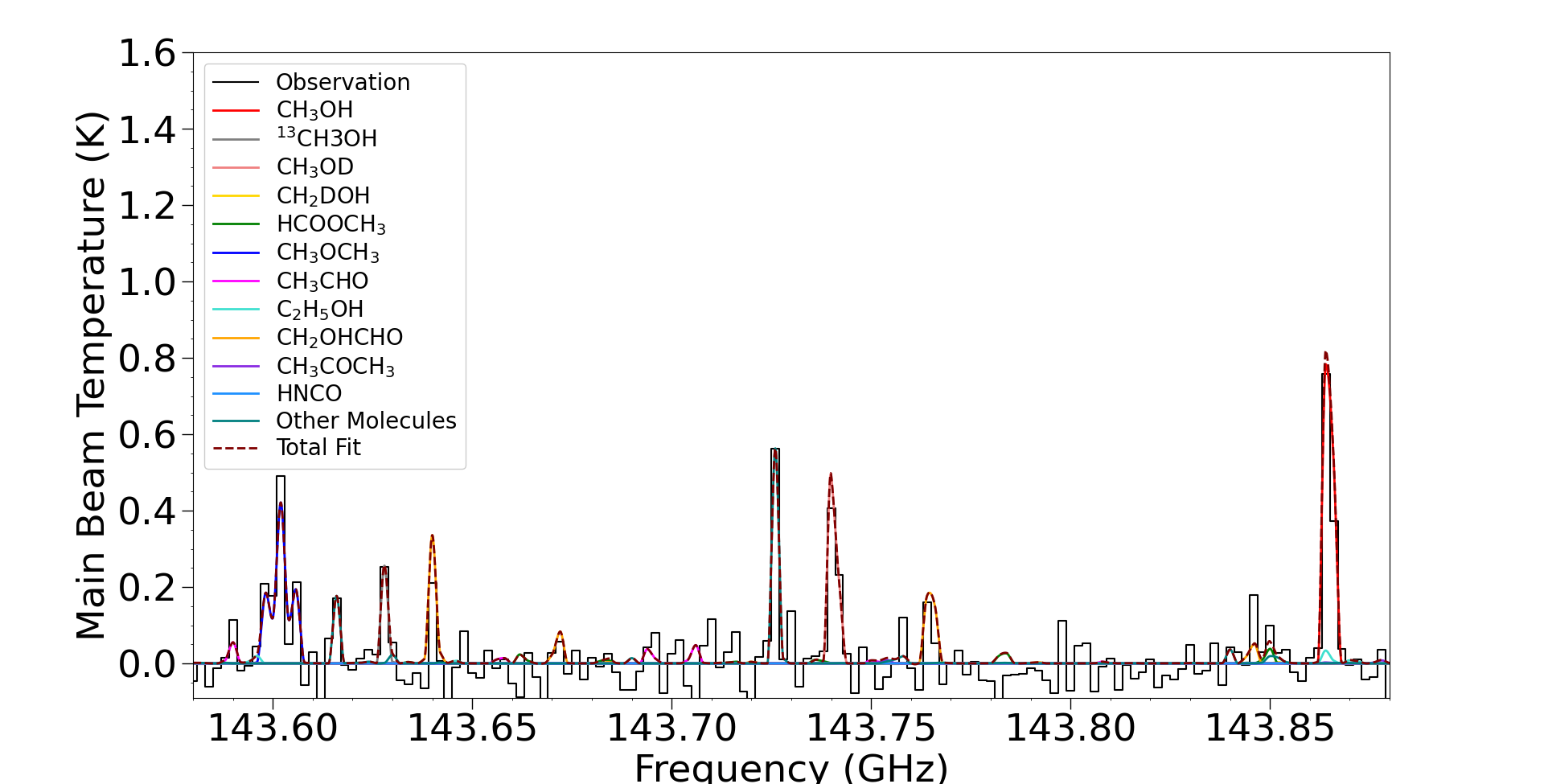}
    \caption{(Top) GOBASIC spectral fit of IRAS 4B containing 11 well fitting molecules. The extracted spectrum is in black, while the fitted spectrum is reflected on the y-axis. Methanol (\ch{CH3OH}) in red is included for completeness, but note numerous transitions are optically thick or are not in local thermodynamic equilibrium. (Bottom) Zoomed-in panels of selected regions show the reliability of the fit. Under-predicted lines may indicate additional unknown molecules that are not currently included.}
    \label{fig:4B-fit}
\end{figure}

A total of 21 molecules and isotopologues has been detected with at least three lines over a 3 rms threshold toward the inner envelope surrounding IRAS 4B. This includes  \ch{CH3OH} and its isotopologues \ch{CH3OD}, \ch{^{13}CH3OH}, \ch{CH2DOH} and \ch{CH3^{18}OH}, \ch{CH3OCH3}, \ch{HCOOCH3}, \ch{CH3CHO}, a methyl formate isotopologue (\ch{^{13}CH3CHO}), \ch{C2H5OH}, \ch{CH2OHCHO}, \ch{HNCO}, acetone (\ch{CH3COCH3}), singly deuterated formaldehyde (\ch{HDCO}), sulfur dioxide (\ch{SO2}), \ch{CH3CN} ($v_t$=0 and $v_t$=8), $trans$-formic acid (\ch{$trans$-HCOOH}), \ch{C2H5CN}, formamide (\ch{NH2CHO}), and methoxymethanol (\ch{CH3OCH2OH}). Still more molecular carriers are present with lines over a 3 rms detection but with less than three lines within the spectral region to claim a firm detection. If these molecules have lines that fall within the high-resolution spectral windows, they are referenced in Subsection \ref{subsec:highres}. Out of these molecules, physical parameters were determined for 11 molecules that had numerous lines that allowed for a full analysis including \ch{CH3OH}, \ch{^{13}CH3OH}, \ch{CH3OD}, \ch{CH2DOH}, \ch{HCOOCH3}, \ch{CH3OCH3}, \ch{CH3CHO}, \ch{C2H5OH}, \ch{CH2OHCHO}, \ch{CH3COCH3}, and \ch{HNCO}. For other molecules detected in this region, calculating physical parameters proved difficult due to too few unblended lines with intensities above 3 rms. These molecules were fit iteratively in the same way as described above with fixed parameters based on other molecules. The calculated parameters of rotational temperature, column density, spectral line width, and velocity shift for the molecules that were fit are listed in Table \ref{tab:gobasic-table} and the simulated spectrum including those molecules compared to the observed is shown in Figure \ref{fig:4B-fit}. \footnote{The extracted spectrum towards the peak continuum pixel of IRAS 4B and the GOBASIC fit is available on Zenodo at 10.5281/zenodo.24118600.}

\begin{deluxetable}{ccccc}[h]
        \tablecaption{GOBASIC calculated physical parameters of 11 molecules in IRAS 4B with associated error.\label{tab:gobasic-table}}
        \tablehead{\colhead{\textbf{Molecule}} & \colhead{\textbf{Temperature}}& \colhead{\textbf{Column Density}} & \colhead{\textbf{FWHM}} & \colhead{\textbf{Velocity Shift}} \\
        \colhead{}    &  \colhead{(K)} &  \colhead{(cm$^{-2}$)} &  \colhead{(km s$^{-1}$)} &  \colhead{(km s$^{-1}$)}
        }
        \startdata
    \ch{CH3OH} & $270\pm5$ & $(2.62\pm0.07) \times 10^{16}$ & $3.94\pm0.08$ & $1.93\pm0.03$ \\
    \ch{^{13}CH3OH} & 210$\pm40$ & (3.8$\pm0.6) \times 10^{15}$ & $3.1\pm0.4$ & $2.0\pm0.1$ \\
    \ch{CH3OD} & $260\pm40$ & $(4\pm1) \times 10^{15}$ & $3.5\pm0.2$ & $1.86\pm0.10$ \\
    \ch{CH2DOH} & $201\pm8$ & $(1.28\pm0.06) \times 10^{16}$ & $3.9\pm0.1$ & $1.65\pm0.05$ \\
    \ch{HCOOCH3} & $210\pm10$ & $(7.4\pm0.6) \times 10^{15}$ & $4.2\pm0.1$ & $1.91\pm0.05$ \\
    \ch{CH3OCH3} & $111\pm4$ & $(7.9\pm0.4) \times 10^{15}$ & $4.2\pm0.2$ & $1.96\pm0.09$ \\
    \ch{CH3CHO} & $250\pm10$ & $(3.9\pm0.3) \times 10^{15}$ & $4.4\pm0.1$ & $2.07\pm0.06$ \\
    \ch{C2H5OH} & $140\pm10$ & $(2.7\pm0.3) \times 10^{15}$ & $4.5\pm0.4$ & $1.9\pm0.2$ \\
    \ch{CH2OHCHO} & $190\pm20$ & $(6.2\pm0.8) \times 10^{14}$ & $4.0\pm0.4$ & $1.8\pm0.2$ \\
    \ch{CH3COCH3} & $180\pm30$ & $(1.4\pm0.9) \times 10^{15}$ & $5.4\pm0.8$ & $1.5\pm0.4$ \\
    \ch{HNCO} & $320\pm60$ & $(9\pm2) \times 10^{14}$ & $4.5\pm0.3$ & $1.7\pm0.2$ \\
    \enddata
    \tablecomments{\ch{CH3OH} is included for completeness since it was required to provide an adequate fit for the other molecules, but is optically thick and thus these values are not physically accurate.}
    \vspace{-20pt}
\end{deluxetable}

Although this line analysis focuses on molecular emission, Figure \ref{fig:extracted-lsb-usb-rc-lc} shows that there is molecular absorption within the observed spectral range. Based on the low energy of these transitions, these features come from the lower density envelope surrounding the YSOs. Three of these absorption features in IRAS 4B with inverse P Cygni line profiles are from \ch{H2CO} and \ch{CS} at 145602.9, 150498.3, and 146969.0 MHz, which has been observed from the same species in a previous study \citep{di_francesco_2001}. These absorption features are also seen in IRAS 4B$'$, albeit without the emission to make it a P Cygni line profile. A single weak absorption line of \ch{DCN} is observed in 4B$'$ at 144828.0 MHz while seen in emission instead in 4B. An unknown absorption feature is located at 144077.3 MHz in both cores with a probable identification as \ch{DCO+}. The rest of these absorption features are from \ch{CH3OH}, the majority with inverse P Cygni line profiles in 4B and Gaussian absorption line shapes in 4B$'$. The Tables in Section \ref{subsec:highres} make note of these features if these transitions fall within the high-resolution spectral windows. 

It became clear during the fitting process towards the core of IRAS 4B that most of the \ch{CH3OH} transitions present in the spectral range were optically thick, as the best fit predicted spectrum consistently over-predicted the observed spectrum. A consistent over-prediction of line intensities signifies that some transitions should be stronger than they appear in the spectrum, indicating that these transitions are optically thick. GOBASIC assumes transitions are optically thin and under LTE conditions, so calculated parameters for \ch{CH3OH} toward the envelope of IRAS 4B are not physically accurate. However, for the purposes of this work, \ch{CH3OH} was included in the fit to account for its line blending with other molecules. It should be noted that \cite{busch_prodige_2025} has completed spectral-line analysis under LTE, with optically thin(ner) lines, and with radiative transfer considerations for \ch{CH3OH}, \ch{CH3CN}, and their isotopologues toward IRAS 4B using NOEMA around 214.7--222.8 GHz (LSB) and 230.2--238.3 GHz (USB). They report a column density for \ch{^{13}CH3OH} of $(8\pm1)\times10^{16}$ cm$^{-2}$, which is an order of magnitude greater than that reported here, but with the same rotational temperature within uncertainties. \cite{busch_prodige_2025} used an optical depth correction calculated from Weeds \citep{weeds}, originating from iterating between the spectral-line modeling software and derived population diagrams \citep{Goldsmith1999}. They excluded blended lines with multiple molecular components besides the molecule of interest, under-predicted lines, and may have had unresolved \ch{^{13}CH3OH} emission that could affect column density calculations (the low spectral resolution channel width is 2 MHz; lines with FWHM $\leq$ channel width would be unresolved in their spectral region). Given that their analysis did not account for line blending, it is possible that their analysis gave a column density that is higher than the actual value because blended lines give an overestimate of integrated intensity. Conversely, the calculations here could be underestimating column density due optically thick transitions, unresolved substructures, or an overestimation of source size due to limited angular resolution \citep{busch_prodige_2025}. The \ch{^{13}CH3OH} was repeatably analyzed with different initial conditions and with suspected optically thick transitions removed, and always returned the same column density within uncertainty as the one listed in Table \ref{tab:gobasic-table}. The presence of unresolved substructures is a possible avenue of discussion and will be explored later in section \ref{sec:results-and-discussion}. Follow up higher-resolution broadband observations are needed to enable a thorough spectral analysis. Once available, a full treatment of line blending and optical depth corrections should be conducted.

Of the molecules included in Table \ref{tab:gobasic-table} and Figure \ref{fig:4B-fit}, all but \ch{HNCO} are iCOMs. The calculated FWHM and velocity shift values toward 4B are rather uniform; FWHM values range from $3.1 \pm 0.4$ to $5.4 \pm 0.8$ km s$^{-1}$ and velocity shift values range from $1.65 \pm 0.05$ to $2.07 \pm 0.06$ km s$^{-1}$. The highest column density is \ch{CH2DOH} at $(1.29 \pm 0.06) \times 10^{16}$ cm$^{-2}$, which is expected since \ch{CH3OH} is optically thick. The lowest column density is \ch{CH2OHCHO} at $(6.2 \pm 0.8) \times 10^{14}$ cm$^{-2}$. Temperatures in the core range from $111 \pm 4$ K for \ch{CH3OCH3} to $320 \pm 60$ K for \ch{HNCO}.  Additionally, there are  numerous outflows surrounding 4B and 4B$'$ with two molecules with consistent lines above 3 rms: \ch{CH3OH} and \ch{H2CO}. These fits are discussed in more detail in the next section.

\subsection{Results: Physical Parameter Maps}\label{subsec:parameter-maps}

In order to determine the physical and chemical conditions across the region of IRAS 4B and 4B$'$, parameter maps of temperature, column density, and velocity shift of well fitted molecules were made. To do this, the spectra from each sideband were extracted, combined, and baseline corrected to be analyzed with GOBASIC as described above and in \cite{will_w3_2023} and \cite{Morgan-2025}. In total, 17,573 pixels were iteratively fit with GOBASIC to determine physical parameters. This was done in two phases: 1) in the region surrounding 4B with the 11 molecules from Table \ref{tab:gobasic-table} and 2) in the regions surrounding 4B and 4B$'$ for \ch{CH3OH} and \ch{H2CO} in order to map the outflows.

\begin{figure}[ht]
    \centering
    \includegraphics[width=0.49\linewidth]{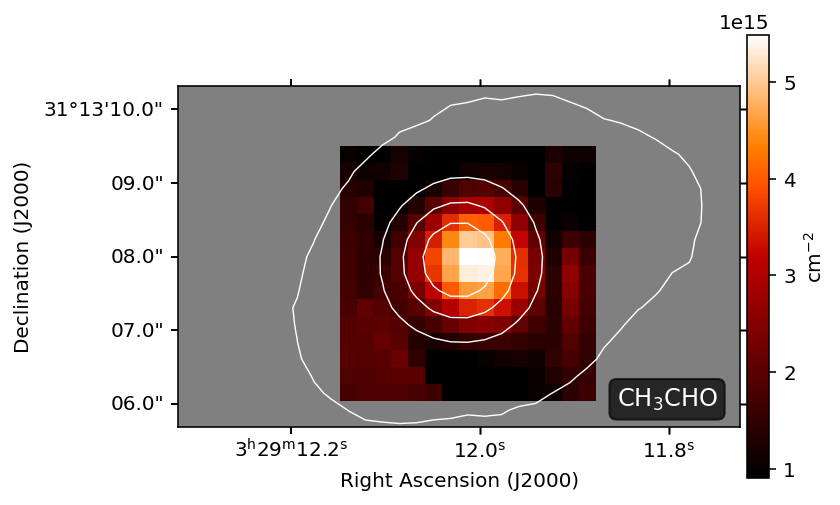}
    \includegraphics[width=0.49\linewidth]{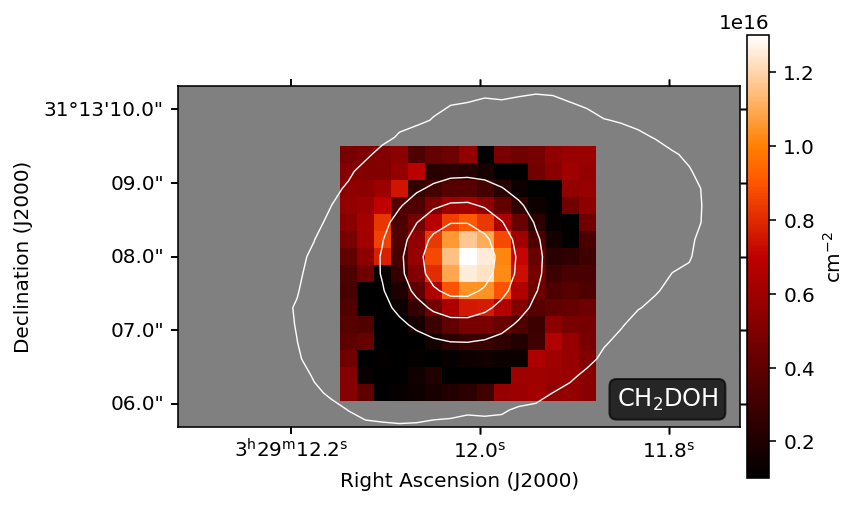}
    \includegraphics[width=0.49\linewidth]{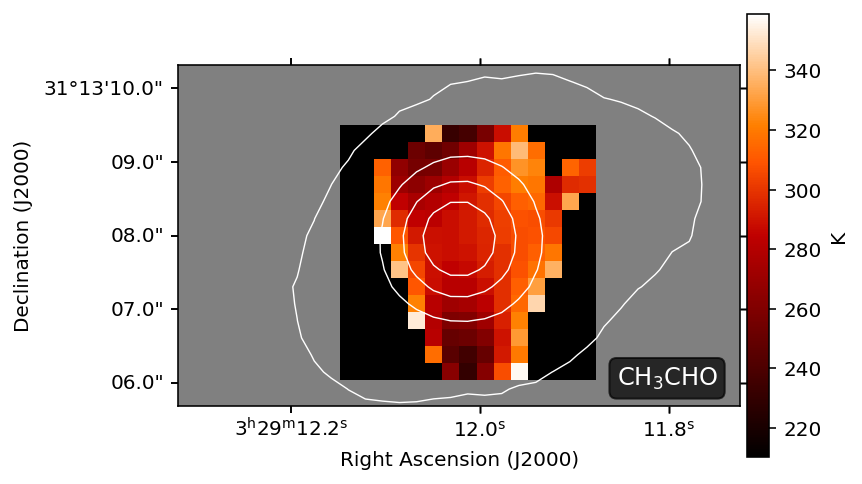}
    \includegraphics[width=0.49\linewidth]{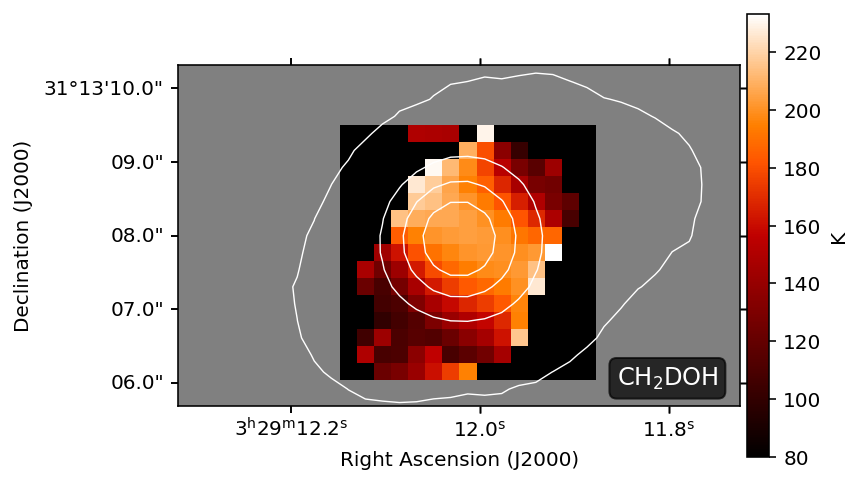}
    \includegraphics[width=0.49\linewidth]{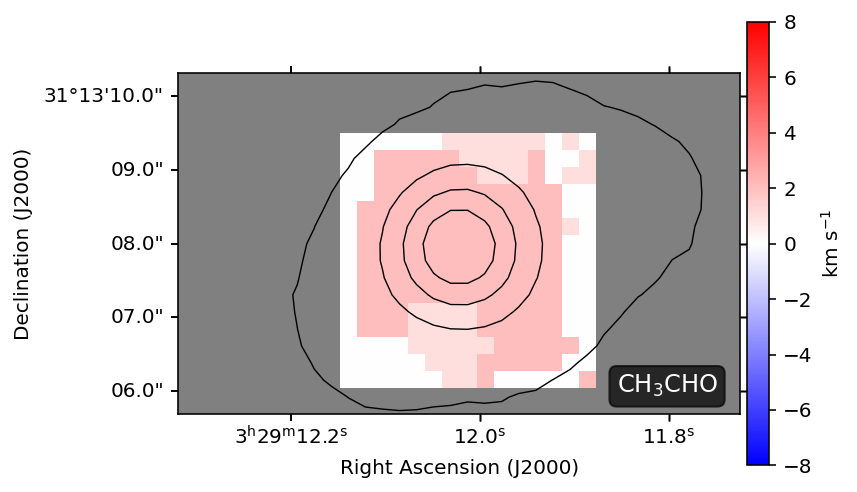}
    \includegraphics[width=0.49\linewidth]{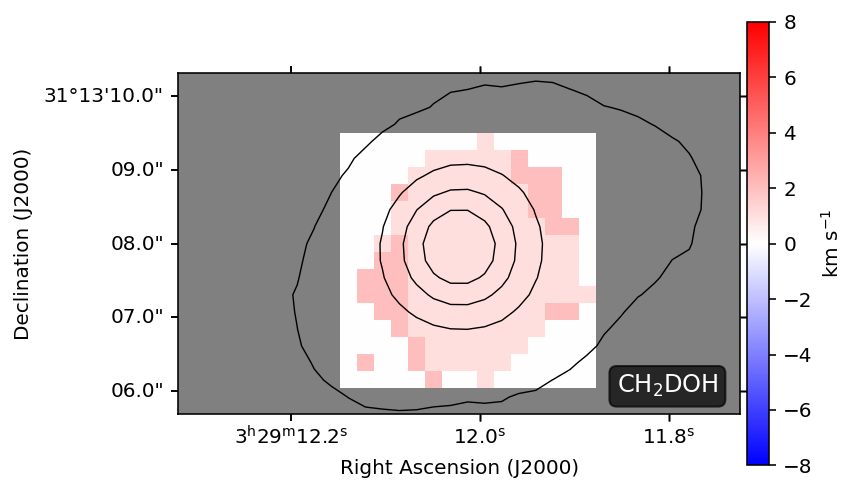}
    \caption{Parameter maps of \ch{CH3CHO} (left) and \ch{CH2DOH} (right) toward IRAS 4B, in order from top to bottom: column density (cm$^{-2}$), rotational temperature (K), and velocity shift (km s$^{-1}$). White and black contours are 5 evenly spaced levels (1 level $=0.0599$ Jy beam$^{-1}$) from $3 \times \sigma_{cont}$ ($\sigma_{cont} = 0.0013$ Jy beam$^{-1}$) to the max continuum flux at 0.2436 Jy beam$^{-1}$. }
    \label{fig:parameter-maps-4b}
\end{figure}

Figure \ref{fig:parameter-maps-4b} shows the parameter maps created for \ch{CH3CHO} and \ch{CH2DOH}, two of the best constrained molecules out of the 11 from Table \ref{tab:gobasic-table}. The parameter maps for \ch{^{13}CH3OH}, \ch{CH3OD}, \ch{HCOOCH3}, \ch{CH3OCH3}, \ch{C2H5OH}, \ch{CH2OHCHO}, and \ch{HNCO} are found in Appendix \ref{sec:appendix-parameter-maps}. The associated uncertainty for each calculated parameter across all pixels for all maps is also found in Appendix \ref{sec:appendix-parameter-maps}. Uncertainty values correspond to derived parameter divided by the determined uncertainty plotted on a logarithmic scale, with a higher logarithmic value corresponding to higher ratio between the derived value and the uncertainty. To save computational time, the other molecules that are present in the spectra but do not have enough lines above 3 rms for a robust fit were not included in the fitting for the parameter maps.  Because of this, the uncertainty values were greater than calculated values for most pixels evaluated for \ch{CH3COCH3} from Table \ref{tab:gobasic-table} and were not reported here. \ch{CH3OH} is also not included in Figure \ref{fig:parameter-maps-4b} or Appendix \ref{sec:appendix-parameter-maps}, but is included in the outflow parameter maps in Figure \ref{fig:parameter-maps-outflows} for completeness. Pixels that were not extracted due to too few lines under the 3 rms threshold are in gray.

Overall, the parameter maps show a peak in column density and temperature with constant velocity shift for all molecules in 4B that are almost concentric with the peak continuum flux, consistent with a protostar. The peak column density pixel matches cospatially with the peak continuum flux pixel for all molecules except \ch{CH3OD} and \ch{^{13}CH3OH}. The peak column density for \ch{CH3OD} and \ch{^{13}CH3OH} peaks one and two pixels south, respectively (1 pixel is  $\approx0.23''$) as compared to the continuum peak in Figure \ref{fig:continuum-image}. When comparing the highest column density pixels of each map to the continuum, there is a slight offset with most molecules peaking slightly to the west of the peak continuum flux, but this difference is only slight, on order of 1 pixel. For example in Figure \ref{fig:parameter-maps-4b}, the peak column density pixel for \ch{CH3CHO} and \ch{CH2DOH} is actually the same as the peak continuum flux pixel, but the cluster of four `brightest' pixels in the column density maps are shifted one column to the west as compared to the continuum peak in Figure \ref{fig:continuum-image}. It is important here to note the size of the synthesized beam is larger than these offsets, and hence shifts by a few pixels may not be meaningful. The column density maps for most molecules approximately follow the first four continuum contours, with lower densities farther away from the center of 4B. The temperature for all molecules follows the same general trend, with higher temperatures near the center of the protostar and lower temperatures farther away from the protostar. The temperature drops more significantly to the north and south, where the outflows from 4B are located. Interestingly, the temperature drops in the north and south are generally not on the same axis; in Figure \ref{fig:parameter-maps-4b} the temperature map of \ch{CH3CHO} shows the drop in temperature in a relatively north-south orientation, while \ch{CH2DOH} is shifted in a northwest-southeast orientation. This may be indicative of two offset outflows, which is further explored below. The parameter maps of velocity shift also show a constant shift around 2 km s$^{-1}$ for all molecules within 4B, similar to the fit in Table \ref{tab:gobasic-table}. 

\begin{figure}[ht]
    \centering
    \includegraphics[width=0.6\linewidth]{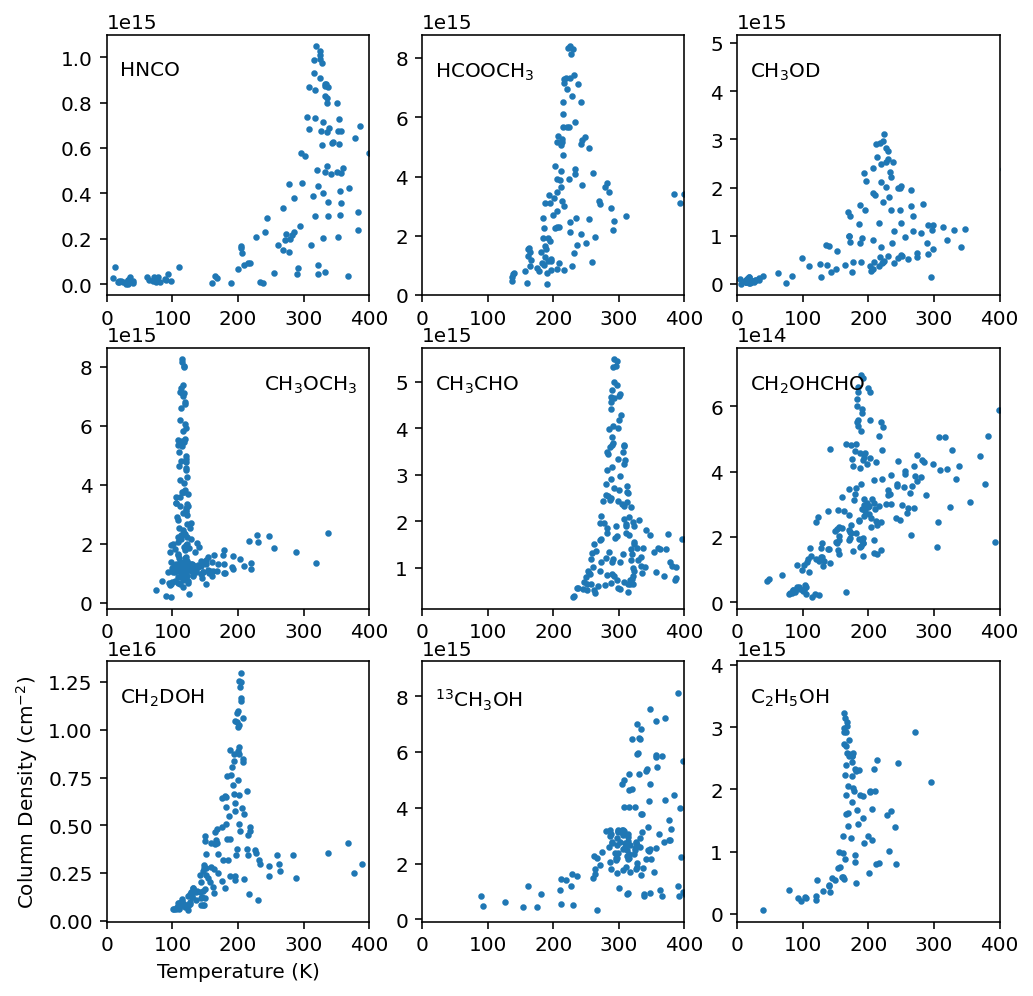}
    \caption{Plots comparing the column density (cm$^{-2}$) on the y-axis to the rotational temperature (K) on the x-axis for each molecule in IRAS 4B from derived parameter maps. For each molecule and physical parameter, the associated uncertainties are determined from the GOBASIC spectral fit and can be found in Appendix \ref{sec:appendix-parameter-maps}.}
    \label{fig:dens-vs-temp-parameter-maps}
\end{figure}

Figure \ref{fig:dens-vs-temp-parameter-maps} shows the trend between rotational temperature and column density for each molecule toward 4B. Overall, the same trend is present between rotational temperature and column density for each molecule. There is a positive relationship, with a higher rotational temperature corresponding to a higher column density, until the mid-temperature regions where there is a peak in column density, before the column density decreases at at higher temperatures. Each individual molecule in Figure \ref{fig:dens-vs-temp-parameter-maps} generally shows a single molecular component, as indicated by the  single peak distribution. The more diffuse trends in \ch{^{13}CH3OH}  and \ch{CH2OHCHO} are either due to noisier column density maps and therefore are less likely to be attributed to true gas conditions, or are indicative of two distinct molecular components towards the inner envelope of IRAS 4B. The possibility of two components towards IRAS 4B is discussed further in Section \ref{subsec:discussion}. 

\begin{figure}[ht]
    \centering
    \includegraphics[width=1\linewidth]{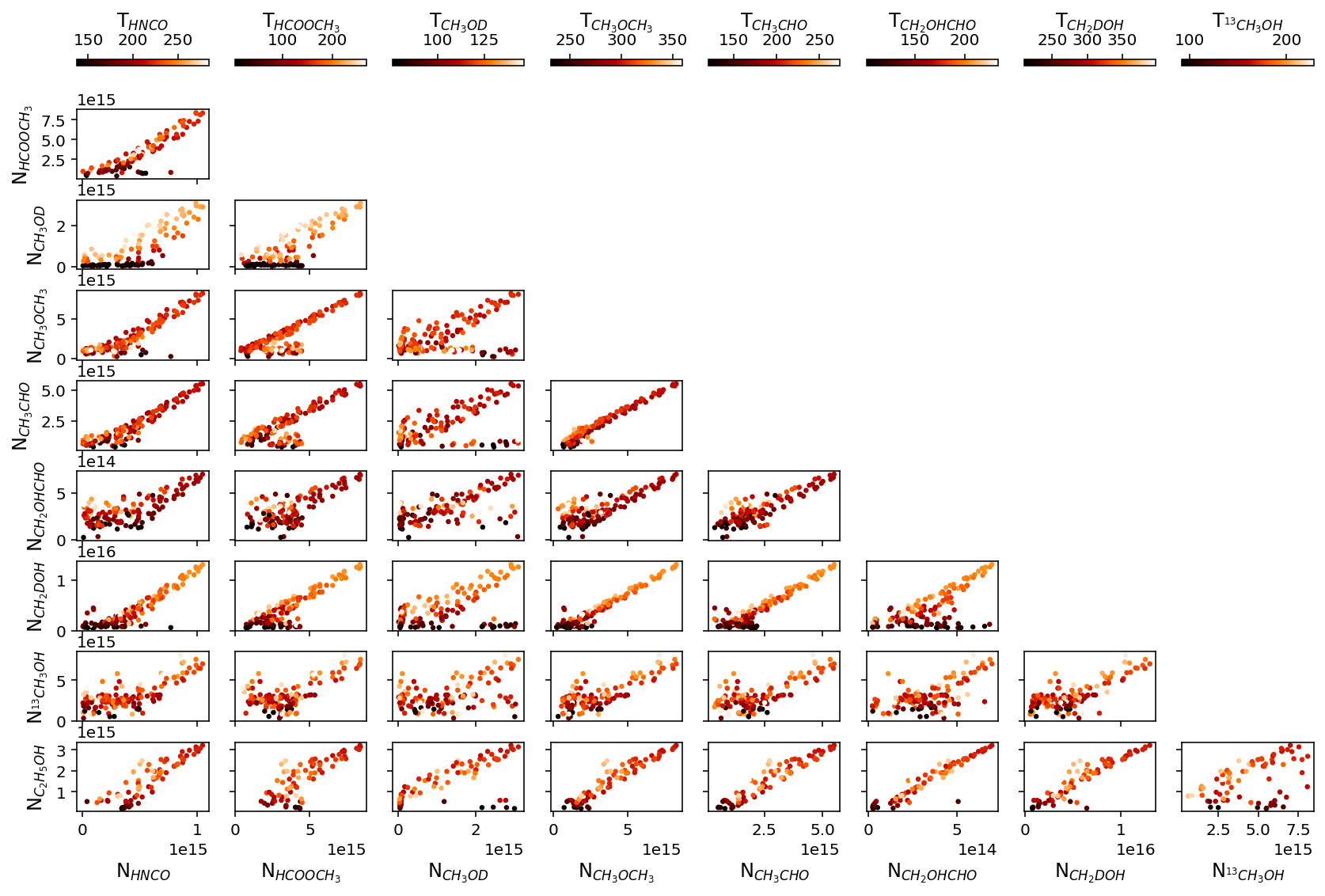}
    \caption{Plots comparing column densities (cm$^{-2}$) between pairs of molecules in IRAS 4B from the parameter maps. The rotational temperature (K) for each molecule is seen in the color bar on the top of each column. For each molecule and physical parameter, the associated uncertainties are determined from the GOBASIC spectral fit and can be found in Appendix \ref{sec:appendix-parameter-maps}.}
    \label{fig:pairs-of-molecules-parameter-maps}
\end{figure}

\begin{figure}[ht]
    \centering
    \includegraphics[width=0.47\linewidth]{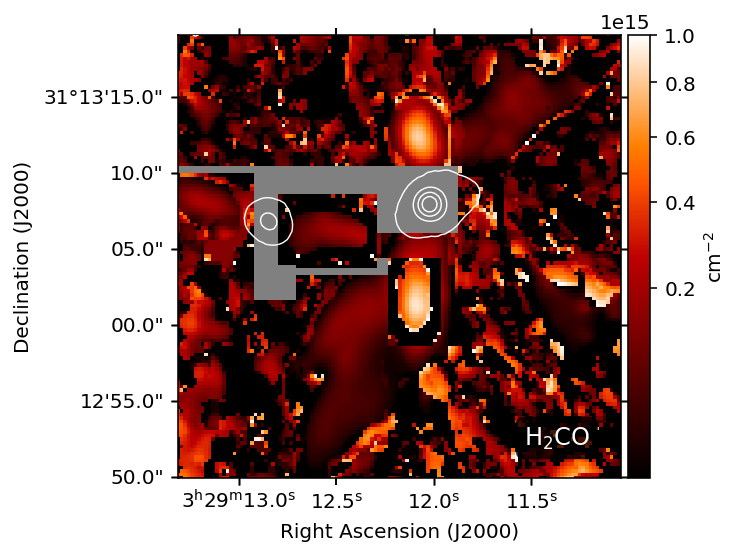}
    \includegraphics[width=0.47\linewidth]{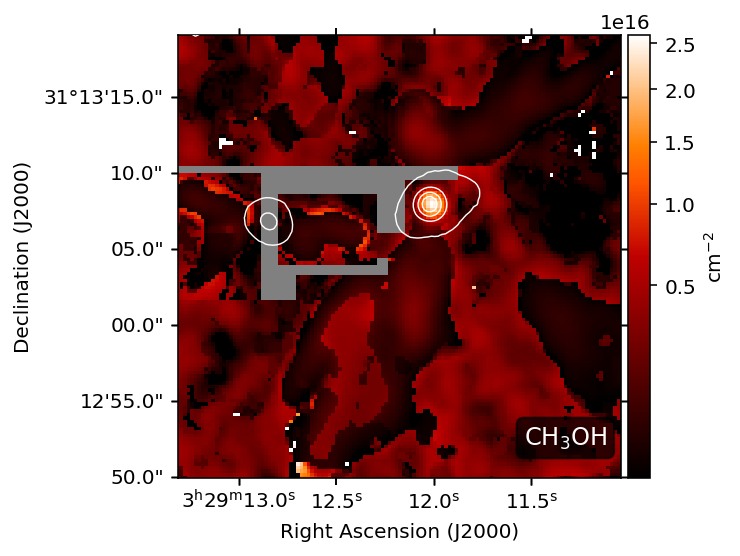}
    \includegraphics[width=0.47\linewidth]{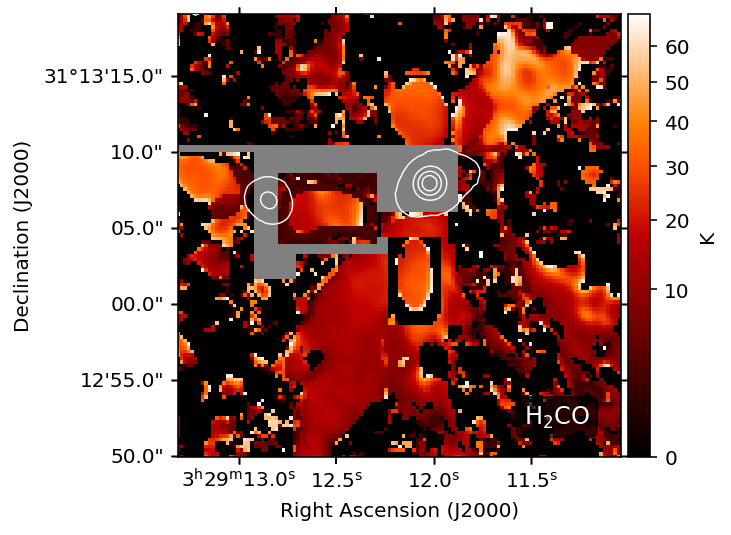}
    \includegraphics[width=0.47\linewidth]{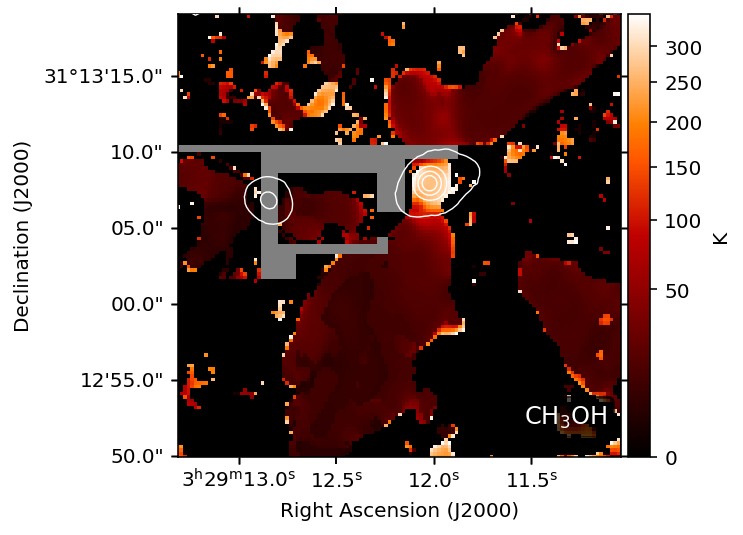}
    \includegraphics[width=0.47\linewidth]{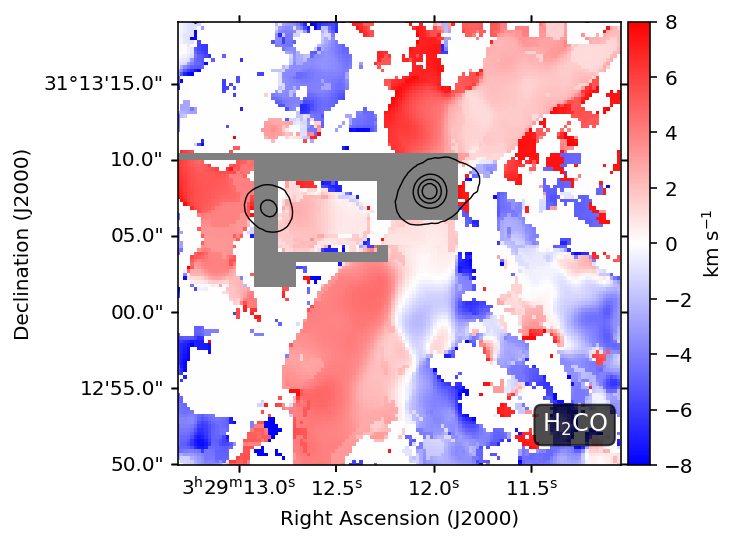}
    \includegraphics[width=0.47\linewidth]{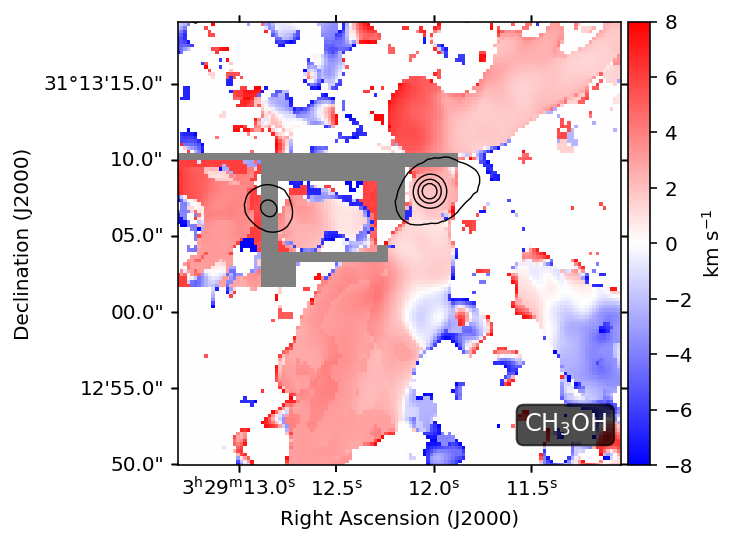}
    \caption{Parameter maps of \ch{H2CO} (left) and \ch{CH3OH} (right) across IRAS 4B and 4B$'$, in order from top to bottom: column density (cm$^{-2}$), rotational temperature (K),and velocity shift (km s$^{-1}$). White and black contours are 5 evenly spaced levels (1 level $=0.0599$ Jy beam$^{-1}$) from $3 \times \sigma_{cont}$ ($\sigma_{cont} = 0.0013$ Jy beam$^{-1}$) to the max continuum flux at 0.2436 Jy beam$^{-1}$.}
    \label{fig:parameter-maps-outflows}
\end{figure}

Understanding the relationship between molecules can give insight into the chemical conditions within the star-forming region. Figure \ref{fig:pairs-of-molecules-parameter-maps} compares the column densities between pairs of molecules toward 4B, with a colored temperature bar for each molecule on the top of each column. In this way, multiple trends that exist within pairs of molecules can be analyzed. Figures \ref{fig:dens-vs-temp-parameter-maps} and \ref{fig:pairs-of-molecules-parameter-maps} enable us to study qualitative relationships between molecules, conduct a preliminary analysis, and provide further avenues of investigation. Drawing any quantitative chemical formation mechanism information -- such as branching ratios and chemical kinetics -- requires detailed theoretical modeling and is beyond the scope of this work. The majority of plots in Figure \ref{fig:pairs-of-molecules-parameter-maps} show a strong positive corelation between molecules that is well described by a single linear relation. This is expected since most of the column density parameter maps exhibited similar spatial characteristics and provide a similar physical environment within the warm inner envelope for iCOM formation. Once again as illustrated in Figure \ref{fig:dens-vs-temp-parameter-maps}, there may be evidence of two distinct components for \ch{^{13}CH3OH} and \ch{CH2OHCHO}, as the comparison plots with the other molecules suggest two separate groupings that are not well described by a single linear relationship compared to the other iCOMs. There seems to be no column density correlation as indicated by broadly scattered points between \ch{^{13}CH3OH} and {\ch{C2H5OH}}. Future broadband observations at higher spectral resolutions would resolve \ch{^{13}CH3OH} and \ch{CH2OHCHO} transitions close to the noise threshold and would distinguish if these correlations are due to noise limits or true associations.

Next, the pixels surrounding IRAS 4B and 4B$'$ were extracted and fit with GOBASIC in a similar manner as the pixels toward 4B. Regions of interest were fit in an iterative process and then subsequently stitched together. Figure \ref{fig:parameter-maps-outflows} shows the parameter maps for column density, rotational temperature, and velocity shift for \ch{H2CO} and \ch{CH3OH}. These two molecules were the only ones selected in the outflow regions since both had numerous lines above 3 rms that allowed for a full analysis.  The associated uncertainty maps are found in Appendix \ref{sec:appendix-parameter-maps}. Multiple outflows were revealed in molecular emission from both \ch{H2CO} and \ch{CH3OH}: one in an east-west orientation from 4B$'$, and at least two outflows from 4B, one in north-south orientation, and one in a northwest-southeast orientation. The velocity shift maps of \ch{H2CO} and \ch{CH3OH} correlate well with each other and show the north outflow lobe of 4B is redshifted and the south outflow lobe is blueshifted. The northwest-southeast outflow of 4B and the west outflow of 4B$'$ are less redshifted in comparison to the north outflow of 4B. The northwest outflow is slightly blueshifted, while the southeast lobe is slightly redshifted. The east outflow of 4B$'$ is redshifted by approximately the same amount as the north outflow of 4B. The column density and rotational temperature maps of \ch{CH3OH} show lower densities and temperatures in the outflows as compared to the central source, even toward the envelope of 4B which is optically thick. Like the \ch{CH3OH} temperature map, \ch{H2CO} also shows low rotational temperatures within all the outflows, with some variation. The north-south outflows of 4B are associated with the highest column densities of \ch{H2CO} as compared to the other outflow regions, and higher rotational temperatures. No \ch{H2CO} emission was observed within the inner envelope of 4B that was not affected by absorption. Overall, the temperatures are lower in the outflows than the densely compact protostar, and there are at least three pairs of outflows identified: east-west from 4B$'$, north-south from 4B, and northwest-southeast from 4B. The outflows are seen again in the integrated intensity maps discussed in the next subsection.

\subsection{Results: High-Resolution Spectral Windows}\label{subsec:highres}
The 28 high-resolution spectral windows were placed strategically within the LSB and USB to capture transitions from iCOMs of interest. Transitions were identified by cross-referencing the high-resolution transitions with the low-resolution fit done with GOBASIC that encompassed all 28 windows. The molecular emission from the high-resolution windows was imaged using the Astropy\footnote{\url{http://www.astropy.org}} python library with moment 0 maps (\hspace*{-4px}\citealt{astropy-2013,astropy-2018,astropy-2022}). Tables \ref{tab:hr-lines-mult} through \ref{tab:hr-lines-UI} list every transition above 3 rms with associated rest frequency (MHz), upper energy level of the transition in Kelvin, and the corresponding high-resolution spectral window for the transition  (refer to Table \ref{tab:table-summary-hr-spw} for spectral ranges and associated rms values). If the transition was blended with another transition anywhere within the spatial range or frequency range of the transition, it is noted in the Table. For every non-blended line listed within the Tables, channels with emission over 3 rms were integrated and plotted as a function of position to produce moment 0 maps. For molecules that had multiple transitions, integrated intensity maps were stacked according to \cite{van-der-walt} with a 1/rms$^2$ weighting scheme, see Figure \ref{fig:moment-0-mult}. The colorbar scales for all moment 0 maps were set so the minimum value corresponded to 3 rms from the high-resolution spectral windows. Maximum values in Figure \ref{fig:moment-0-mult} are $f_{max}$, the weighted maximum emission flux for each molecule. Molecules with multiple transitions were only stacked if they had the same spatial distribution, which was true for all molecules except \ch{CH3OH}. \ch{CH3OH} transitions were separated into two integrated intensity maps, \ch{CH3OH}-compact, in which molecular emission was only observed toward the inner envelope of 4B, and \ch{CH3OH}-extended, in which molecular emission was observed in outflows surrounding 4B and 4B$'$, which may or may not have included compact emission as well. The \ch{CH3OH} transitions that were included in the compact or extended integrated intensity maps are noted in Table \ref{tab:hr-lines-mult} and are generally separated by upper energy state; the outflows display transitions with low E$_u$ and the inner envelope displays transitions with high E$_u$.

\begin{figure}[ht]
    \centering
    \includegraphics[width=0.49\linewidth]{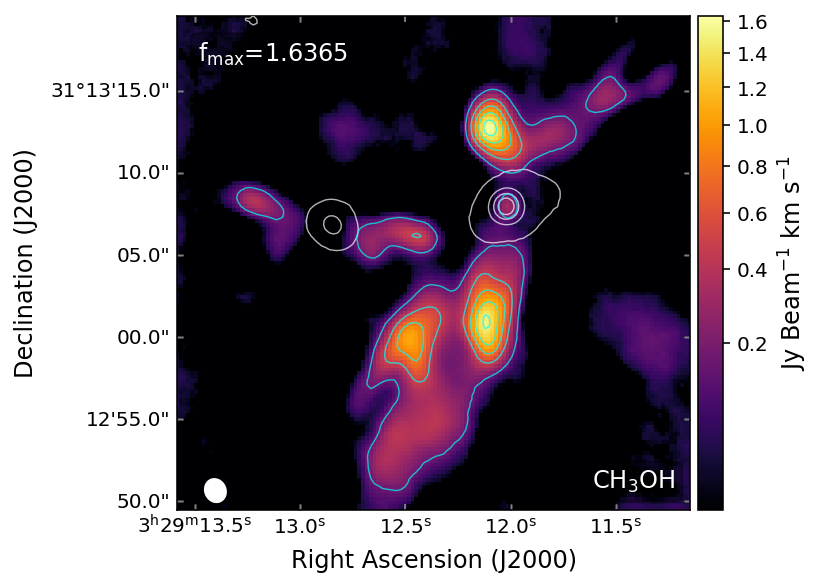}
    \includegraphics[width=0.49\linewidth]{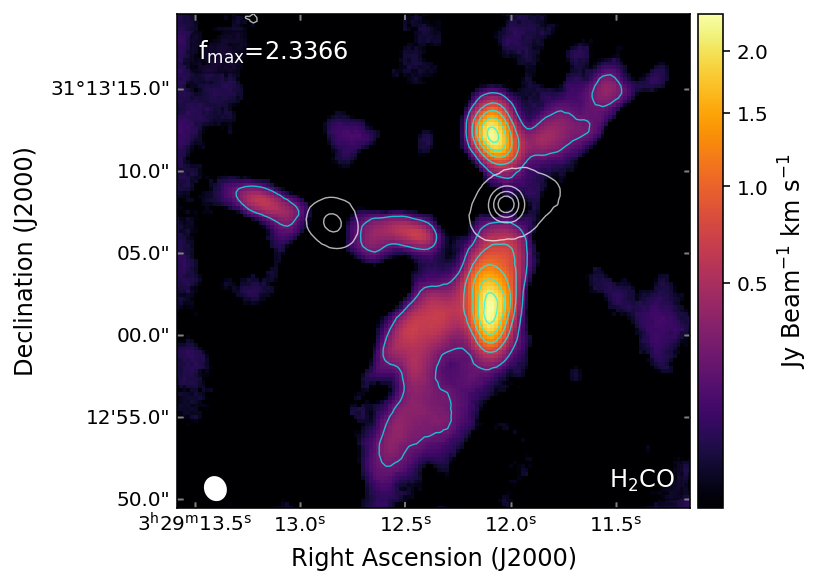}
    \includegraphics[width=0.49\linewidth]{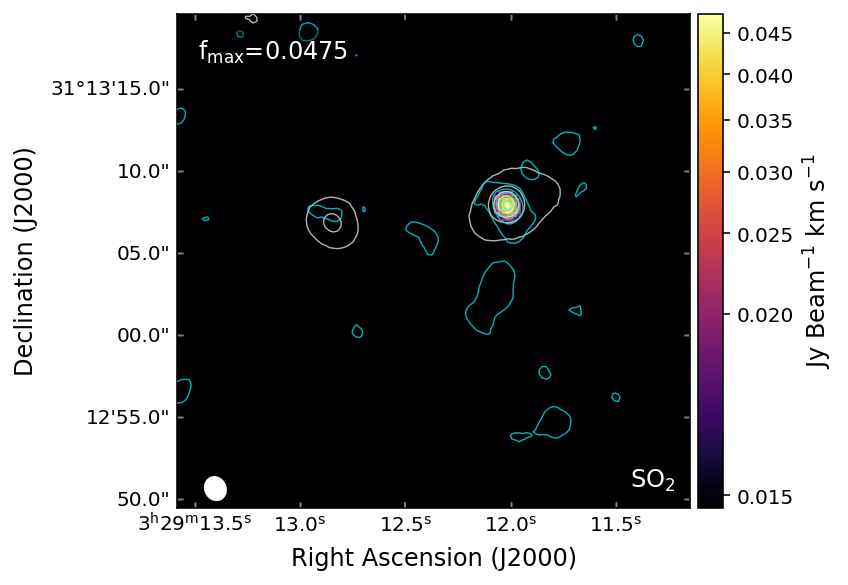}
    \includegraphics[width=0.49\linewidth]{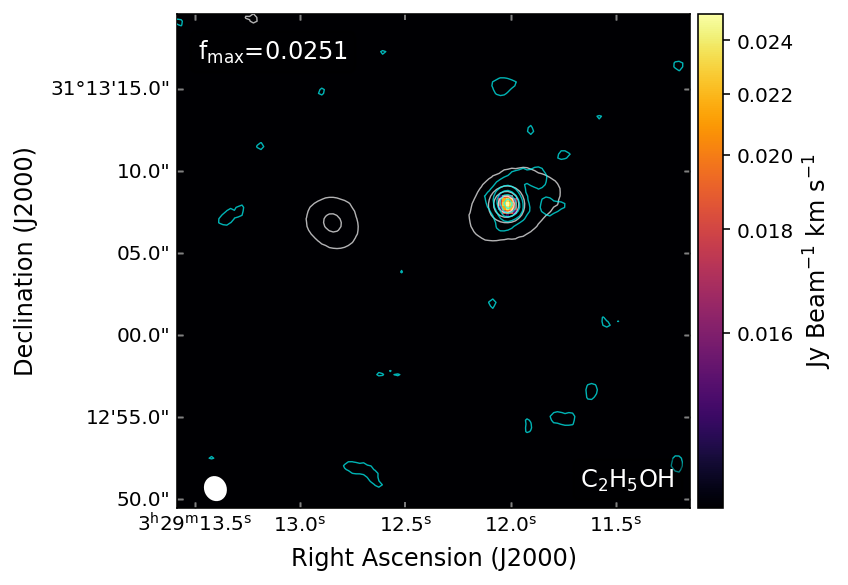}
    \includegraphics[width=0.49\linewidth]{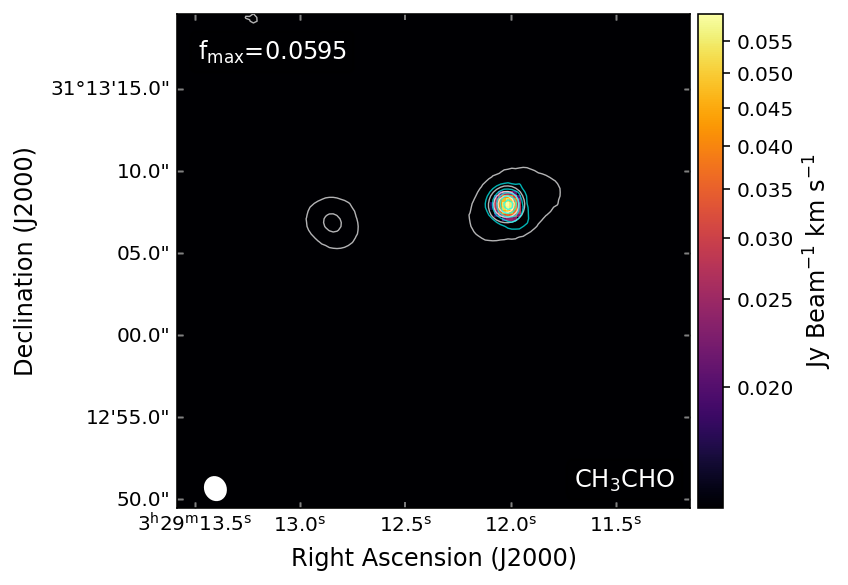}
    \includegraphics[width=0.49\linewidth]{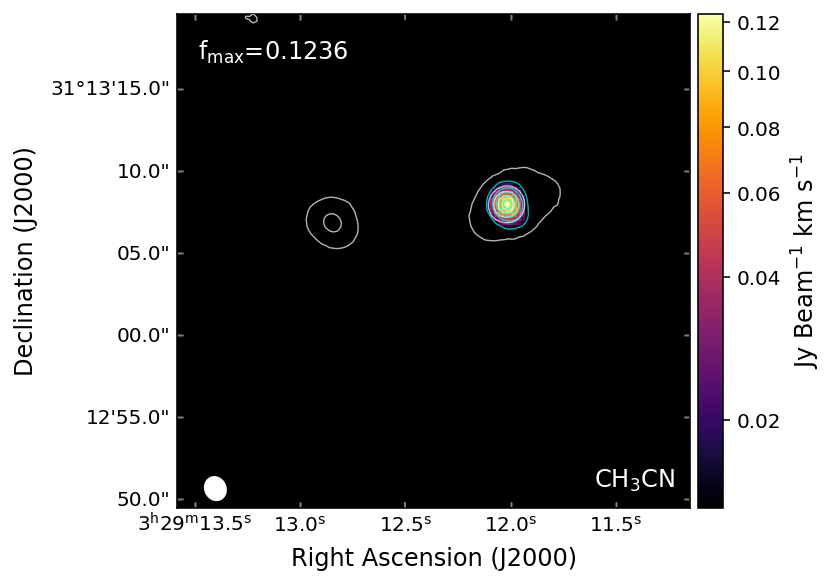}
    \caption{Integrated intensity maps of molecules that had multiple transitions within the high-resolution spectral windows. Maps were stacked according to a $1/rms^2$ weighing scheme. Molecular transitions included in each stacked map are listed in Table \ref{tab:hr-lines-mult}. The weighted maximum emission flux ($f_{max}$)} for each molecule is listed in the top left corner. Blue contours correspond to 10\%, 30\%, 50\%, 70\%, and 90\% of $f_{max}$. White contours are 5 evenly spaced levels (1 level $=0.0599$ Jy beam$^{-1}$) from $3 \times \sigma_{cont}$ ($\sigma_{cont} = 0.0013$ Jy beam$^{-1}$) to the max continuum flux at 0.2436 Jy beam$^{-1}$. The synthesized beam is shown in the lower left corner. \textit{Continued on next page}
\end{figure}

\begin{figure}[ht]
    \centering
    \ContinuedFloat

    \includegraphics[width=0.49\linewidth]{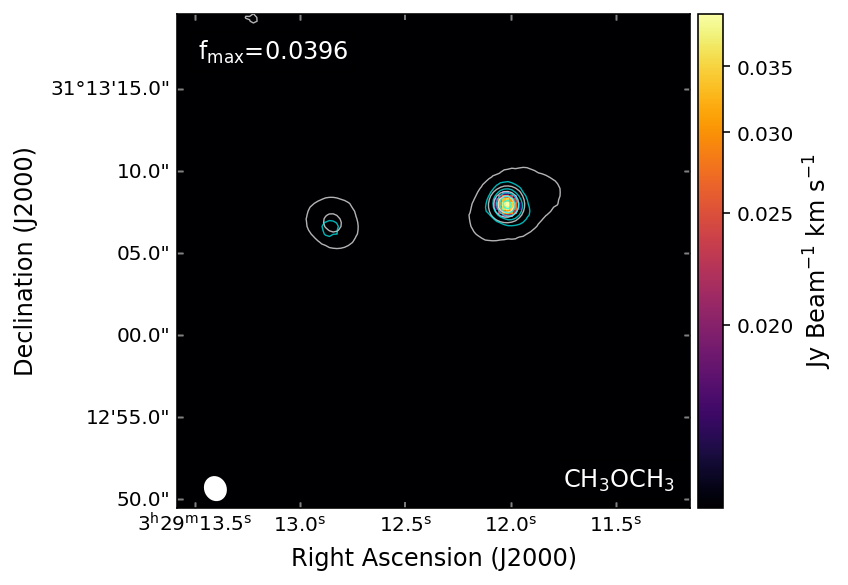}
    \includegraphics[width=0.49\linewidth]{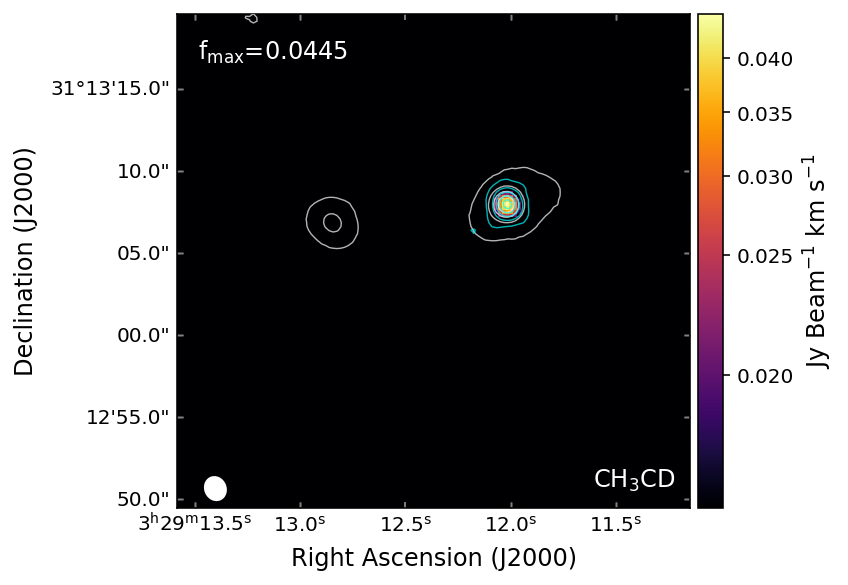}
    \includegraphics[width=0.49\linewidth]{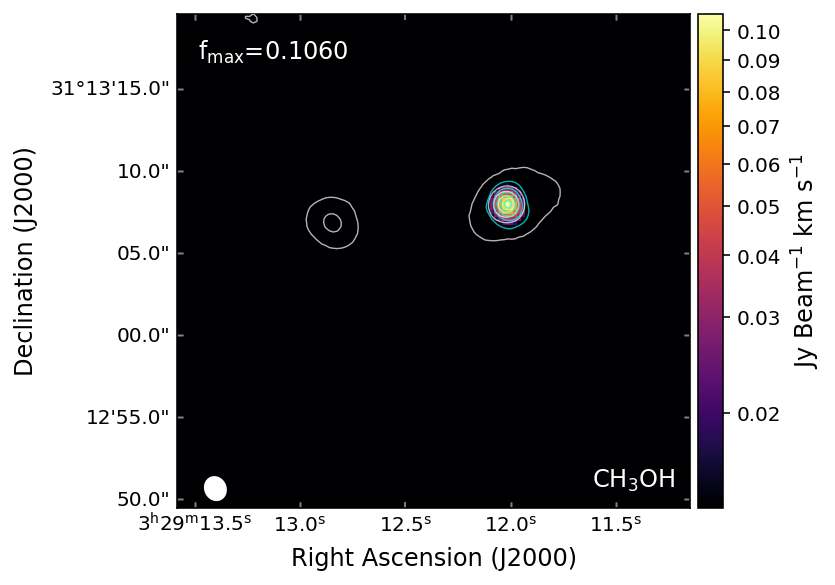}
    \includegraphics[width=0.49\linewidth]{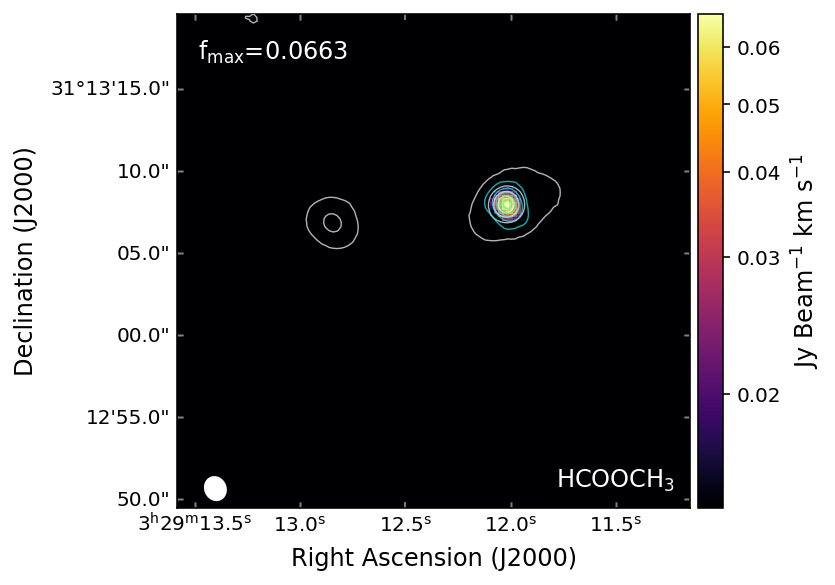}
    \caption{Integrated intensity maps of molecules that had multiple transitions within the high-resolution spectral windows. 
    \textit{Continued on next page}}
\end{figure}  

\begin{figure}[ht]
    \centering
    \ContinuedFloat
    \includegraphics[width=0.48\linewidth]{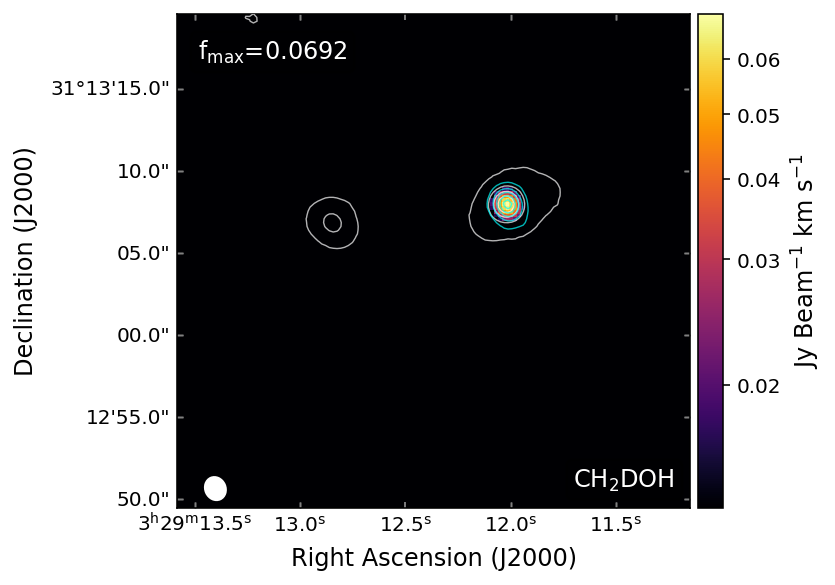}
    \includegraphics[width=0.49\linewidth]{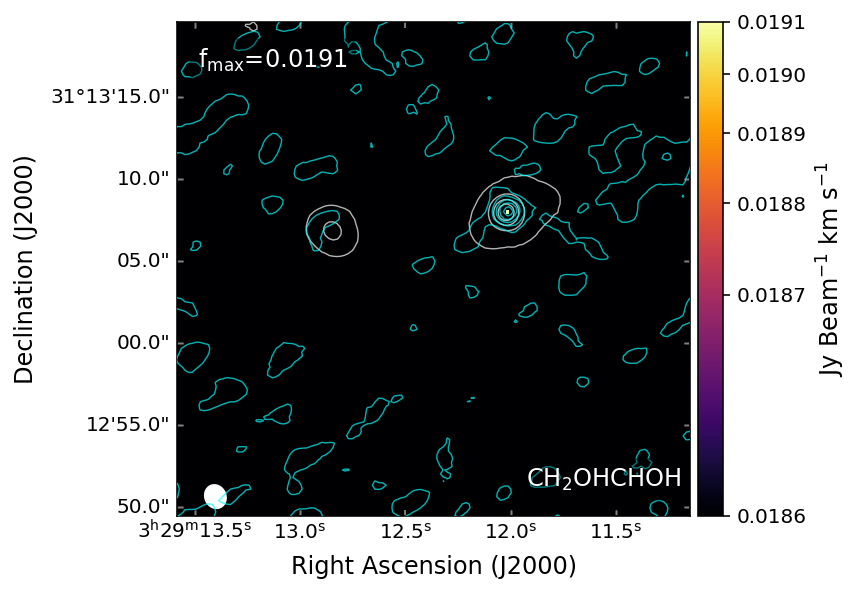}
    \includegraphics[width=0.47\linewidth]{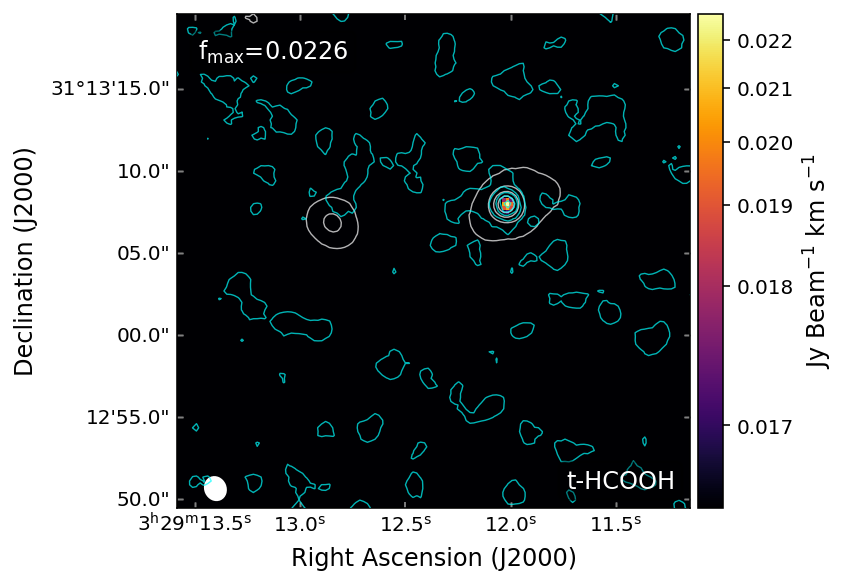}
    \caption{Integrated intensity maps of molecules that had multiple transitions within the high-resolution spectral windows. 
    }
    \label{fig:moment-0-mult}
\end{figure} 
\newpage
Integrated intensity maps of molecules with only a single molecular emission line within the high-resolution spectral windows are seen in Figure \ref{fig:moment-0-single} and found in Appendix \ref{sec:appendix-transitions}. These moment 0 maps were made in a similar manner to that described above. Instead of a weighted stacking scheme, the $f_{max}$ listed in each moment 0 map is the maximum emission flux of each transition. Here the maximum emission flux for each molecules is $f_{max}$, located in the top left corner of each map. Transition specific information can be found in Table \ref{tab:hr-lines-mult} and \ref{tab:hr-lines-single} in Appendix \ref{sec:appendix-transitions}. Molecular transitions that could not be identified are listed in Table \ref{tab:hr-lines-UI} in Appendix \ref{sec:appendix-transitions}.

Figures \ref{fig:correlation-matrix-extended} and \ref{fig:correlation-matrix-compact} show the Pearson cross-correlation coefficients between all pairs of moment 0 maps and the continuum emission, focusing on the inner envelope of IRAS 4B. This linear correlation coefficient measures how spatially similar molecular emission is between two integrated intensity maps, with emission most similar when the correlation value is close to one. This method has been used previously in the analysis of molecular emission of star-forming regions \citep{guzman-2018,law-2021,will_w3_2023,Morgan-2025}. Even though the molecular emission may be spatially similar with a high Pearson cross-corelation coefficient, it does not necessarily mean that there exists a direct chemical reaction mechanism route between the two species.  Instead, a high correlation Pearson cross-correlation coefficient typically suggests both species form in similar environments. The region included in Figure \ref{fig:correlation-matrix-extended} and \ref{fig:correlation-matrix-compact} used pixels within the region of $\alpha$(J2000) = $03^h29^m12^s.04-12^s.13$ and $\delta$(J2000) = 31\textdegree$13'08''.17-10''.04$ that included only the compact emission around 4B, corresponding to the region seen in the parameter maps in Figure \ref{fig:parameter-maps-4b}. There is a strong correlation in IRAS 4B between the continuum emission and the iCOMs. There is also strong correlation between many of the iCOMs, especially between [\ch{CH2DOH}]/[\ch{CH3CHO}], [\ch{CH2DOH}]/[\ch{CH3CN}], [\ch{CH2DOH}]/[\ch{CH3OCH3}], [\ch{CH2DOH}]/[\ch{CH3OH}], [\ch{CH2DOH}]/[\ch{HCOOCH3}], [\ch{CH3CHO}]/[\ch{CH3CN}], [\ch{CH3CN}]/[\ch{CH3OH}], [\ch{CH3CN}]/[\ch{HCOOCH3}], [\ch{CH3OCH3}]/[\ch{CH3OH}], [\ch{CH3OCH3}]/[\ch{HCOOCH3}], and [\ch{CH3OH}]/[\ch{HCOOCH3}] that have a coefficient of unity. Molecules that do not have a strong correlation with the continuum, indicated by a low or negative Pearson cross-corelation coefficient, are those that trace only the outflows and not the envelope such as some sulfur bearing species (\ch{SO}, \ch{C^{34}S}, and \ch{CS}), \ch{SiO}, and \ch{H2CO}, as can be observed in Figure \ref{fig:correlation-matrix-extended}. These low coefficients suggest a different formation environment separate from the warm envelope of the hot corino where iCOMs are formed, such as ice sputtering from icy grain mantles. These molecular tracers and their conditions of formation are discussed further in Section \ref{subsec:discussion}.

\begin{figure}[ht]
    \centering
    \includegraphics[width=0.43\linewidth]{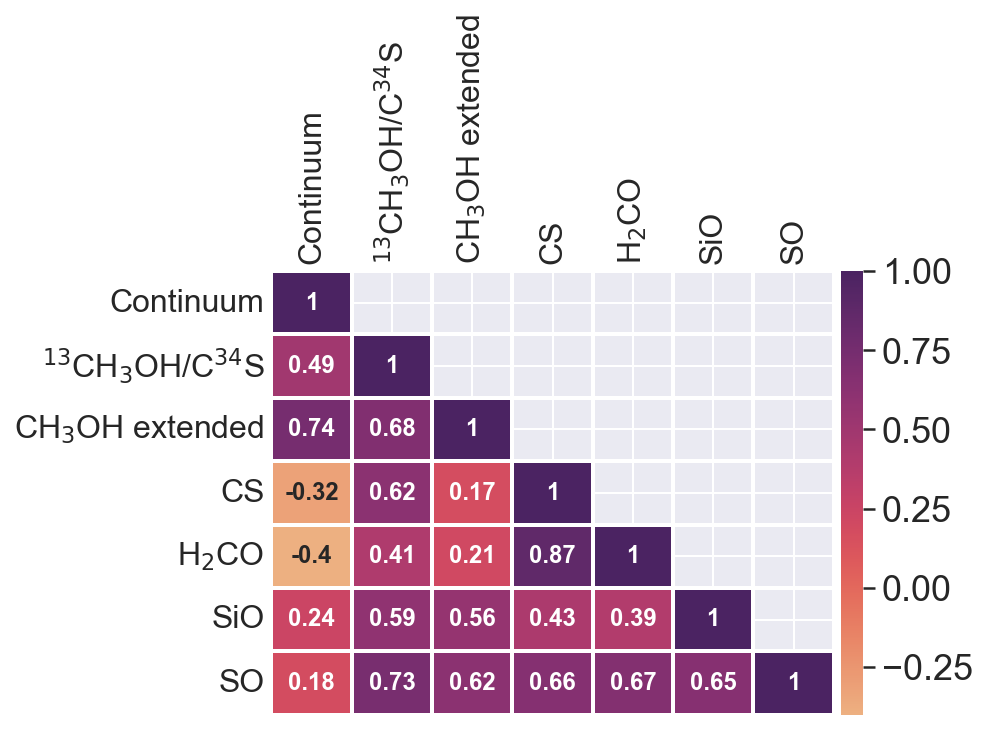}
    \caption{Correlation matrix of molecules toward the inner envelope of IRAS 4B exhibiting extended emission within the 4B region using the integrated intensity maps from Figures \ref{fig:moment-0-mult} and \ref{fig:moment-0-single}. The continuum is at 147 GHz. \ch{^{13}CH3OH}/\ch{C^{34}S} is the only blended moment 0 map presented in this work; \ch{C^{34}S} emission is dominant (30.7 $\sigma$) vs. \ch{^{13}CH3OH} (2.3 $\sigma$).}
    \label{fig:correlation-matrix-extended}
\end{figure}

\begin{sidewaysfigure}[ht]
    \centering
    \includegraphics[width=1\linewidth]{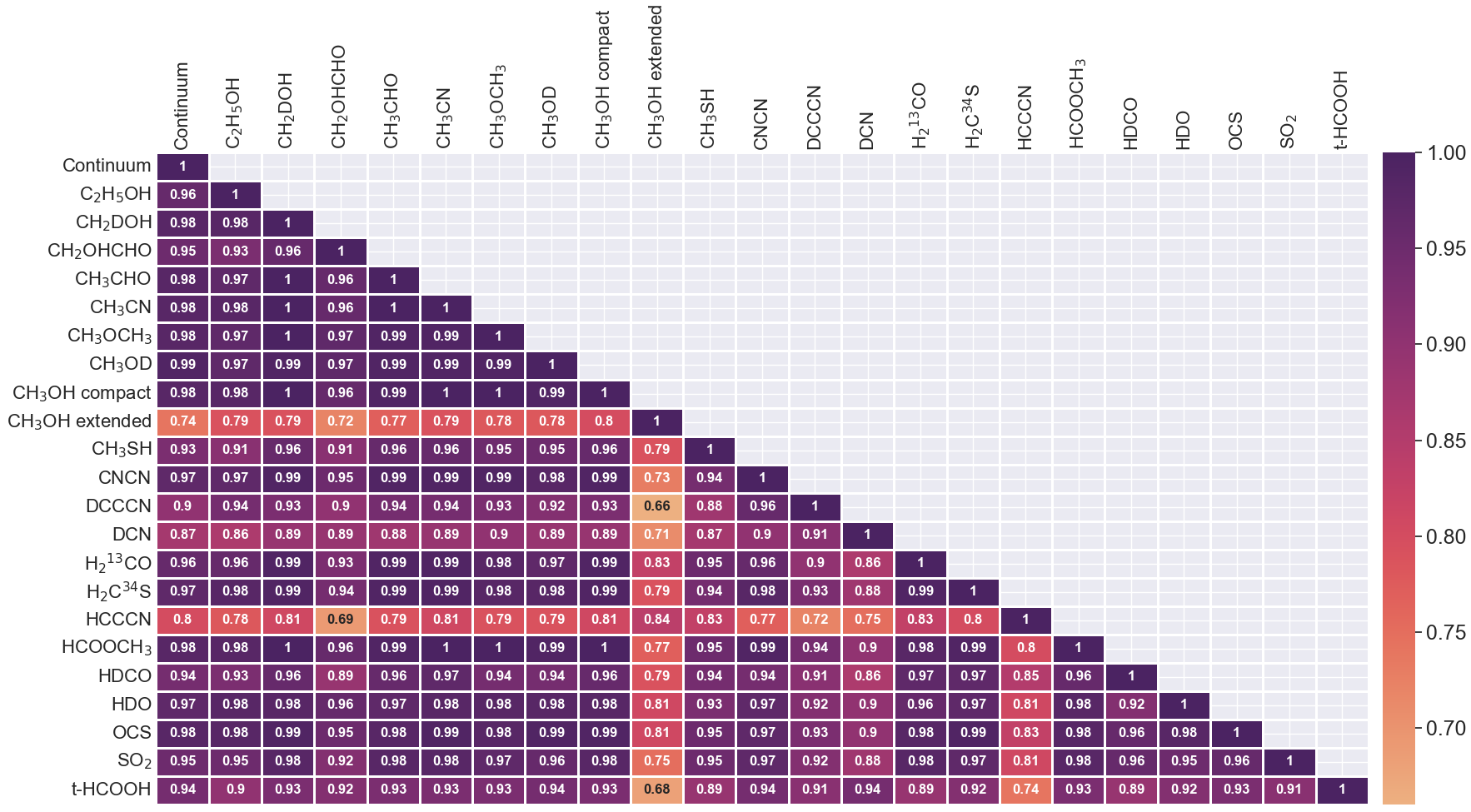}
    \caption{Correlation matrix of molecules toward the inner envelope of IRAS 4B exhibiting compact emission within the 4B region using the integrated intensity maps from Figures \ref{fig:moment-0-mult} and \ref{fig:moment-0-single}. The continuum is at 147 GHz. \ch{CH3OH} is separated into two spatial arrangements, compact and extended.}
    \label{fig:correlation-matrix-compact}
\end{sidewaysfigure}

\clearpage
\subsection{Discussion}\label{subsec:discussion}
\begin{figure}[H]
    \centering
    \includegraphics[width=0.7\linewidth]{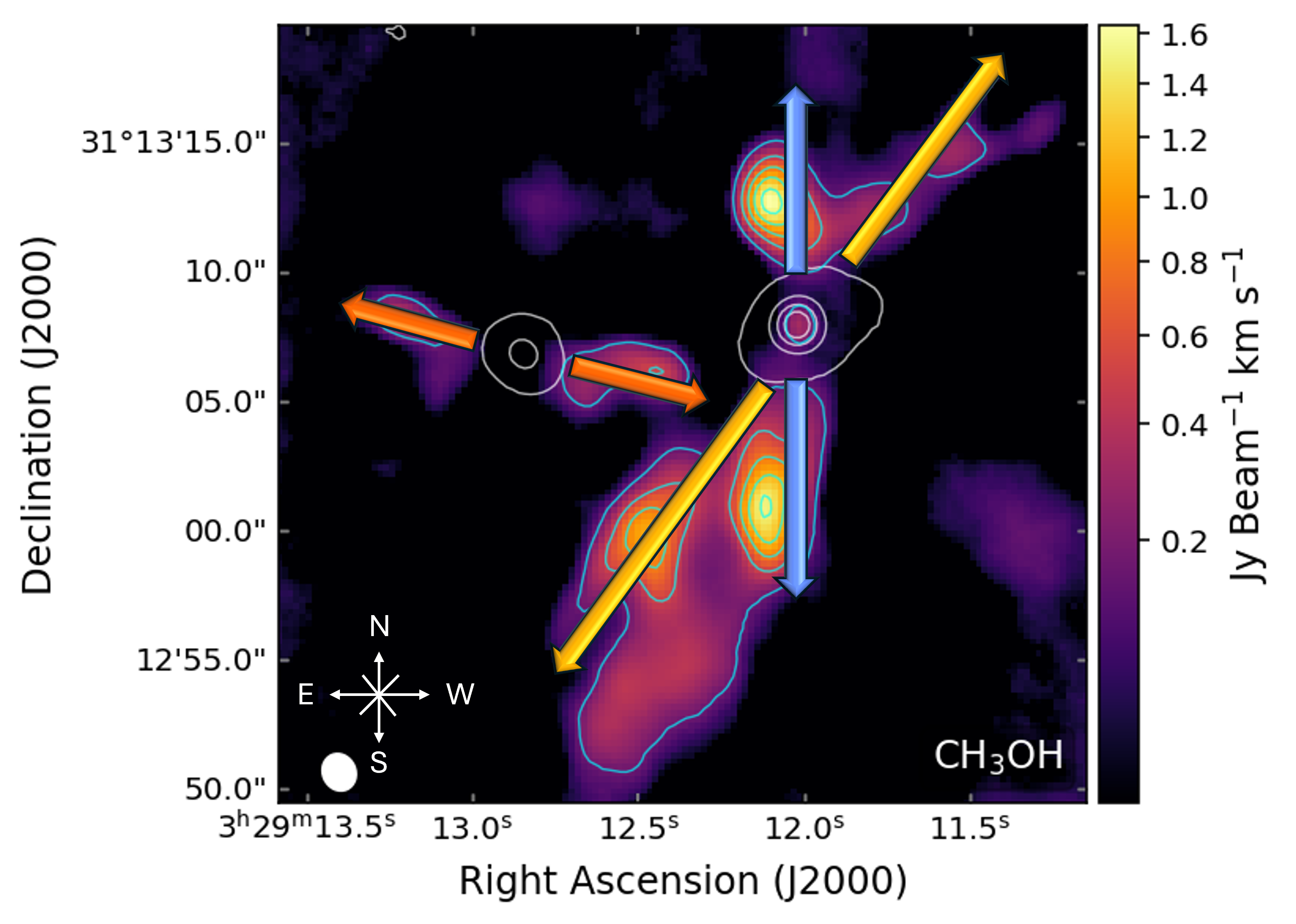}
    \caption{Labeled schematic of 4B, 4B$'$, and the discussed outflows overlaid on the integrated intensity map of the extended \ch{CH3OH} emission. Blue arrows denote the north-south outflows from 4B, yellow arrows denote the northwest-southeast outflows from 4B, and orange arrows denote the east-west outflows from 4B$'$.}
    \label{fig:cartoon-labeled}
\end{figure}

Figure \ref{fig:cartoon-labeled} gives an overview of IRAS 4B and 4B$'$ and the three outflows discussed below. Generally, molecular emission in the integrated intensity maps can be separated into two groups: 1) compact emission toward the warm inner envelope of 4B, cospatial with continuum emission, and 2) extended emission tracing the outflows surrounding 4B and 4B$'$, with or without compact emission. The molecules with compact emission include \ch{SO2}, \ch{C2H5OH}, \ch{CH2DOH}, \ch{CH2OHCHO}, \ch{CH3CHO}, \ch{CH3CN}, \ch{CH3OCH3}, \ch{CH3OD}, compact-\ch{CH3OH}, \ch{HCOOCH3}, \ch{$trans$-HCOOH}, \ch{DCCCN}, \ch{CNCN}, \ch{HDO}, and \ch{CH3SH}. These molecules are released from ice mantles within the warm inner envelope of the protostar where temperatures are sufficient to thermally desorb molecules from the surface of the dust grains. These molecules include the interstellar complex organic molecules identified within this region.  Based on the integrated intensity maps, it can be seen that the molecular emission of \ch{SiO}, \ch{SO}, \ch{CS}, \ch{H2CO}, and \ch{CH3OH}-extended is enhanced in the outflows and depleted in the 4B core, consistent with previous observations and the release of molecules from ice sputtering processes \citep{blake_1995}. Additional details about the outflows can be gleaned from the chemistry observed in these regions.

The north-south outflow from 4B is highlighted by the blue arrows in Figure \ref{fig:cartoon-labeled}. It should be noted that the north and south lobes are slightly misaligned with the peak continuum emission and do not perfectly intersect, indicating there may be underlying structure that is not captured with these observations. Molecules with extended emission that trace the north-south outflows from 4B include \ch{H2CO}, \ch{CH3OH}-extended, blended \ch{^{13}CH3OH}/\ch{C^{34}S}, \ch{DCN}, \ch{H2C^{34}S}, \ch{HCCCN}, \ch{OCS}, \ch{SO}, \ch{SiO}, \ch{CS}, \ch{HDCO}, and \ch{H2^{13}CO}. There is inconsistent terminology in the literature as to whether this is an outflow or a jet. Typically, molecular emission of O-bearing molecules traveling at velocities greater than 20-30 km s$^{-1}$ are classified as jets, with more Class 0 sources having jets than previous believed \citep{what-traces-what-2021}. Jets are distinguished from outflows by the velocity and angle of collimation of the outflowing gas. \cite{jorgensen_prosac_2007} reported molecular detections of \ch{CH3OH}, \ch{H2CO}, \ch{SiO}, and \ch{CS} in ``jetlike outflows" in a north-south orientation from IRAS 4B on an axis close to 0\textdegree. \cite{choi-2011} also observed blue and redshifted lobes of \ch{HCN}, indicating 4B drives a ``bipolar outflow" in roughly the north-south orientation. The typical jet molecular tracers, SiO and SO, are present here tracing the north-south outflow with a fairly collimated opening angle of less than 60\textdegree. The SiO emission is traveling at a velocity of 25 km s$^{-1}$ in the north-south outflow from 4B, while SO is traveling at 20 km s$^{-1}$. Both SiO and SO are also tracers of shocked regions, where shocks moving through outflows release atomic Si and S from the grains through sputtering and grain destruction, then react with OH and O to form SiO and SO \citep{hartquist-1980,millar-williams-1993,Schilke-1997,Caselli-1997,Gusdorf-2008a,Gusdorf-2008b}. This process is enhanced in jets where grains are more easily destroyed \citep{what-traces-what-2021}. Thus, the north-south outflow of IRAS 4B is more likely a higher velocity bipolar jet than a lower velocity bipolar outflow.

\cite{prodige-2026} also observed the north-south outflow of IRAS 4B in molecular emission from simpler molecules such as \ch{SiO}, \ch{CO}, \ch{H2CO}, \ch{D2CO}, \ch{HNCO}, \ch{HCCCN}, \ch{c-C3H2}, \ch{C^{34}S}, and \ch{DCN}, and from iCOMs including \ch{CH3OH}, \ch{CH2DOH}, \ch{CH3CN}, and \ch{CH3CHO}. Comparing with this work, molecules that were also observed in the same outflows include \ch{SiO}, \ch{DCN}, \ch{C^{34}S}, \ch{HCCCN}, \ch{H2CO}, and \ch{CH3OH}. Additional molecules exhibiting extended emission in this work are \ch{SO}, \ch{H2C^{34}S}, \ch{OCS}, \ch{CS}, \ch{HDCO}, and \ch{H2^{13}CO}. \cite{prodige-2026} extracted a single spectrum from the north and south lobe at the peak emission pixel for \ch{CH3OH} and iteratively fit with Weeds and population diagrams to derive physical parameters for \ch{CH3OH}, \ch{CH2DOH}, \ch{CH3CHO}, \ch{CH3CN}, \ch{HC3N}, and \ch{D2CO}. As stated in Subsection \ref{subsec:parameter-maps}, it was determined that only \ch{CH3OH} and \ch{H2CO} had at least 3 lines over 3 rms with which to determine physical parameters. As such, only the results from \ch{CH3OH} can be directly compared between the two studies. At the two pixel locations reported in \cite{prodige-2026}, \ch{CH3OH} has a column density of $(5\pm2)\times10^{15}$ cm$^{-2}$ and a rotational temperature of $79 \pm 9$ K in the northern outflow and $(4\pm1)\times10^{15}$ cm$^{-2}$ and $80$ K (no uncertainty reported) in the southern outflow. Comparing with the same location in the \ch{CH3OH} parameter maps in Figure \ref{fig:parameter-maps-outflows} yields a column density of $(3.8\pm0.2)\times10^{15}$ cm$^{-2}$ and a rotational temperature of $25\pm2$ K in the north and $(3.0\pm0.1)\times10^{15}$ cm$^{-2}$ and $17\pm1$ K in the south. Thus, the derived column densities agree within uncertainty, while the rotational temperatures in this work are lower by $54\pm7$ K in the northern outflow and approximately 63 K in the southern outflow. The differences in rotational temperature between the two studies likely arise from several factors, the most significant of which is that the two studies probe different frequency ranges and therefore separate sets of spectral lines with distinct upper state energies; this work observes upper level energies up to 104.41 K (Table \ref{tab:hr-lines-mult}) while \citet{prodige-2026} observes up to approximately 260 K. As such, it is possible that this source has two temperature components, one probed by the observations presented here, and one probed by the observations presented by \cite{prodige-2026}. Additional factors such as optical depth corrections and line blending considerations may also contribute to the differences between the two studies. Nonetheless, the results produced by the best fit to the current work are those presented here.

The northwest-southeast outflow from 4B is shown by the yellow arrows in Figure \ref{fig:cartoon-labeled} to be offset from the north-south axis by approximately -40\textdegree. \ch{CH3OH} and \ch{H2CO} trace this outflow; Table \ref{tab:hr-lines-mult} lists the \ch{CH3OH} and \ch{H2CO} transitions that are included in the molecular outflows tracers. Low energy (low E$_u$ K) transitions of these molecules trace molecular outflows and jets where non-thermal desorption mechanisms are responsible for releasing the molecules from the ice grains. The transitions included in the extended emission integrated intensity maps of \ch{CH3OH} and \ch{H2CO} range from 13.93 -- 104.41 K and 10.43 -- 22.62 K, respectively. The \ch{CH3OH} transitions that do not overlap with the continuum emission and are exclusively extended are the three listed with E$_u \leq 27.05$ K. These lines are the same lines used for the analysis in the outflows in Section \ref{subsec:parameter-maps}. Thus, these tracers are likely not from thermal desorption like that occurring in the warm inner envelope, and are instead from sputtering occurring in the outflow cavity walls. Both \ch{CH3OH} and \ch{H2CO} are ice mantle tracers and are found in shock-released ice mantles \citep{what-traces-what-2021}. \ch{H2CO} can also be effectively formed in jets through gas-phase chemistry. The northwest-southeast outflow of IRAS 4B is also consistent with \ch{H2CO} outflows observed in \cite{di_francesco_2001}, who observed the same outflow at approximately -35\textdegree. However, this angle also corresponds with the large-scale emission from \ch{N2D^+}, which traces the filament connecting IRAS 4B and 4A  that is located northwest out of the field of view for these observations. \cite{prodige-2026} also postulate that this is an arc of unknown origin. However, the molecular emission maps of \ch{CH3OH} and \ch{H2CO} presented here are better tracers of outflows rather than large-scale protostar envelope tracers like \ch{N2D+}, \ch{C^{18}O}, and \ch{DCO+} \citep{what-traces-what-2021}. \cite{sakai-ch3oh-2012} also observed both pairs of outflows from IRAS 4B in \ch{CH3OH} emission, but noted the northwest-southeast outflow was slightly shifted off axis, which is also corroborated here. The cause of this asymmetry is currently unknown, but it is possible that the northwest slightly blueshifted lobe is affected by the outflows from IRAS 4A \citep{sakai-ch3oh-2012}. Therefore it can be concluded that IRAS 4B has a northwest-southeast outflow, in addition to the north-south jet.

There is also some discrepancy between these findings and the literature as to whether IRAS 4B is a multi-component or a single source based on continuum observations. Some suggest that 4B shows evidence of a multi-component system.  This was first mentioned by \cite{lay-1995}, who stated that their CSO-JCMT (James Clerk Maxwell Telescope) observations of 4B could not be well described with a single source model and were better described with a triplet or higher order source model. It should be noted that those single-baseline observations did not resolve IRAS 4B$'$. However, follow-up $\sim1''$ interferometric observations could never resolve more than a single continuum source from 4B, although elongation of the continuum to the northwest or southeast direction is common, which is the same as seen here in Figure \ref{fig:continuum-image} \citep{looney-2001,di_francesco_2001,jorgensen_prosac_2007,choi-2011}. The double bipolar outflows of 4B suggests that it is indeed a binary protostar, with one protostar driving the north-south jet and one protostar driving the northwest-southeast outflow. The triplet order multiplicity that \cite{lay-1995} concluded could be explained by one young stellar object in 4B$'$ (since it was unknown at that point) and the binary protostar system in 4B. Thus we concur with \cite{sakai-ch3oh-2012}, that a stronger, younger protostar drives the north-south jet, and a older, more evolved protostar drives the northwest-southeast outflow.  This suggests that IRAS 4B is a binary protostar system with two bipolar outflows.

Newly characterized in this work in its entirety is the east-west outflow driven by IRAS 4B$'$. The west portion of the outflow in the \ch{H2CO} emission can been seen in the maps reported by \cite{di_francesco_2001} and \cite{jorgensen_prosac_2007}, although the field of view for both do not include the east lobe of outflow. \cite{jorgensen_prosac_2007} also showed the west outflow in \ch{CS}, \ch{CH3OH}, and \ch{H2CO}, which is consistent with observations presented here. \cite{prodige-2026} captured the outflow in its entirety with molecular emission from \ch{SiO}, \ch{CO}, \ch{H2CO}, and \ch{CH3OH}, but did not discuss it further. As stated above, \ch{CH3OH} and \ch{H2CO} are enhanced in shocks, when ice mantle tracers are released from grains through sputtering. In Figure \ref{fig:moment-0-single}, the east-west outflow is also seen in \ch{^{13}CH3OH}/\ch{C^{34}S}, \ch{SO}, \ch{SiO}, \ch{CS}, and faintly in \ch{HDCO}. The velocity of outflowing \ch{SiO} gas in the east-west outflow is around 10 km s$^{-1}$, less than the north-south jet of 4B. Besides the outflows shown in the \ch{CH3OH} and \ch{H2CO} parameter maps, and the multiple molecular emission outflow tracers seen in the moment 0 maps, IRAS 4B$'$ does not show any other activity. Within the spectral region of these observations, a single emission line of SiO was observed. No  iCOM emission was observed. In the literature, the only other molecules claimed to be detected in 4B$'$ include \ch{HCN} and \ch{^{13}CO} \citep{choi-2001}. This source has no detections in the infrared, so the SED is incomplete for a YSO classification. \cite{choi-2011} suggests IRAS 4B$'$ could either be a very low luminosity object (VeLLO) or the first hydrostatic core (FHSC). Thus, these observations show that IRAS 4B$'$ drives a west-east outflow, but has no observable hot iCOM emission in the observed frequencies cospatial with the warm inner envelope of the continuum, suggesting it is a very cold young stellar object with heavy dust extinction and low molecular column densities. 

\subsection{Previous NOEMA Studies of CSO Sources}
This work is a part of a series of NOEMA follow-up observations done on the northern most star-forming regions from a set of 30 targets previously observed with the CSO \citep{cso}.  \cite{will_w3_2023} and \cite{giese_mapping_2024} observed W3(OH) and W3(\ch{H2O}) within the high-mass star-forming region W3, and \cite{Morgan-2025} analyzed W75N(B).  Both studies were conducted with NOEMA using the same spectral setup used here. W3 and W75N(B) both have multiple cores, similar to IRAS 4B and 4B$'$. Table \ref{tab:sources-comparison} summarizes the derived molecular parameters across the star-forming regions analyzed so far.

\begin{deluxetable}{cccc}[ht]
        \tablecaption{Comparison of directly comparable molecules across the different star-forming regions analyzed so far with NOEMA based on \cite{cso}. \label{tab:sources-comparison}}
         \tablehead{\colhead{\textbf{Molecule}} & \colhead{\textbf{Source}}& \colhead{\textbf{Rotational Temperature}}&\colhead{\textbf{Column Density}}\\
         \colhead{}& \colhead{}& \colhead{\textbf{(K)}}& \colhead{\textbf{(cm$^{-2}$)}}
         }
         \startdata
         \ch{CH3OH} & IRAS 4B & - & $(2.6\pm0.7) \times 10^{17}$ $\dagger$ \\
          & W3(\ch{OH}) & $209\pm8$ & $(2.4\pm0.1) \times 10^{17}$ \\
          & W3(\ch{H2O}) & $187\pm2$ & $(2.41\pm0.03) \times 10^{17}$ \\
          & W75N(B) MM1a & $160\pm10$ & $(8.4\pm0.8) \times 10^{16}$ \\
          & W75N(B) MM1b & $205\pm3$ & $(1.17\pm0.02) \times 10^{18}$ \\
          & W75N(B) MM3 & - & - \\
         \hline
         \ch{HCOOCH3} & IRAS 4B & $210\pm10$ & $(7.4\pm0.6) \times 10^{15}$ \\
          & W3(\ch{OH}) & $110\pm30$ & $(1.9\pm0.6) \times 10^{16}$ \\
          & W3(\ch{H2O}) & $164\pm6$ & $(6.7\pm0.3) \times 10^{16}$ \\
          & W75N(B) MM1a & - & - \\
          & W75N(B) MM1b & $89\pm5$ & $(7.3\pm0.5) \times 10^{16}$ \\
          & W75N(B) MM3 & - & - \\
          \hline
         \ch{CH3OCH3} & IRAS 4B & $111\pm4$ & $(7.9\pm0.4) \times 10^{15}$ \\
          & W3(\ch{OH}) & $120\pm20$ & $(5.7\pm0.1) \times 10^{16}$ \\
          & W3(\ch{H2O}) & $108\pm3$ & $(1.07\pm0.03) \times 10^{17}$ \\
          & W75N(B) MM1a & - & - \\
          & W75N(B) MM1b & -- & - \\
          & W75N(B) MM3 & - & - \\
          \hline
         \ch{HNCO} & IRAS 4B & $320\pm60$ & $(9\pm2) \times 10^{14}$ \\
         & W3(\ch{OH}) & - & - \\
          & W3(\ch{H2O}) & - & - \\
         & W75N(B) MM1a & $200\pm10$ & $(3.5\pm0.3) \times 10^{16}$ \\
         & W75N(B) MM1b & $110\pm20$ & $(1.5\pm0.2) \times 10^{16}$ \\
         & W75N(B) MM3 & $200\pm100$ & $(1.0\pm0.7) \times 10^{15}$ \\
         \enddata
    \tablecomments{Data from W3 is from \cite{giese_mapping_2024} while W75N(B) is from \cite{Morgan-2025}. Derived parameters are from the peak continuum flux pixel of each source. $\dagger$ indicates the \ch{CH3OH} column density is an upper limit based on the \ch{^{13}CH3OH} column density from Table \ref{tab:gobasic-table} and assuming a \ch{^{12}C}/\ch{^{13}C} $=68\pm15$ \citep{Milam_2005}. \cite{busch_prodige_2025} suggests the \ch{^{12}C}/\ch{^{13}C} could be as low as $4\pm1$.}
\end{deluxetable}

In W3(OH) and W3(\ch{H2O}), the six most abundant molecules detected that accounted for the majority of the spectral features were \ch{C2H5CN}, \ch{CH3CN}, \ch{CH3OCH3}, \ch{CH3OH}, \ch{HCOOCH3}, and \ch{SO2}, all of which were also detected in IRAS 4B.  W75N(B) has multiple cores including MM1a, MM1b, MM2, and MM3. The MM1b core is the most similar to IRAS 4B, with well fitting parameters determined for \ch{CH3CN}, \ch{CH3OCH3}, \ch{CH3OH}, \ch{HCOOCH3}, \ch{HNCO}, and \ch{SO2}, all of which were also detected in IRAS 4B. MM1b has well fitting parameters for \ch{CH3CN}, \ch{CH3OH}, \ch{HCCCNv7}, \ch{HNCO}, \ch{NH2CHO}, \ch{OCS}, \ch{SO}, \ch{SO2}, and \ch{^{34}SO2}, most of which were also detected in IRAS 4B.  \ch{HCCCNv7}, \ch{^{34}SO2}, and \ch{OCS} are the exceptions, though  a single transition of \ch{OCS} $J$= 12--11 did fall within a high-resolution window and was imaged (see Figure 9). MM3 had only \ch{CH3CN}, \ch{HNCO}, and \ch{SO2}, which again have all been detected in IRAS 4B. 

It is difficult to draw quantitative comparisons with this list of sources because only a handful of molecules are detected in all three regions, and not all of these molecules are detected in all cores. As more sources in the larger survey based on the 30 sources in \cite{cso} are analyzed, a more complete picture of the chemical inventory and trends across star-forming regions will become apparent. Statistical analysis will hopefully reveal driving characteristics between chemical formation pathways and the physical conditions of these astronomical environments that are the birthplaces of star and planet formation. 

\section{Conclusions}\label{sec:conclusions}
This work analyzes observational results from the NOEMA interferometer to investigate the spatial distribution of interstellar complex organic molecules within the star-forming region NGC 1333 IRAS 4B and 4B$'$. Results from these observations include spectral analysis, molecular line detections, integrated intensity maps, correlation matrices, as well as derived physical values for molecules such as rotational temperature, column density, velocity shift, and spectral line width (FWHM). Conclusions are summarized as follows:
\begin{enumerate}
    \item High-resolution NOEMA observations revealed hundreds of molecular transitions and multiple interstellar complex organic molecules toward the warm inner envelope of IRAS 4B, marking 4B as a rich source in organic material. A total of 46 molecules were identified while 21 were firmly detected via the identification of multiple transitions. A single molecular emission line of \ch{SiO} was observed toward the continuum source of IRAS 4B$'$. Derived parameters such as rotational temperature (K), column density (cm$^{-2}$), spectral line width (FWHM in km s$^{-1}$), and velocity shift (km s$^{-1}$) were calculated for 11 molecules toward the inner envelope of 4B using the Global Optimization and Broadband Analysis Software for Interstellar Chemistry (GOBASIC), including \ch{CH3OH} and its isotopologues (\ch{^{13}CH3OH}, \ch{CH2DOH}, \ch{CH3OD}), \ch{HCOOCH3}, \ch{CH3OCH3}, \ch{CH3CHO}, \ch{C2H5OH}, \ch{CH2OHCHO}, \ch{CH3COCH3}, and \ch{HNCO}.
    \item Calculated parameter maps of column density, rotational temperature, and velocity shift for nine molecules (excluding \ch{CH3OH} due to optical depth and \ch{CH3COCH3} due to line blending) toward 4B and two molecules (\ch{CH3OH} and \ch{H2CO}) of the larger region surrounding 4B and 4B$'$ were derived. 
    \item Most molecular emission, especially for the iCOMs, trace the warm inner envelope of IRAS 4B, and is cospatial with the peak continuum flux. There is a strong spatial correlation between the continuum emission and organics in 4B and also between various iCOMs. Molecules that do not have a strong correlation with the continuum and instead trace the outflows include some sulfur bearing species, \ch{SiO}, and extended emission from low E$_u$ (K) \ch{CH3OH} and \ch{H2CO} transitions.
    \item Parameter maps of \ch{CH3OH} and \ch{H2CO} and integrated intensity maps of outflow molecular tracers revealed three bipolar outflows from IRAS 4B and 4B$'$. IRAS 4B is a binary protostar system that drives a pair of bipolar outflows. Based on the molecular tracers and velocity of collimated gas, the north-south molecular emission is most likely a higher velocity jet driven by a younger protostar within 4B. The northwest-southeast lower velocity outflow is driven by a second protostar, older and more evolved within 4B. IRAS 4B$'$ is shown to have a bipolar outflow in a east-west orientation, which is characterized for the first time in this work. No molecular iCOM emission was observed to trace the warm inner envelope. IRAS 4B$'$ is likely a very low luminosity object or the first hydrostatic core stage of a Class 0 protostar. 
\end{enumerate}

\section*{Disclaimer}
No generative artificial intelligence (AI) was used in the writing of this work or in the analysis. The authors declare no competing financial interest.

\begin{acknowledgments}
The authors acknowledge support from SLWW's startup funding from the University of Wisconsin-Madison. UW-Madison has a partnership with NOEMA, through which the observing time was allocated. This work is based on observations carried out under project number S22AE with the IRAM NOEMA Interferometer. IRAM is supported by INSU/CNRS (France), MPG (Germany) and IGN (Spain). We thank Dr. Laure Bouscasse for her assistance in data calibration and reduction. Support for this research was provided by the University of Wisconsin-Madison, Office of the Vice Chancellor for Research with funding from the Wisconsin Alumni Research Foundation. Additional computing resources were provided by the\cite{chtc} at the University of Wisconsin-Madison. Part of this research was carried out at the Jet Propulsion Laboratory, California Institute of Technology, under a contract with the National Aeronautics and Space Administration (80NM0018D0004).
\end{acknowledgments}

\begin{contribution}
C.C.S. conducted the data reduction, data analysis, modified existing python scripts, wrote new python scripts, and wrote the manuscript.
M.M.G. wrote the original data analysis python scripts, provided valuable input during the data analysis process, and edited the manuscript.
S.L.W.W. provided the original research framework, provided valuable input during the data analysis process, and edited the manuscript.
D.C.L contributed to the research framework, provided valuable input during the interpretation of the results, and edited the manuscript.
\end{contribution}

\newpage
\begin{appendix} 
\section{Physical Parameter Maps}\label{sec:appendix-parameter-maps}
Physical derived parameter maps and their associated uncertainty maps of rotational temperature (K), column density (cm$^{-2}$), and velocity shift (km s$^{-1}$) of the 7 other molecules from Subsection \ref{subsec:parameter-maps} toward the inner envelope of IRAS 4B. These molecules include \ch{^{13}CH3OH}, \ch{CH3OD}, \ch{HCOOCH3}, \ch{CH3OCH3}, \ch{C2H5OH}, \ch{CH3OCH3}, and \ch{HNCO}. Uncertainty maps of \ch{CH3CHO} and \ch{CH2DOH} from Figure \ref{fig:parameter-maps-4b} are presented first. 

\begin{figure}[ht]
    \centering
    \includegraphics[width=0.49\linewidth]{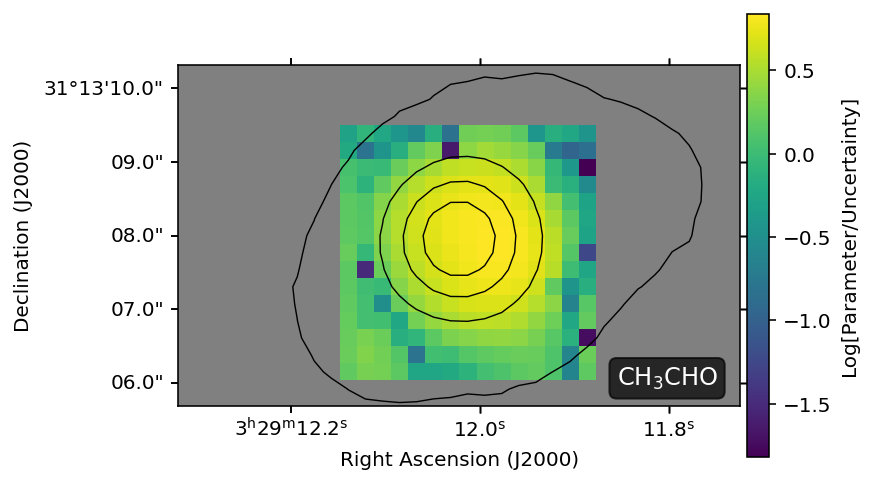}
    \includegraphics[width=0.49\linewidth]{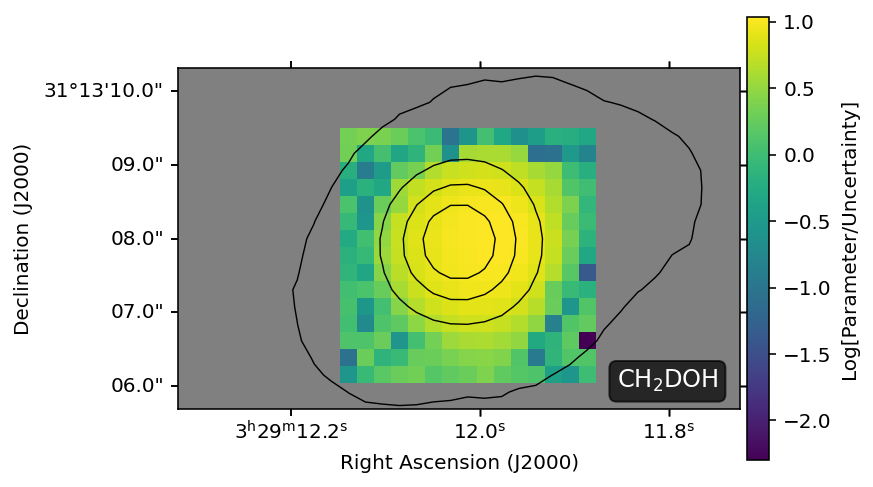}
    \includegraphics[width=0.49\linewidth]{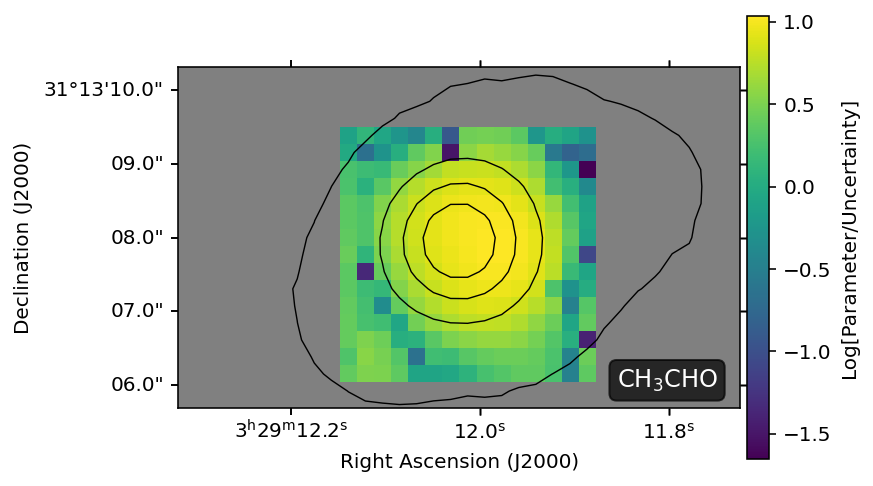}
    \includegraphics[width=0.49\linewidth]{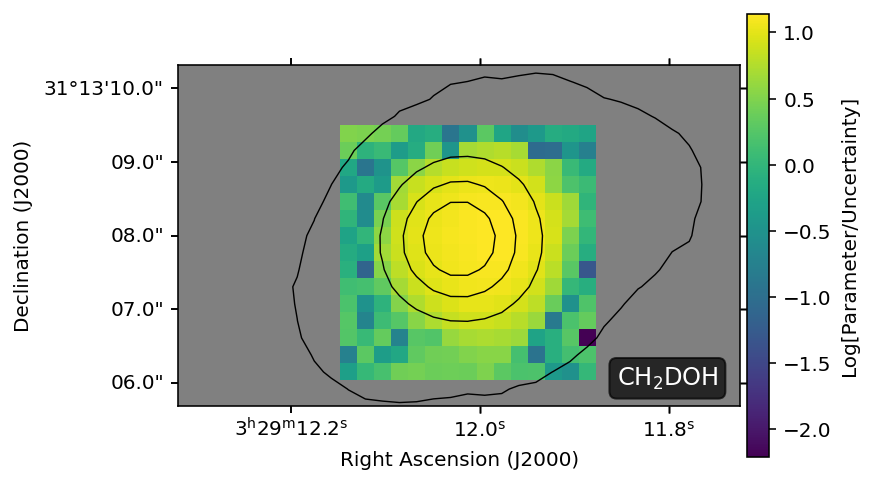}
    \includegraphics[width=0.49\linewidth]{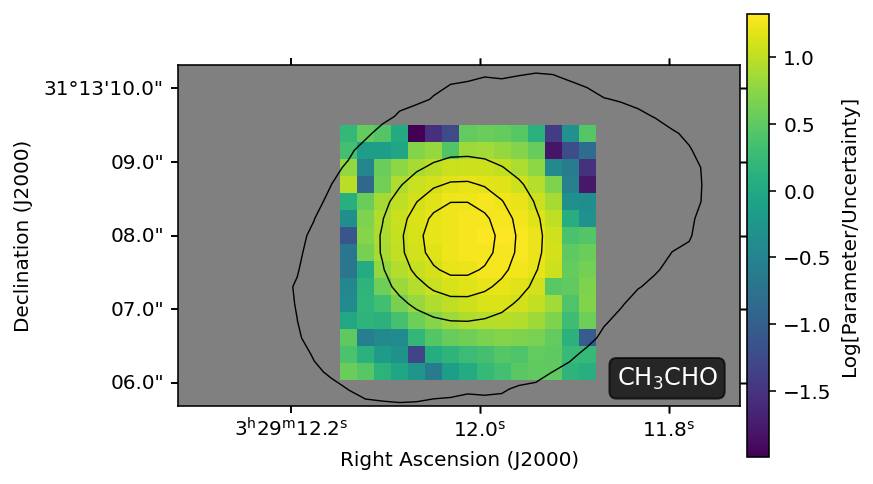}
    \includegraphics[width=0.49\linewidth]{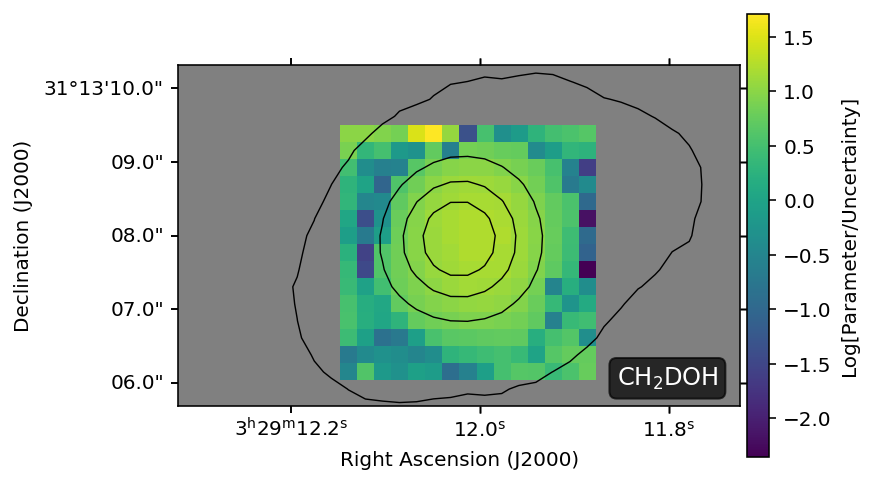}
    \caption{Uncertainty maps of \ch{CH3CHO} (left) and \ch{CH2DOH} (right) toward IRAS 4B, in order from top to bottom: column density (cm$^{-2}$), rotational temperature (K), and velocity shift (km s$^{-1}$). White contours are 5 evenly spaced levels (1 level $=0.0599$ Jy beam$^{-1}$) from $3 \times \sigma_{cont}$ ($\sigma_{cont} = 0.0013$ Jy beam$^{-1}$) to the max continuum flux at 0.2436 Jy beam$^{-1}$. The color bar corresponds to the derived parameter divided by the determined uncertainty plotted on a logarithmic scale. A ratio of 1 between these values would be 0 on the scale, and a ratio of 3 would be 0.48.}
    \label{fig:uncertainty-maps-ch2doh-ch3cho}
\end{figure}

\begin{figure}[ht]
    \centering
    \includegraphics[width=0.49\linewidth]{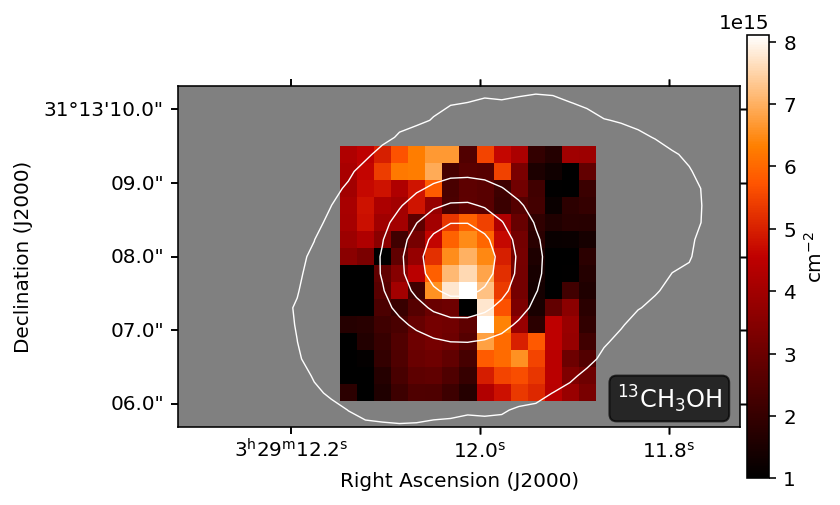}
    \includegraphics[width=0.49\linewidth]{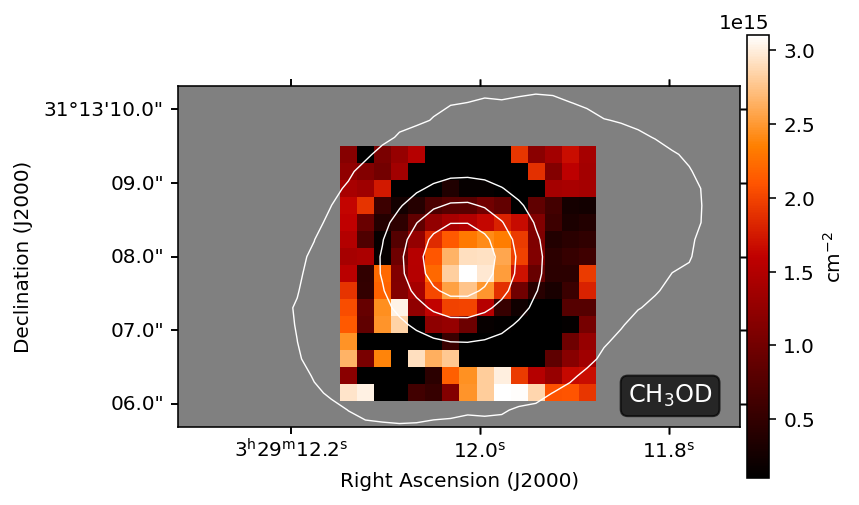}
    \includegraphics[width=0.49\linewidth]{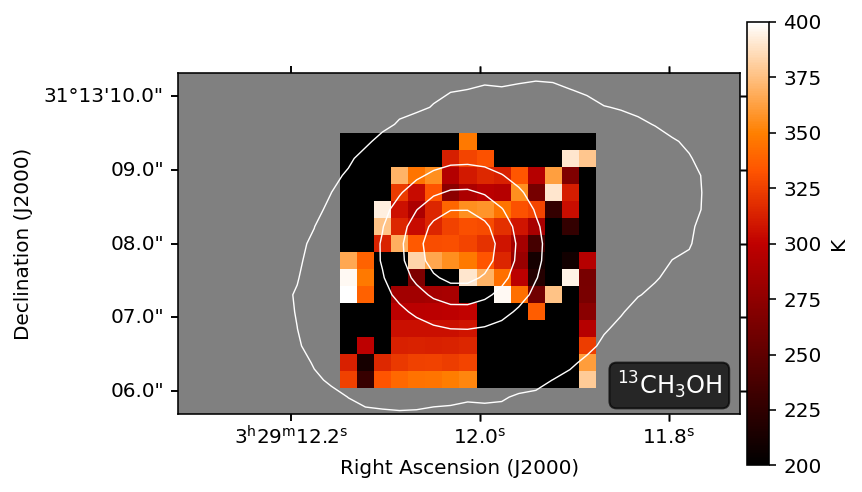}
    \includegraphics[width=0.49\linewidth]{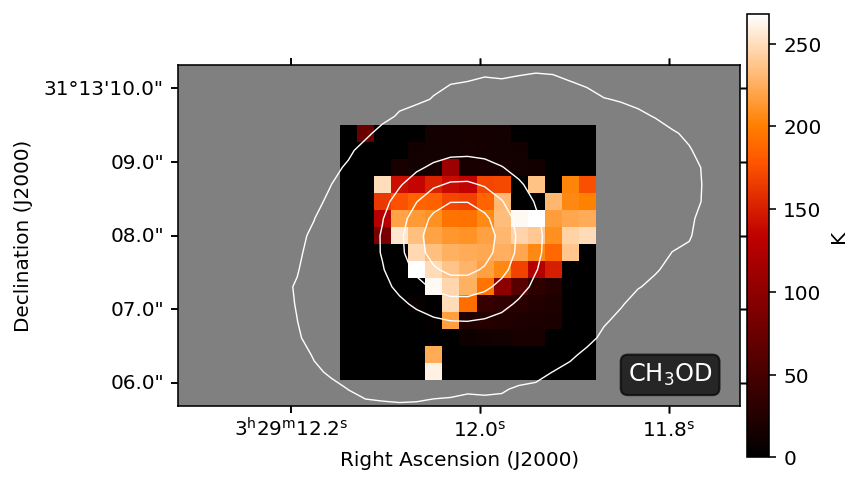}
    \includegraphics[width=0.49\linewidth]{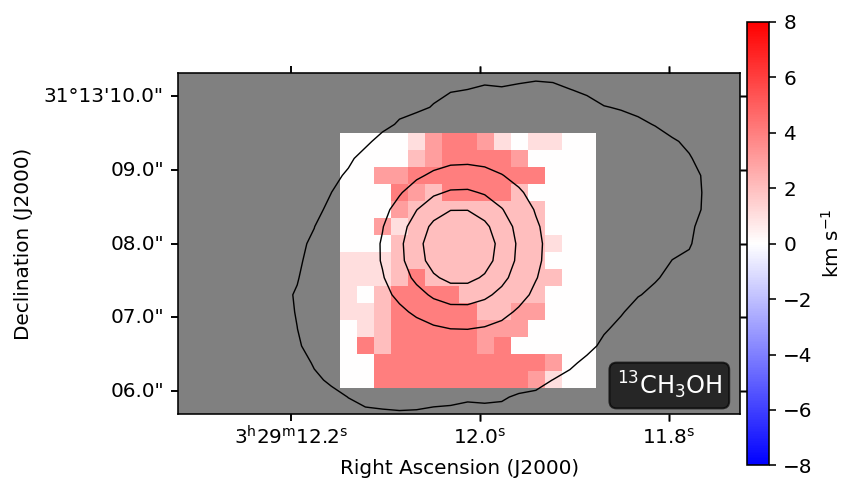}
    \includegraphics[width=0.49\linewidth]{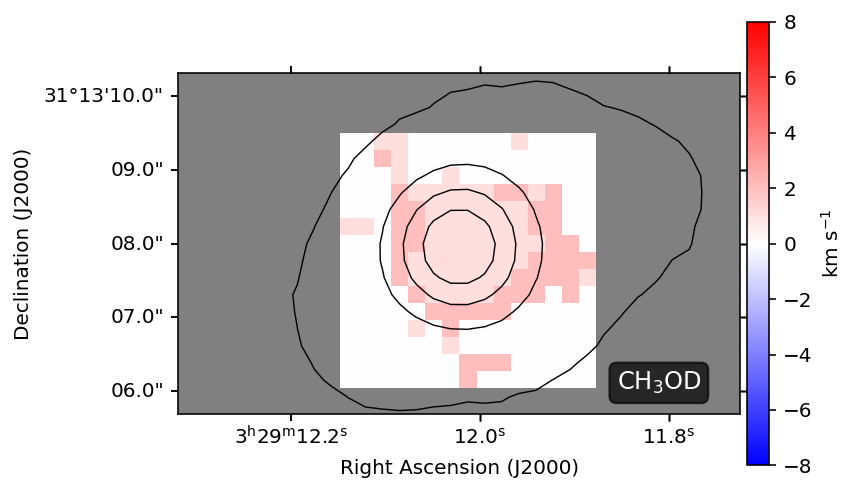}
    \caption{Parameter maps of \ch{^{13}CH3OH} (left) and \ch{CH3OD} (right) toward IRAS 4B, in order from top to bottom: column density (cm$^{-2}$), rotational temperature (K), and velocity shift (km s$^{-1}$). White and black contours are 5 evenly spaced levels (1 level $=0.0599$ Jy beam$^{-1}$) from $3 \times \sigma_{cont}$ ($\sigma_{cont} = 0.0013$ Jy beam$^{-1}$) to the max continuum flux at 0.2436 Jy beam$^{-1}$. }
    \label{fig:parameter-maps-13ch3oh-ch3od}
\end{figure}

\begin{figure}[ht]
    \centering
    \includegraphics[width=0.49\linewidth]{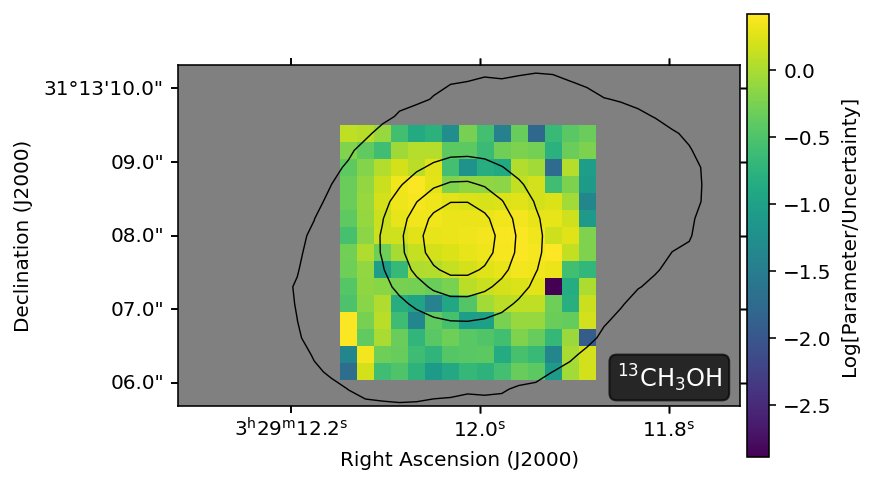}
    \includegraphics[width=0.49\linewidth]{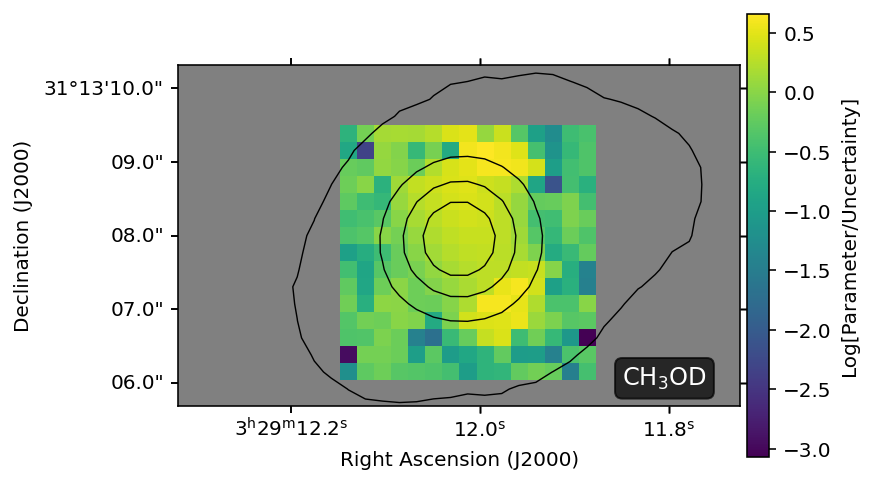}
    \includegraphics[width=0.49\linewidth]{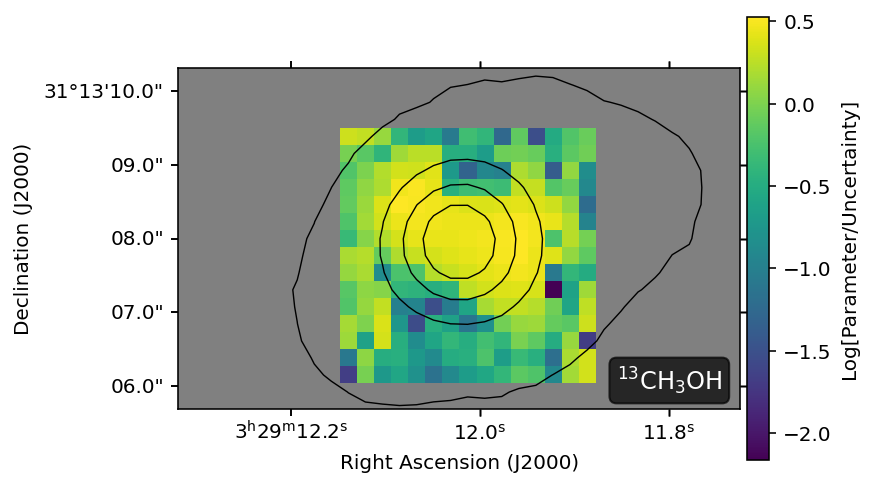}
    \includegraphics[width=0.49\linewidth]{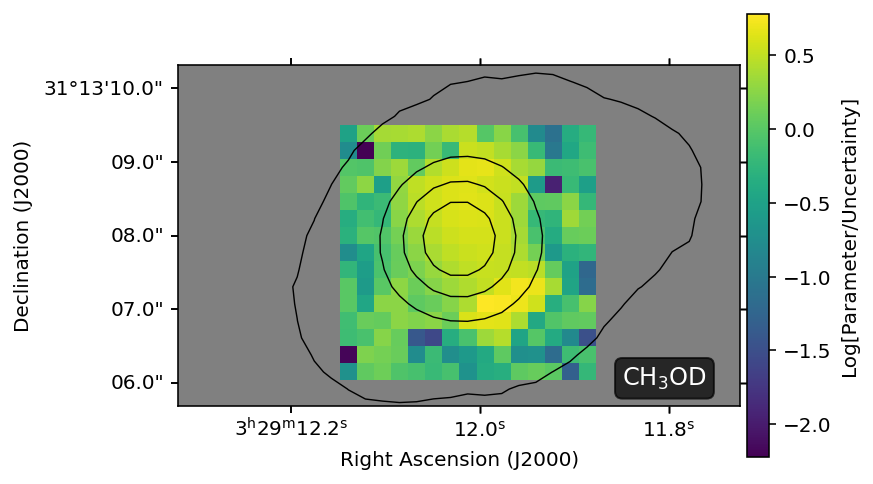}
    \includegraphics[width=0.49\linewidth]{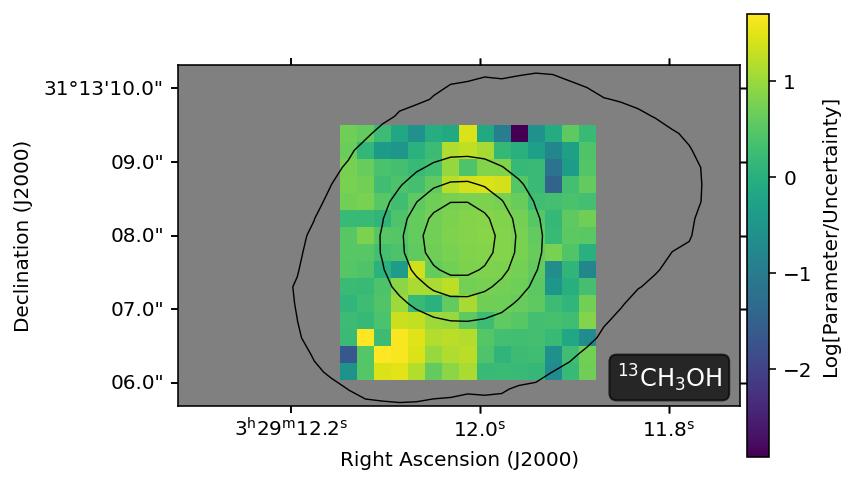}
    \includegraphics[width=0.49\linewidth]{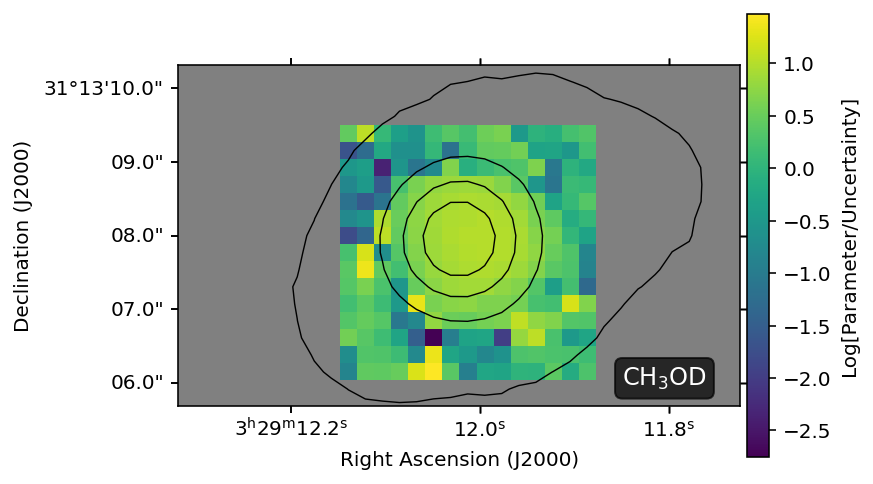}
    \caption{Uncertainty maps of \ch{^{13}CH3OH} (left) and \ch{CH3OD} (right) toward IRAS 4B, in order from top to bottom: column density (cm$^{-2}$), rotational temperature (K), and velocity shift (km s$^{-1}$). White contours are 5 evenly spaced levels (1 level $=0.0599$ Jy beam$^{-1}$) from $3 \times \sigma_{cont}$ ($\sigma_{cont} = 0.0013$ Jy beam$^{-1}$) to the max continuum flux at 0.2436 Jy beam$^{-1}$. The color bar corresponds to the derived parameter divided by the determined uncertainty plotted on a logarithmic scale. A ratio of 1 between these values would be 0 on the scale, and a ratio of 3 would be 0.48.}
    \label{fig:uncertainty-maps-13ch3oh-ch3od}
\end{figure}

\begin{figure}[ht]
    \centering
    \includegraphics[width=0.49\linewidth]{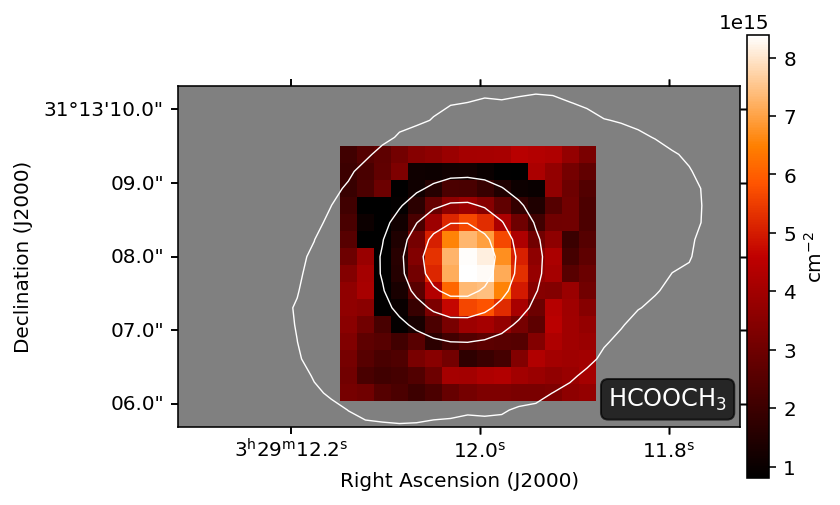}
    \includegraphics[width=0.49\linewidth]{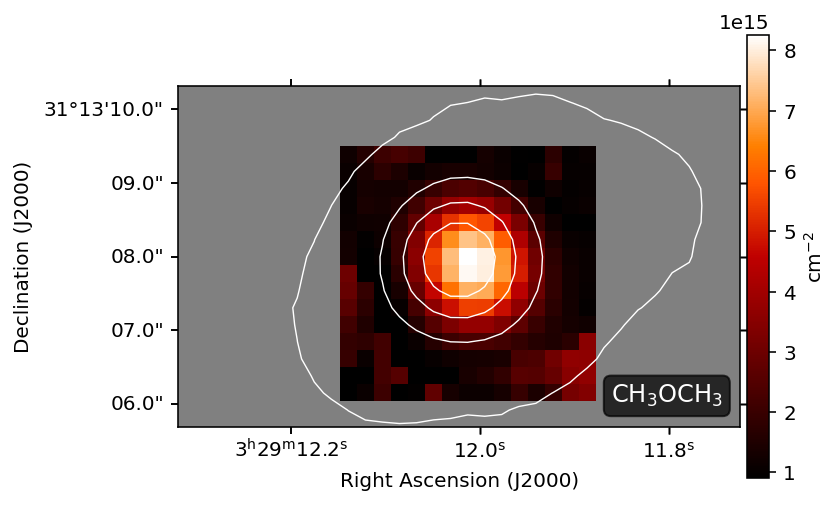}
    \includegraphics[width=0.49\linewidth]{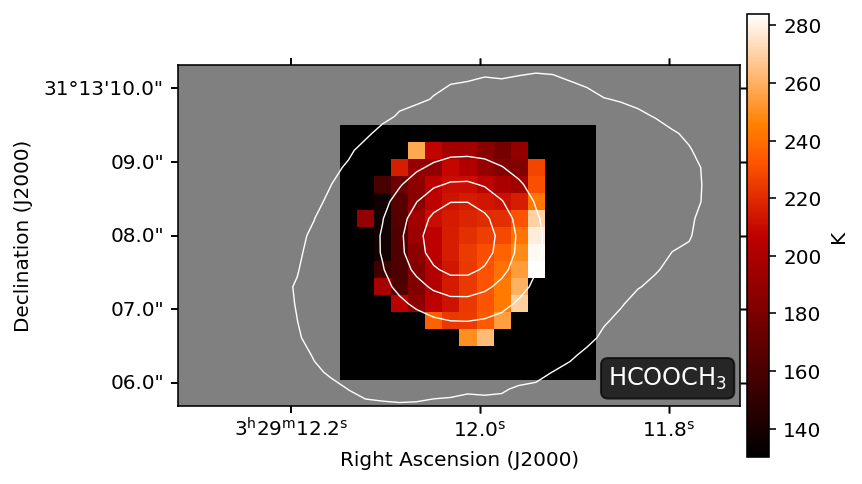}
    \includegraphics[width=0.49\linewidth]{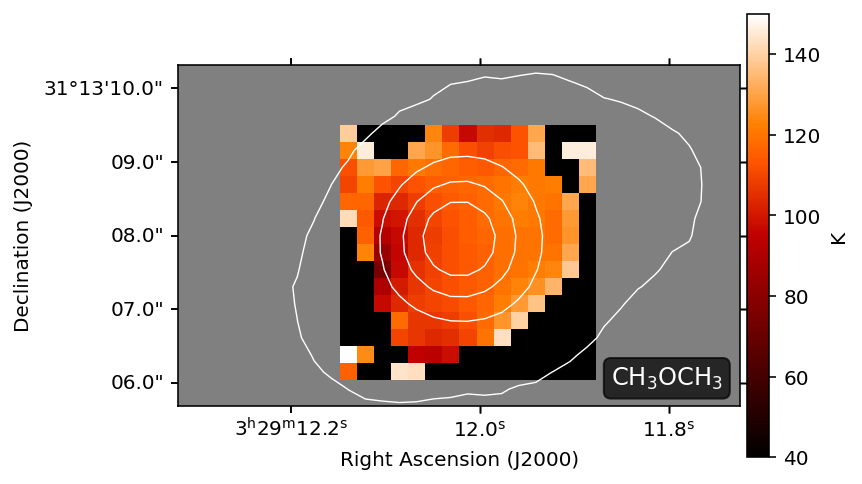}
    \includegraphics[width=0.49\linewidth]{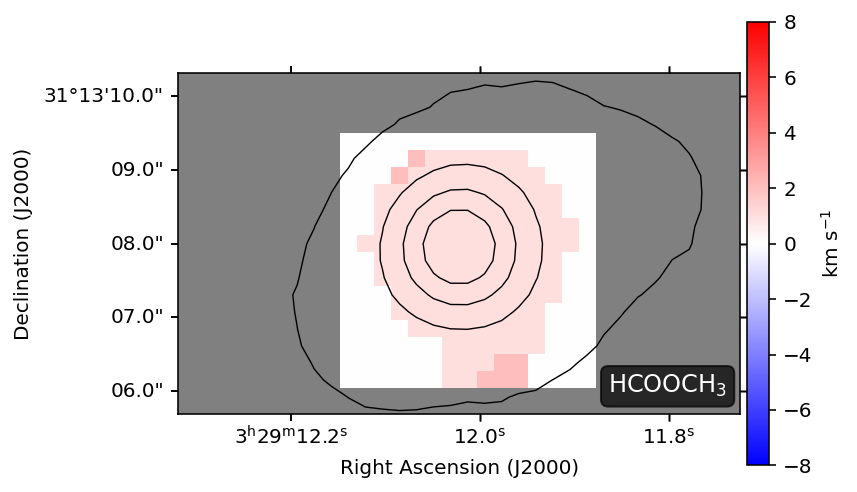}
    \includegraphics[width=0.49\linewidth]{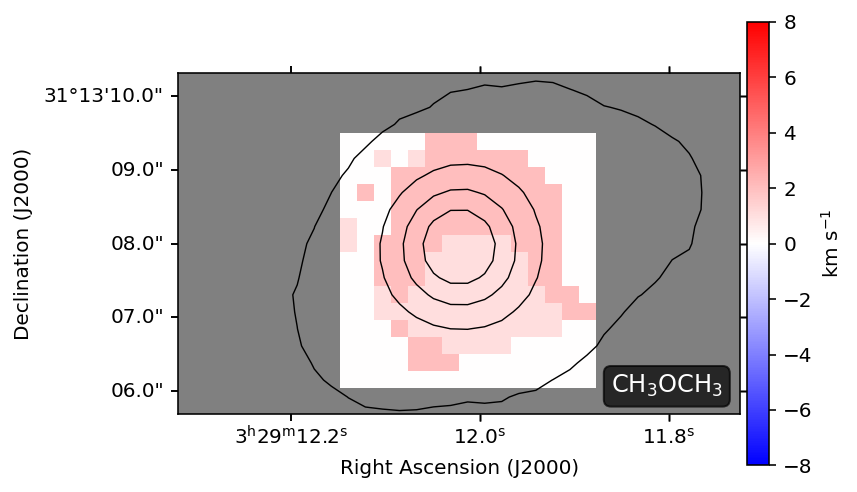}
    \caption{Parameter maps of \ch{HCOOCH3} (left) and \ch{CH3OCH3} (right) toward IRAS 4B, in order from top to bottom: column density (cm$^{-2}$), rotational temperature (K), and velocity shift (km s$^{-1}$). White and black contours are 5 evenly spaced levels (1 level $=0.0599$ Jy beam$^{-1}$) from $3 \times \sigma_{cont}$ ($\sigma_{cont} = 0.0013$ Jy beam$^{-1}$) to the max continuum flux at 0.2436 Jy beam$^{-1}$. }
    \label{fig:parameter-maps-hcooch3-ch3och3}
\end{figure}

\begin{figure}[ht]
    \centering
    \includegraphics[width=0.49\linewidth]{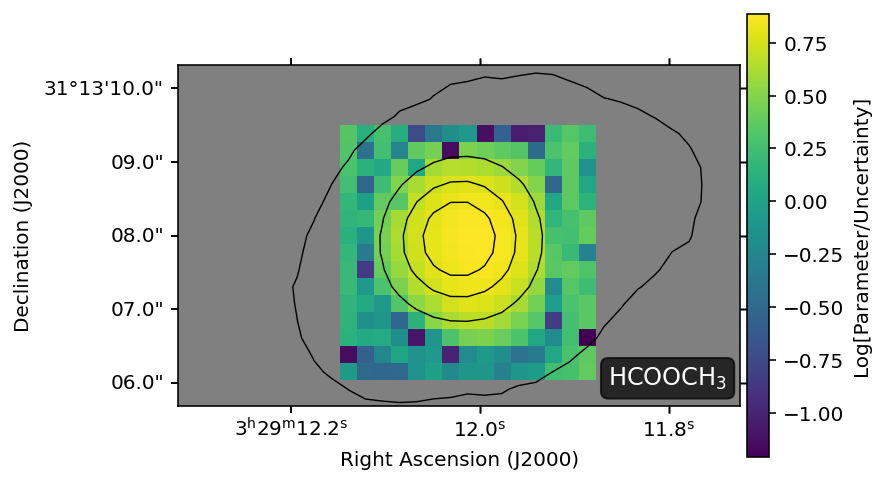}
    \includegraphics[width=0.49\linewidth]{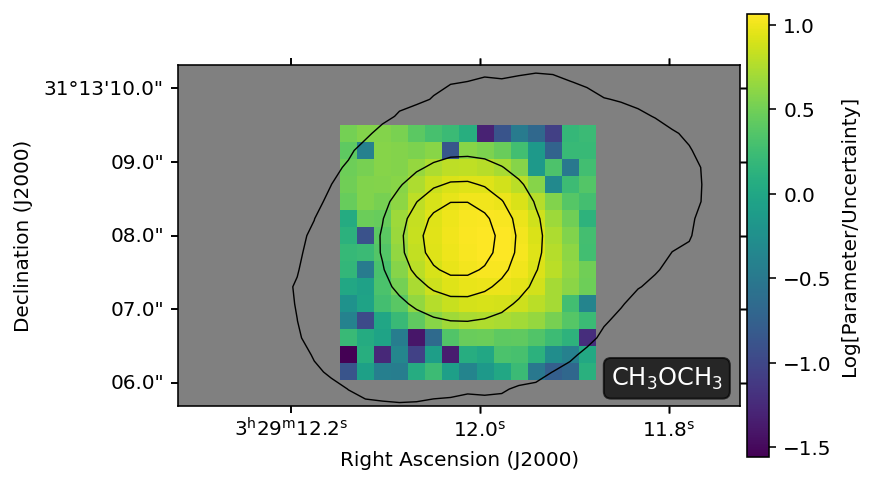}
    \includegraphics[width=0.49\linewidth]{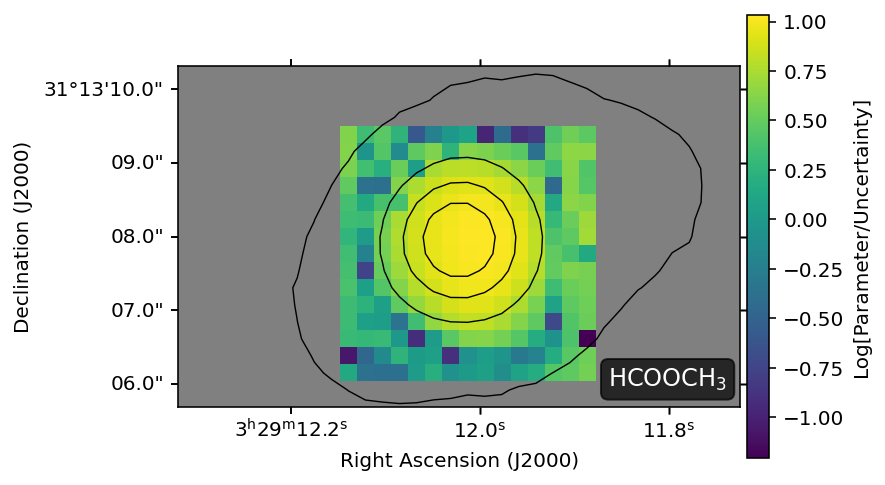}
    \includegraphics[width=0.49\linewidth]{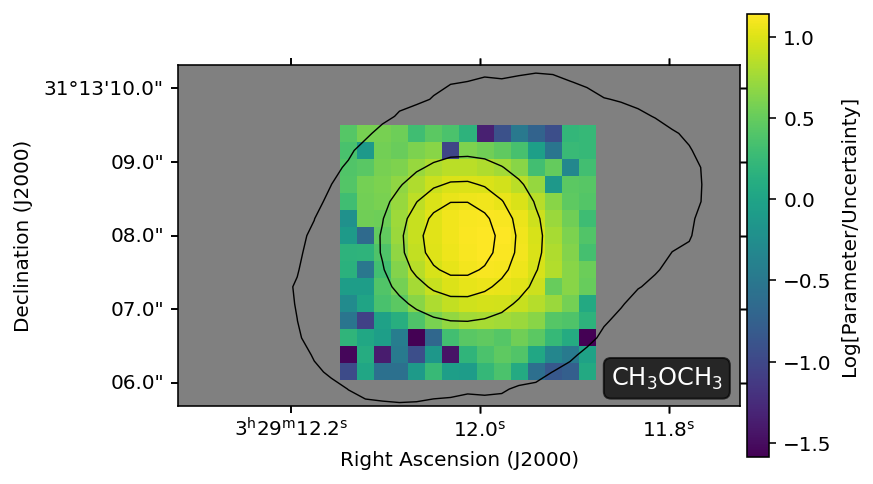}
    \includegraphics[width=0.49\linewidth]{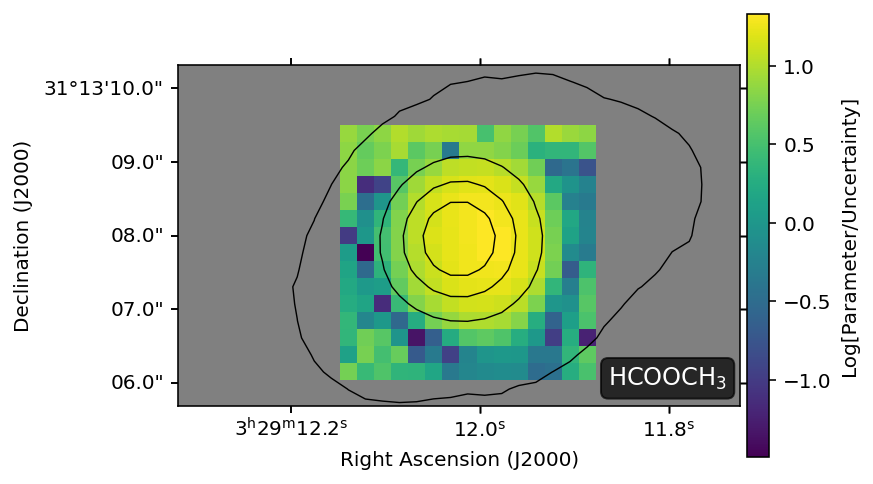}
    \includegraphics[width=0.49\linewidth]{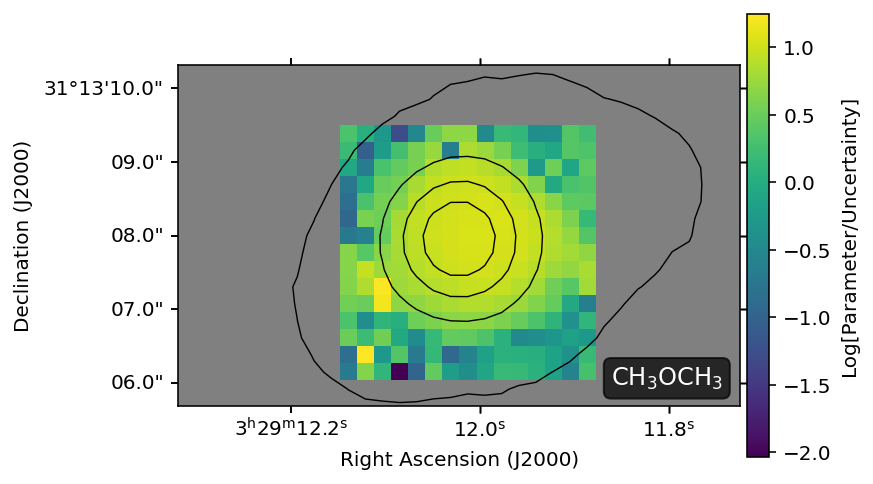}
    \caption{Uncertainty maps of \ch{HCOOCH3} (left) and \ch{CH3OCH3} (right) toward IRAS 4B, in order from top to bottom: column density (cm$^{-2}$), rotational temperature (K), and velocity shift (km s$^{-1}$). White contours are 5 evenly spaced levels (1 level $=0.0599$ Jy beam$^{-1}$) from $3 \times \sigma_{cont}$ ($\sigma_{cont} = 0.0013$ Jy beam$^{-1}$) to the max continuum flux at 0.2436 Jy beam$^{-1}$. The color bar corresponds to the derived parameter divided by the determined uncertainty plotted on a logarithmic scale. A ratio of 1 between these values would be 0 on the scale, and a ratio of 3 would be 0.48. }
    \label{fig:uncertainty-maps-hcooch3-ch3och3}
\end{figure}

\begin{figure}[ht]
    \centering
    \includegraphics[width=0.49\linewidth]{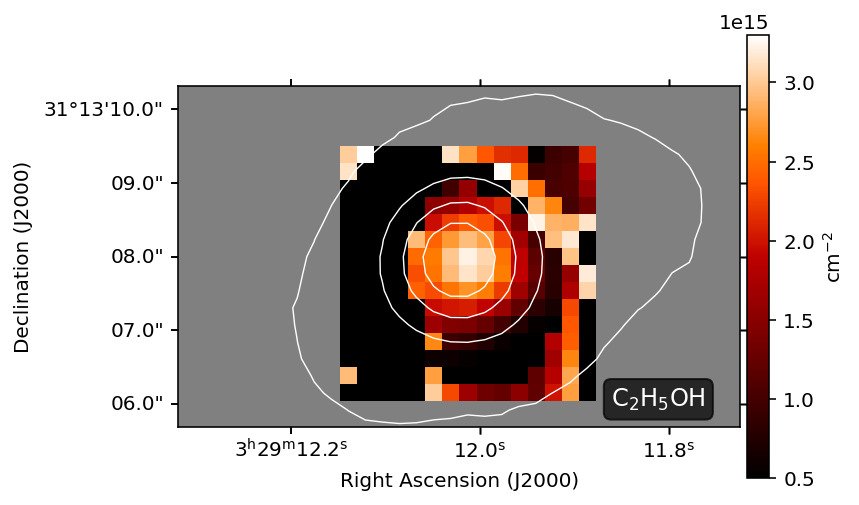}
    \includegraphics[width=0.49\linewidth]{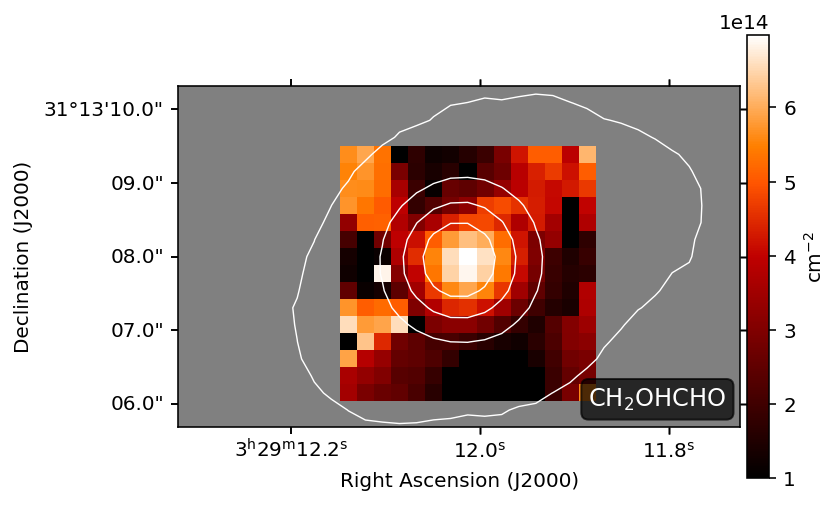}
    \includegraphics[width=0.49\linewidth]{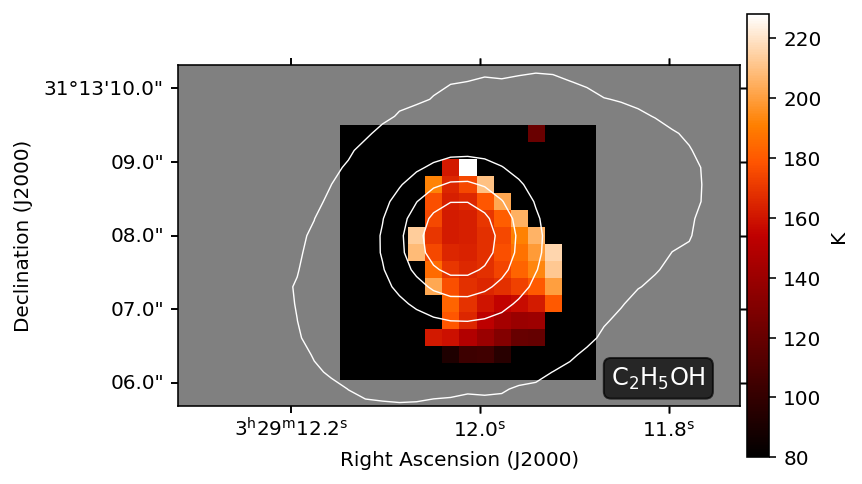}
    \includegraphics[width=0.49\linewidth]{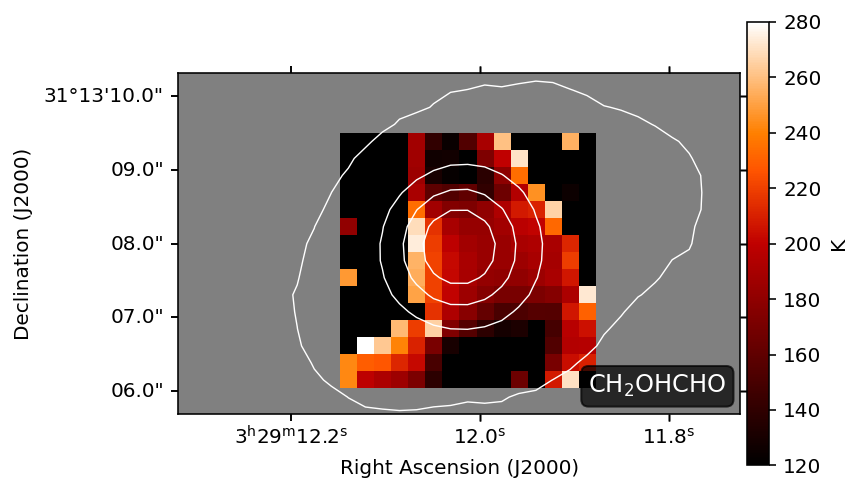}
    \includegraphics[width=0.49\linewidth]{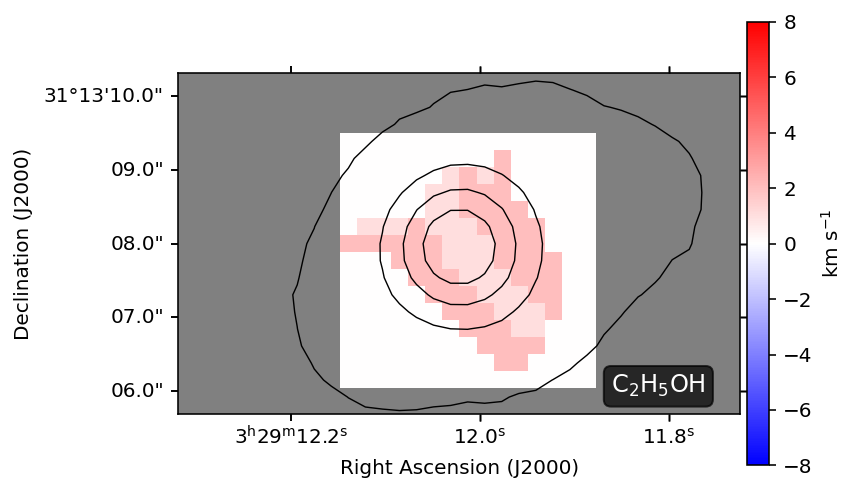}
    \includegraphics[width=0.49\linewidth]{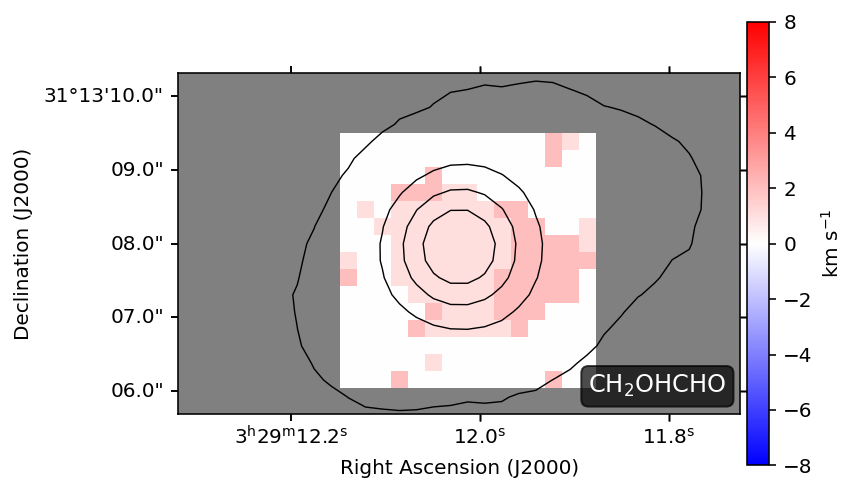}
    \caption{Parameter maps of \ch{C2H5OH} (left) and \ch{CH2OHCHO} (right) toward IRAS 4B, in order from top to bottom: column density (cm$^{-2}$), rotational temperature (K), and velocity shift (km s$^{-1}$). White and black contours are 5 evenly spaced levels (1 level $=0.0599$ Jy beam$^{-1}$) from $3 \times \sigma_{cont}$ ($\sigma_{cont} = 0.0013$ Jy beam$^{-1}$) to the max continuum flux at 0.2436 Jy beam$^{-1}$. }
    \label{fig:parameter-maps-c2h5oh-ch2ohcho}
\end{figure}

\begin{figure}[ht]
    \centering
    \includegraphics[width=0.49\linewidth]{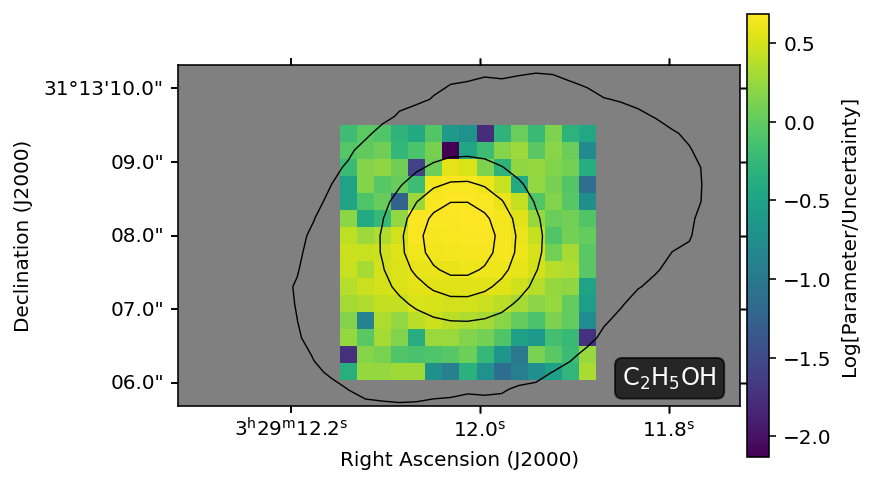}
    \includegraphics[width=0.49\linewidth]{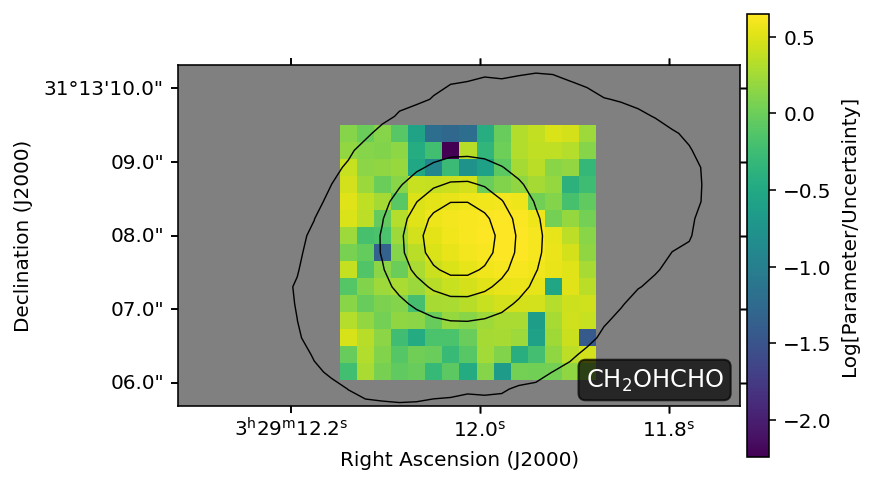}
    \includegraphics[width=0.49\linewidth]{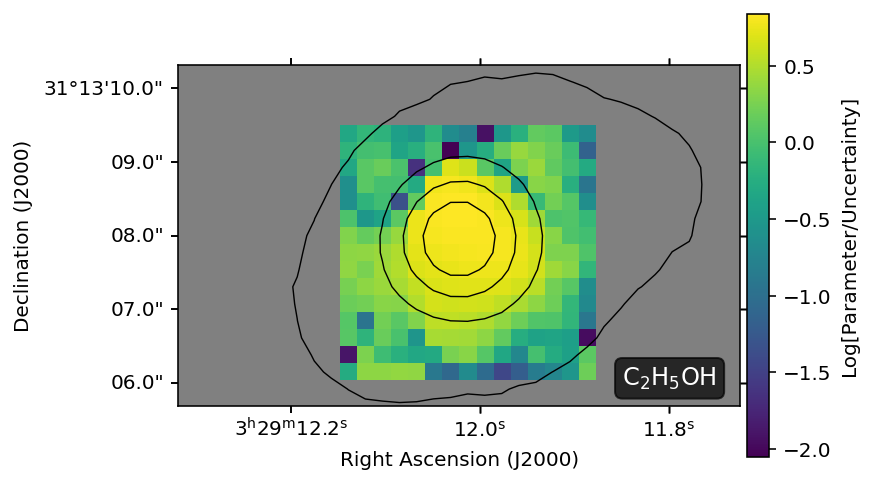}
    \includegraphics[width=0.49\linewidth]{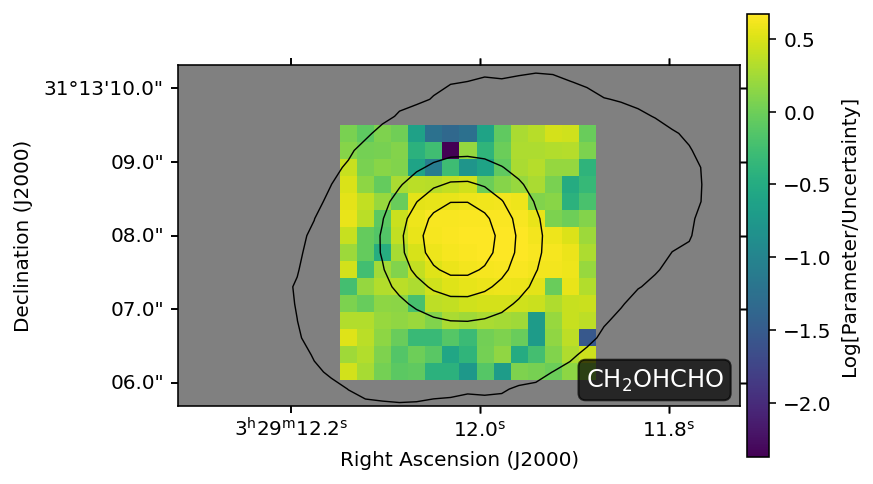}
    \includegraphics[width=0.49\linewidth]{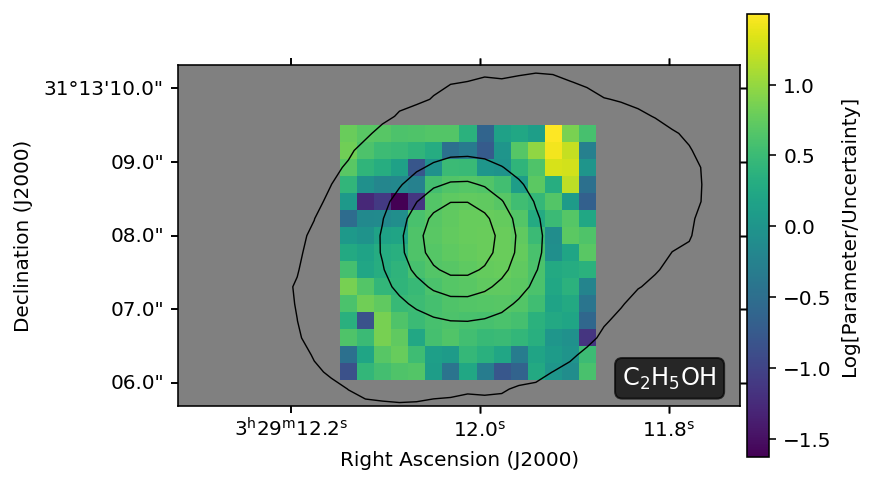}
    \includegraphics[width=0.49\linewidth]{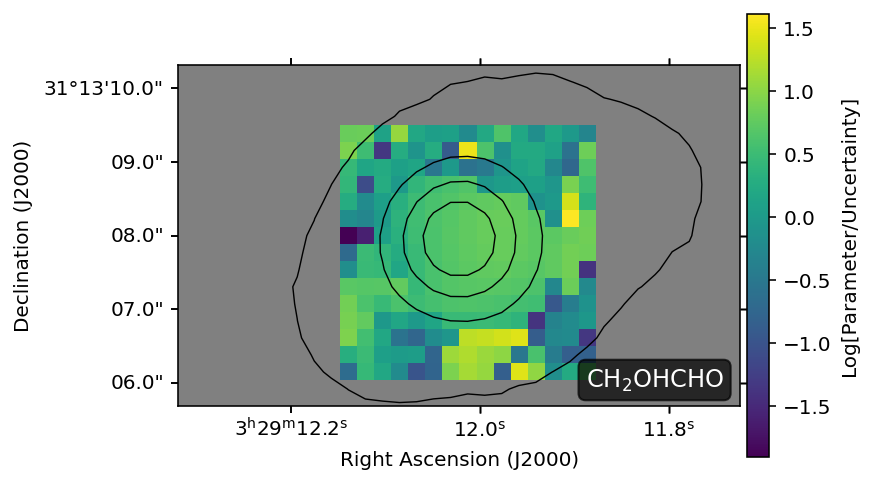}
    \caption{Uncertainty maps of \ch{C2H5OH} (left) and \ch{CH2OHCHO} (right) toward IRAS 4B, in order from top to bottom: column density (cm$^{-2}$), rotational temperature (K), and velocity shift (km s$^{-1}$). White contours are 5 evenly spaced levels (1 level $=0.0599$ Jy beam$^{-1}$) from $3 \times \sigma_{cont}$ ($\sigma_{cont} = 0.0013$ Jy beam$^{-1}$) to the max continuum flux at 0.2436 Jy beam$^{-1}$. The color bar corresponds to the derived parameter divided by the determined uncertainty plotted on a logarithmic scale. A ratio of 1 between these values would be 0 on the scale, and a ratio of 3 would be 0.48.}
    \label{fig:uncertainty-maps-c2h5oh-ch2ohcho}
\end{figure}

\begin{figure}[ht]
    \centering
    \includegraphics[width=0.49\linewidth]{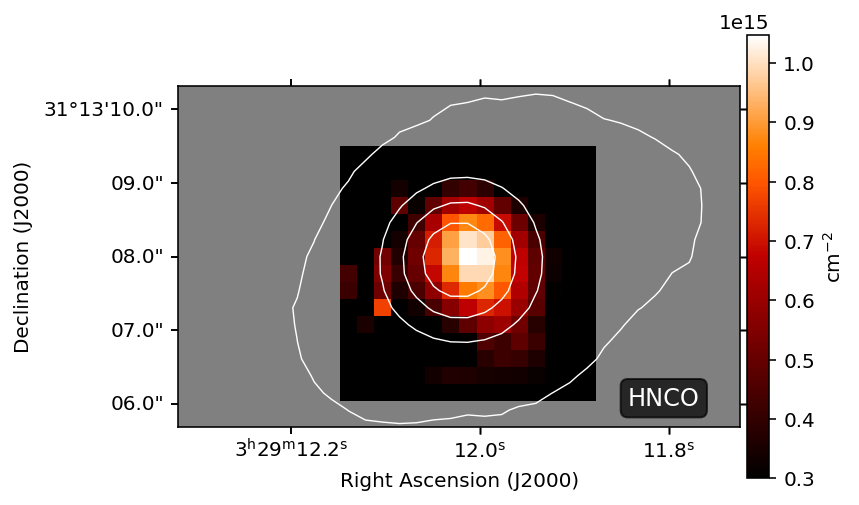}
    \includegraphics[width=0.49\linewidth]{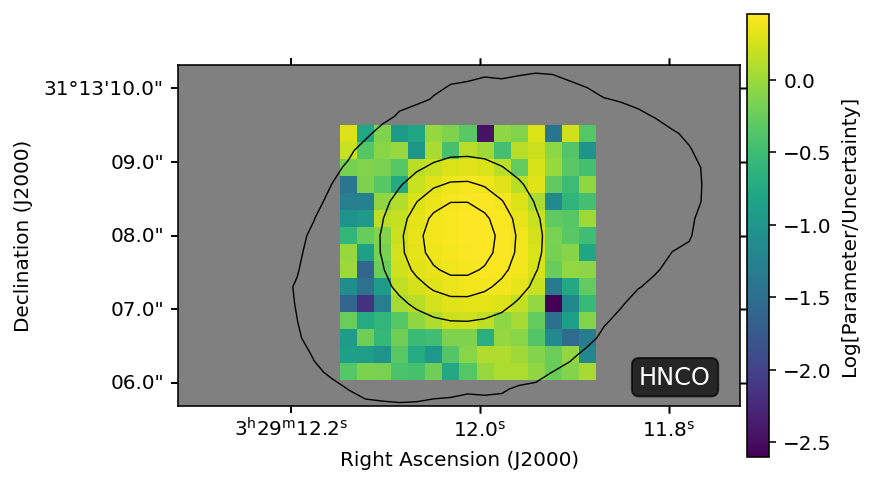}
    \includegraphics[width=0.49\linewidth]{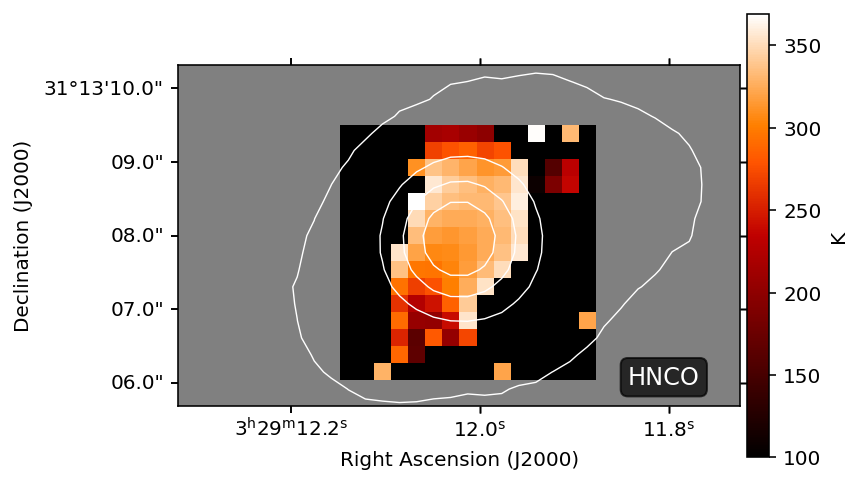}
    \includegraphics[width=0.49\linewidth]{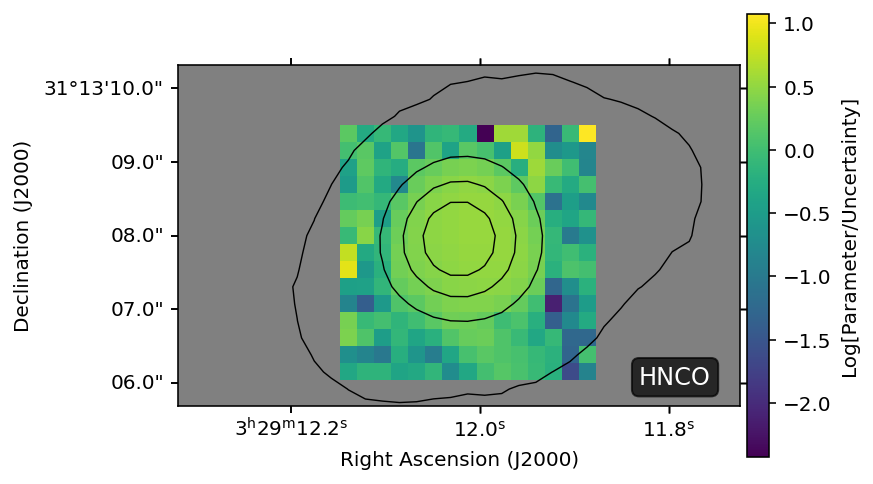}
    \includegraphics[width=0.49\linewidth]{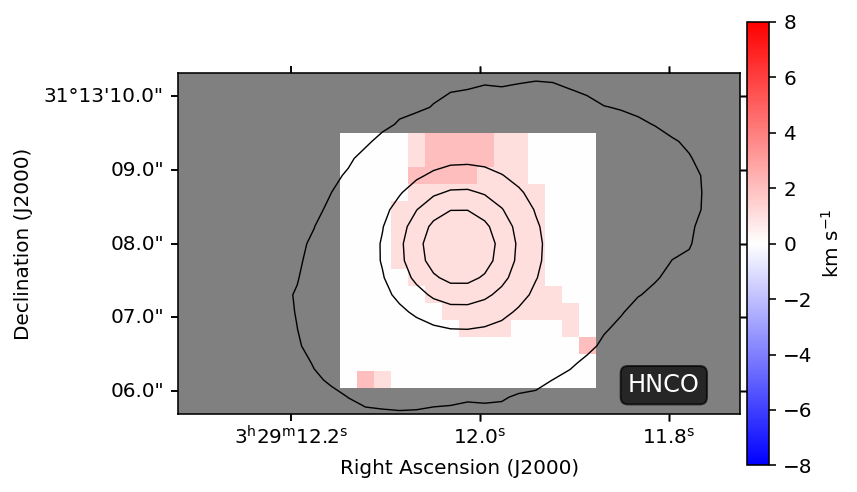}
    \includegraphics[width=0.49\linewidth]{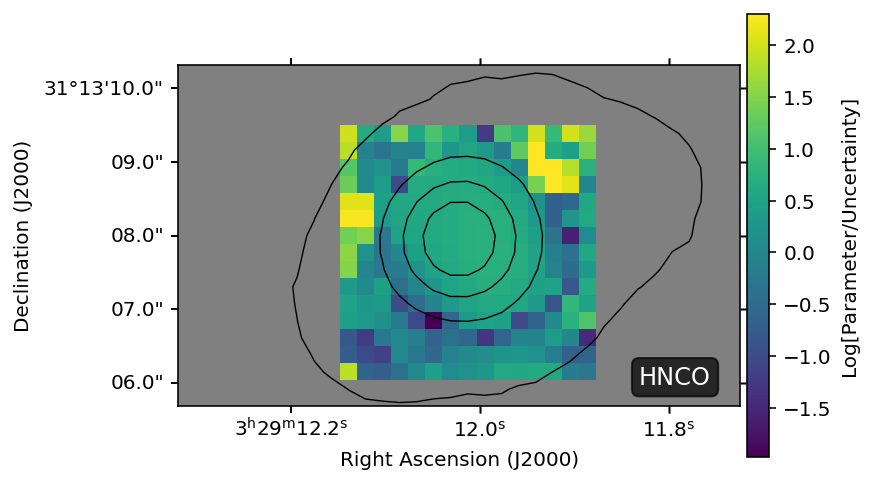}
    \caption{Parameter maps of \ch{HNCO} (left) and its uncertainty (right) toward IRAS 4B. Parameter maps in order from top to bottom: column density (cm$^{-2}$), rotational temperature (K), and velocity shift (km s$^{-1}$). White and black contours are 5 evenly spaced levels (1 level $=0.0599$ Jy beam$^{-1}$) from $3 \times \sigma_{cont}$ ($\sigma_{cont} = 0.0013$ Jy beam$^{-1}$) to the max continuum flux at 0.2436 Jy beam$^{-1}$. The color bar on the uncertainty map corresponds to the derived parameter divided by the determined uncertainty plotted on a logarithmic scale. A ratio of 1 between these values would be 0 on the scale, and a ratio of 3 would be 0.48. }
    \label{fig:parameter-maps-hnco}
\end{figure}

\begin{figure}[ht]
    \centering
    \includegraphics[width=0.49\linewidth]{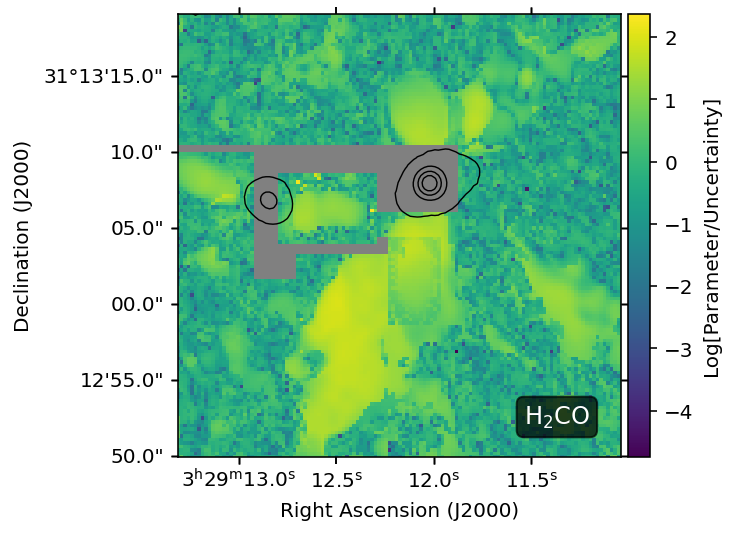}
    \includegraphics[width=0.49\linewidth]{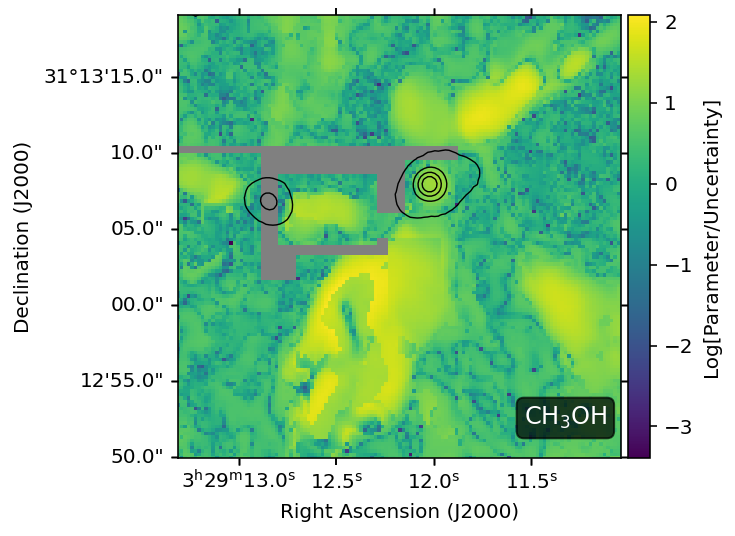}
    \includegraphics[width=0.49\linewidth]{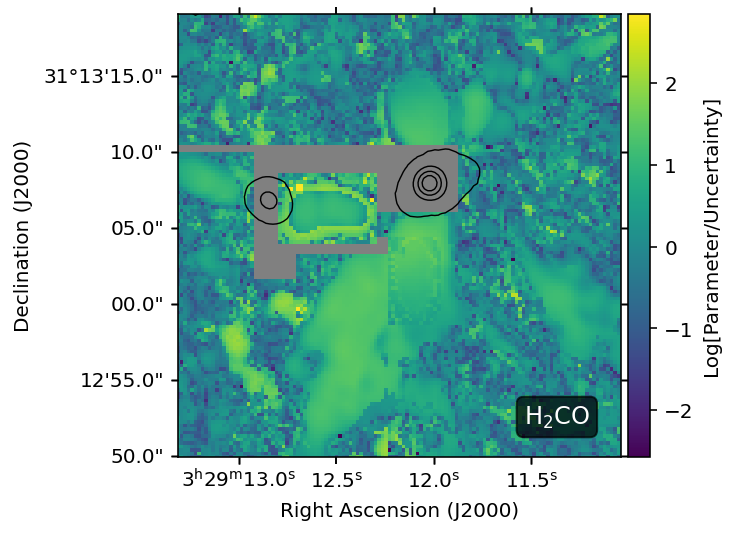}
    \includegraphics[width=0.49\linewidth]{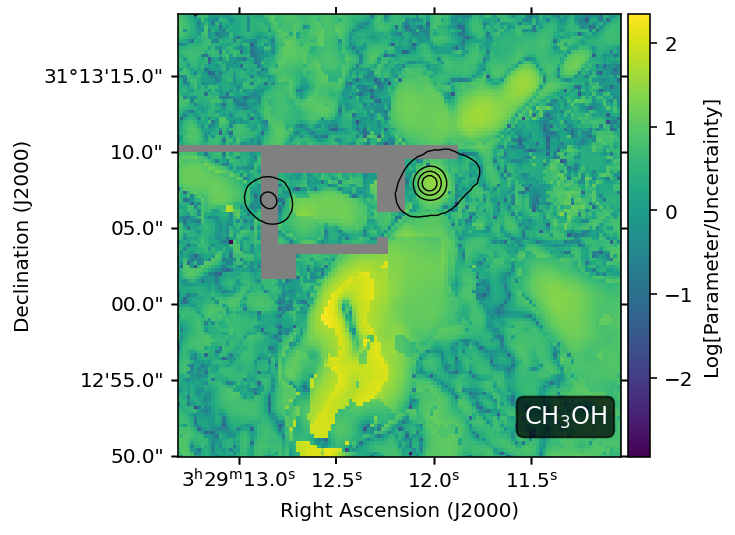}
    \includegraphics[width=0.49\linewidth]{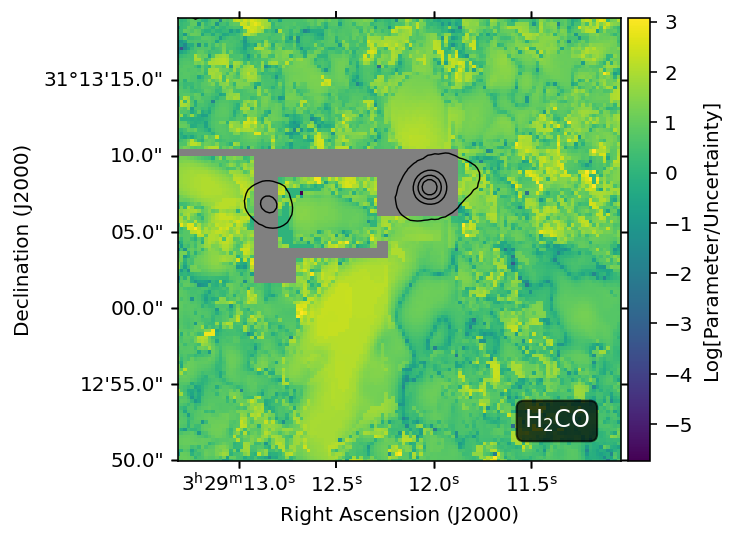}
    \includegraphics[width=0.49\linewidth]{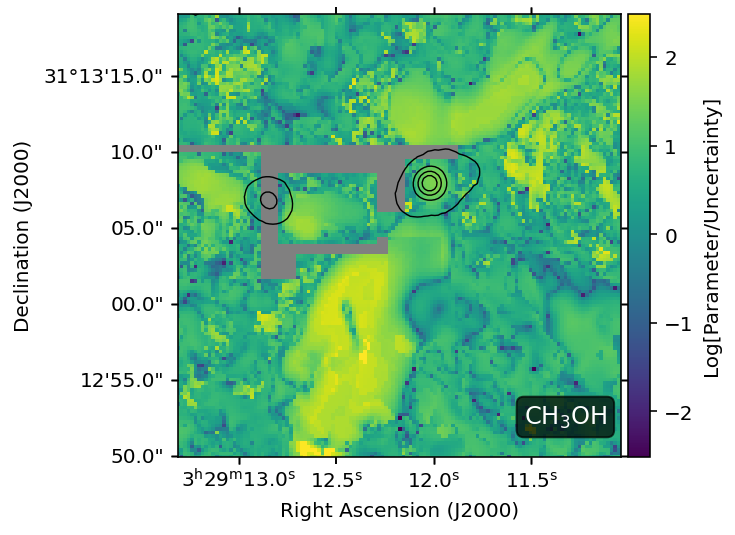}
    \caption{Uncertainty maps of \ch{H2CO} (left) and \ch{CH3OH} (right) across IRAS 4B and 4B$'$, in order from top to bottom: column density (cm$^{-2}$), rotational temperature (K), and velocity shift (km s$^{-1}$). White contours are 5 evenly spaced levels (1 level $=0.0599$ Jy beam$^{-1}$) from $3 \times \sigma_{cont}$ ($\sigma_{cont} = 0.0013$ Jy beam$^{-1}$) to the max continuum flux at 0.2436 Jy beam$^{-1}$. The color bar corresponds to the derived parameter divided by the determined uncertainty plotted on a logarithmic scale. A ratio of 1 between these values would be 0 on the scale, and a ratio of 3 would be 0.48.}
    \label{fig:uncertainty-maps-outflows}
\end{figure}

\clearpage
\section{Transitions within high-resolution windows}\label{sec:appendix-transitions}
Figure \ref{fig:moment-0-single} shows the integrated intensity maps of molecules with a single transition within the high-resolution spectral windows. Transition specific information including quantum numbers, rest frequency, upper energy (K) of the transition state, and weighted $f_{max}$ from Figure \ref{fig:moment-0-mult} is listed in Table \ref{tab:hr-lines-mult}. Table \ref{tab:hr-lines-single} contains the same information as Table \ref{tab:hr-lines-mult} but for Figure \ref{fig:moment-0-single}. Finally, Table \ref{tab:hr-lines-UI} contains the unidentified lines within the high-resolution windows.  

\begin{figure}[ht]
    \centering
    \includegraphics[width=0.49\linewidth]{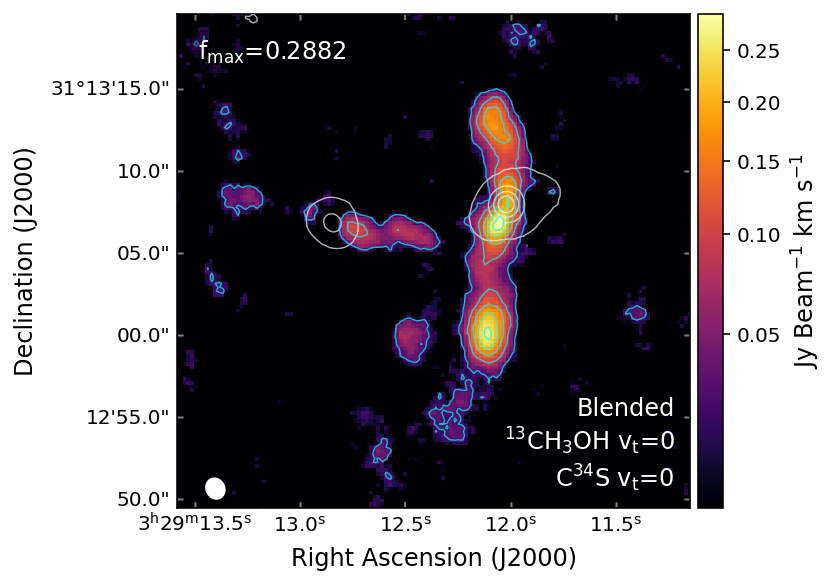}
    \includegraphics[width=0.49\linewidth]{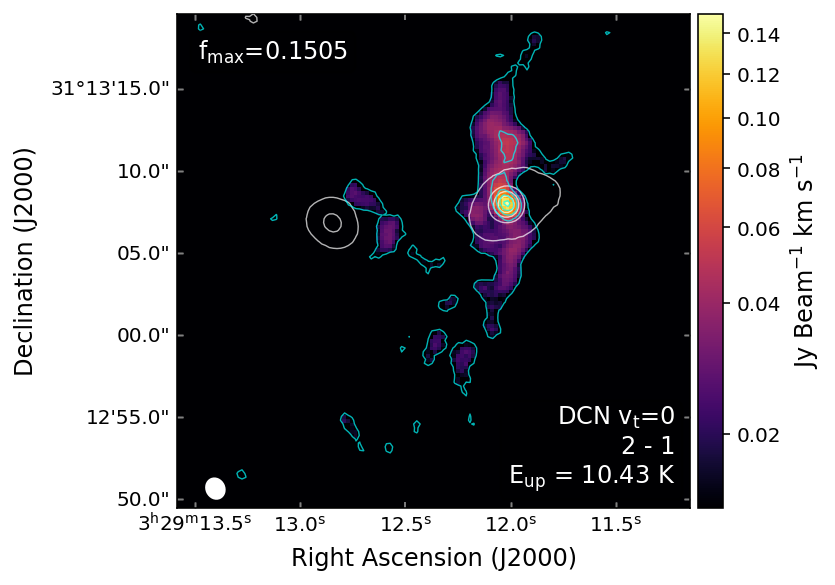}
    \includegraphics[width=0.49\linewidth]{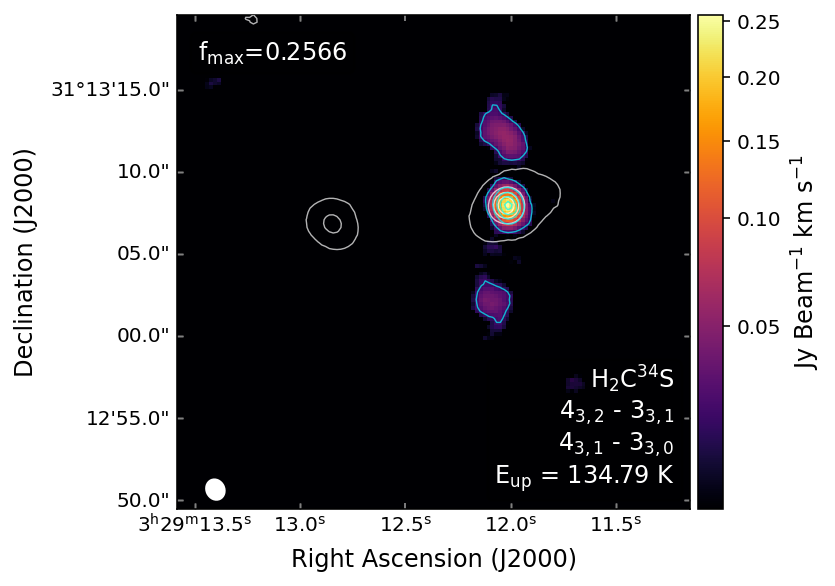}
    \includegraphics[width=0.49\linewidth]{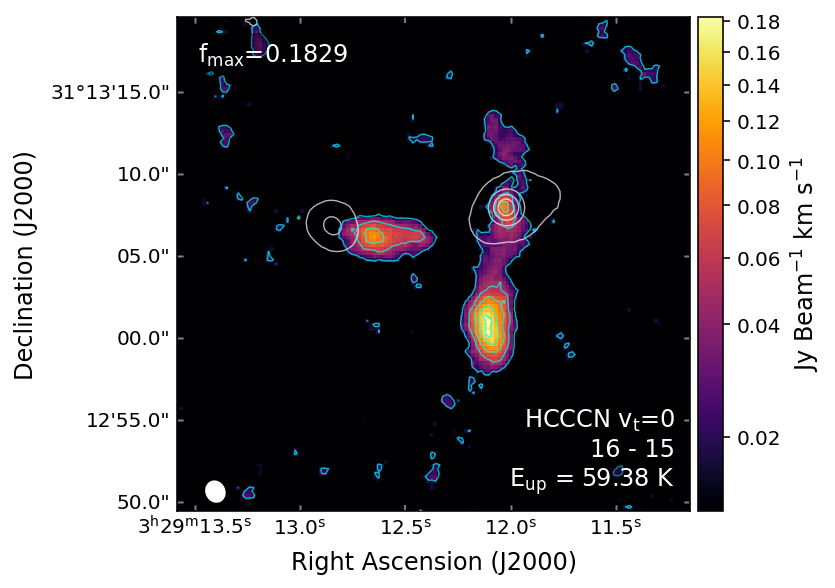}
    \caption{Integrated intensity maps of molecules with only a single transition within the high-resolution spectral windows. The max emission flux ($f_{max}$) for each molecule is listed in the top left corner. Blue contours correspond to 10\%, 30\%, 50\%, 70\%, and 90\% of $f_{max}$. White contours are 5 evenly spaced levels (1 level $=0.0599$ Jy beam$^{-1}$) from $3 \times \sigma_{cont}$ ($\sigma_{cont} = 0.0013$ Jy beam$^{-1}$) to the max continuum flux at 0.2436 Jy beam$^{-1}$. The synthesized beam is shown in the lower left corner. \textit{Continued on next page.}}
\end{figure}

\begin{figure}[ht]
    \centering
    \ContinuedFloat
    \includegraphics[width=0.49\linewidth]{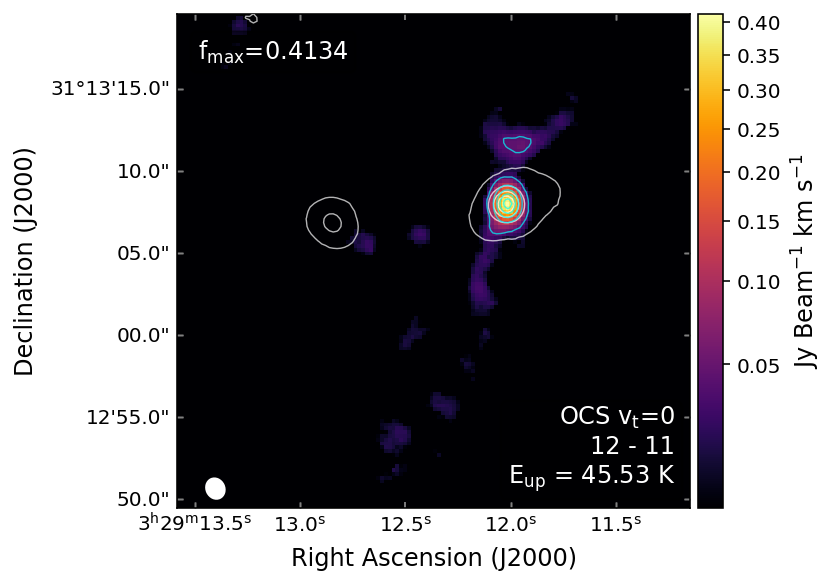}
    \includegraphics[width=0.49\linewidth]{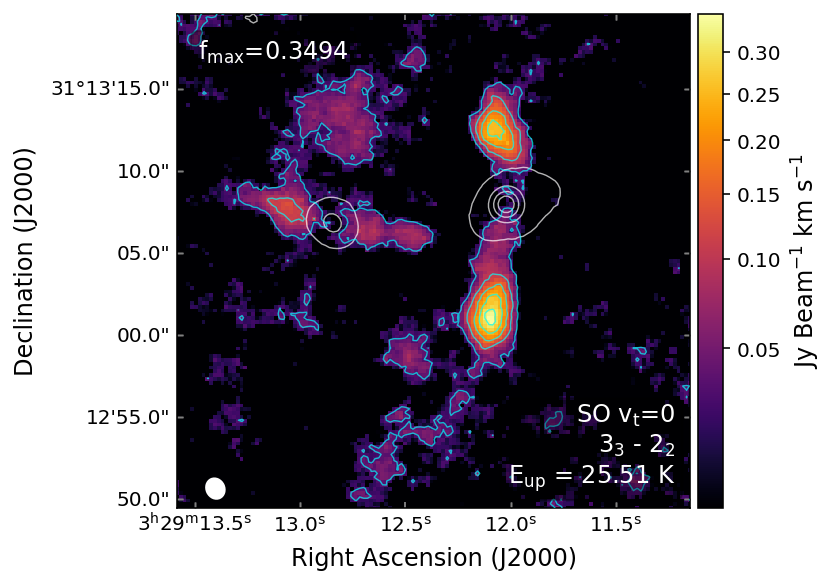}
    \includegraphics[width=0.49\linewidth]{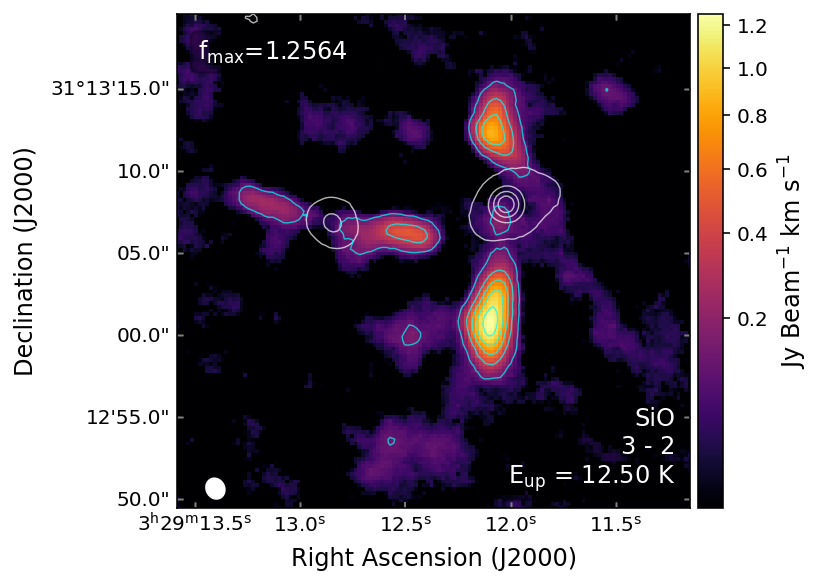}
    \includegraphics[width=0.49\linewidth]{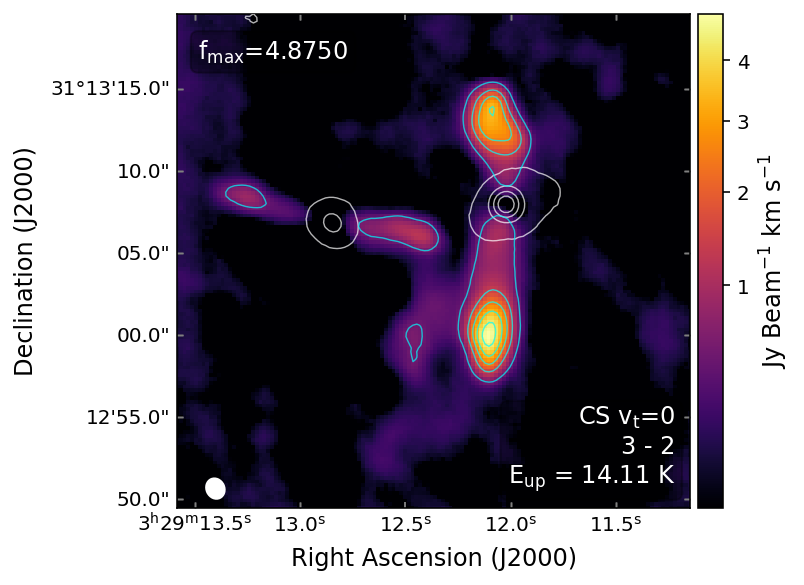}
    \includegraphics[width=0.49\linewidth]{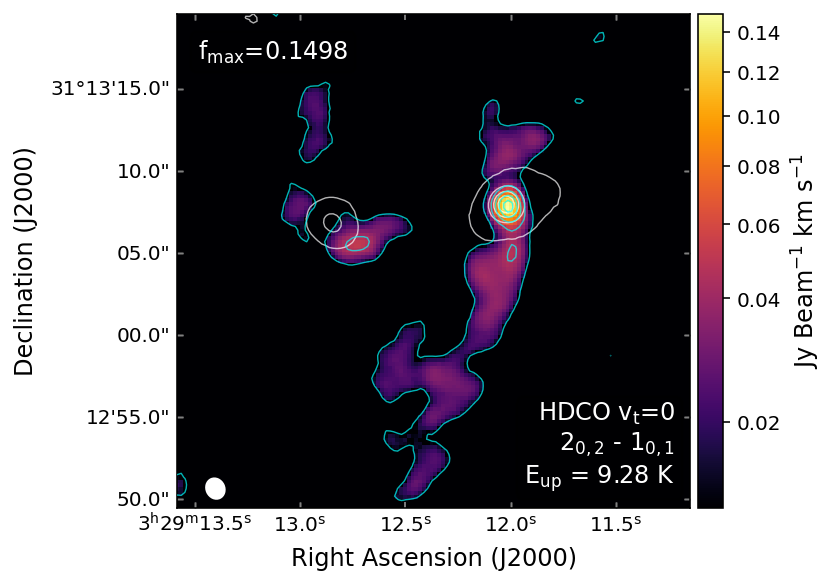}
    \includegraphics[width=0.49\linewidth]{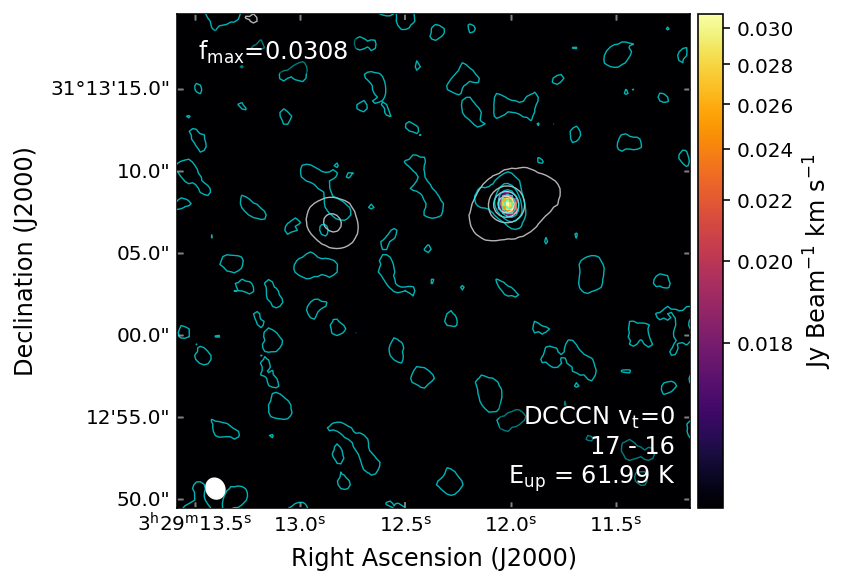}
    \caption{Integrated intensity map of molecules with only a single transition within the high-resolution spectral windows. \textit{Continued on next page.}}
\end{figure}

\begin{figure}[ht]
    \centering
    \ContinuedFloat
    \includegraphics[width=0.49\linewidth]{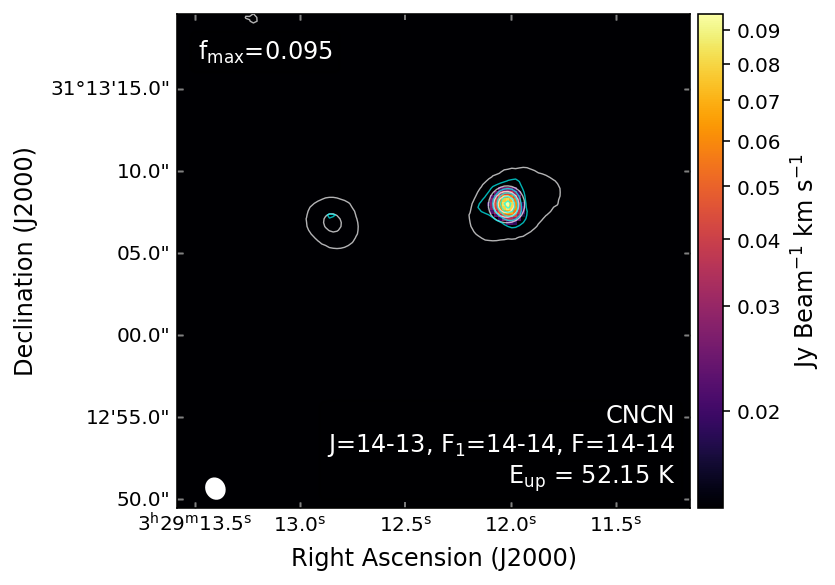}
    \includegraphics[width=0.49\linewidth]{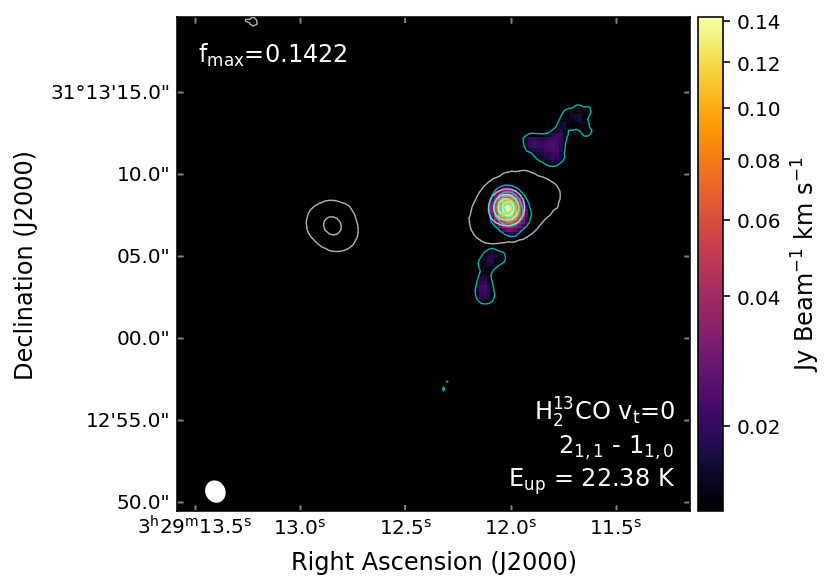}
    \includegraphics[width=0.49\linewidth]{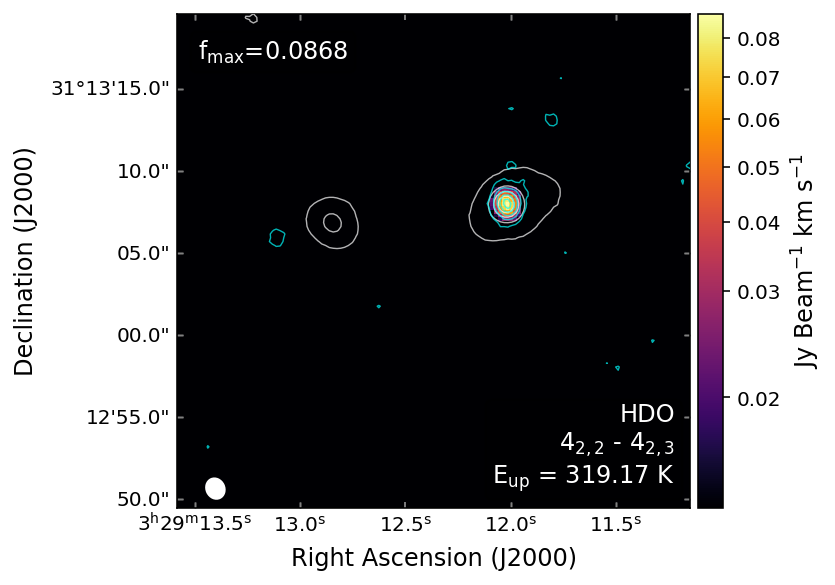}
    \includegraphics[width=0.49\linewidth]{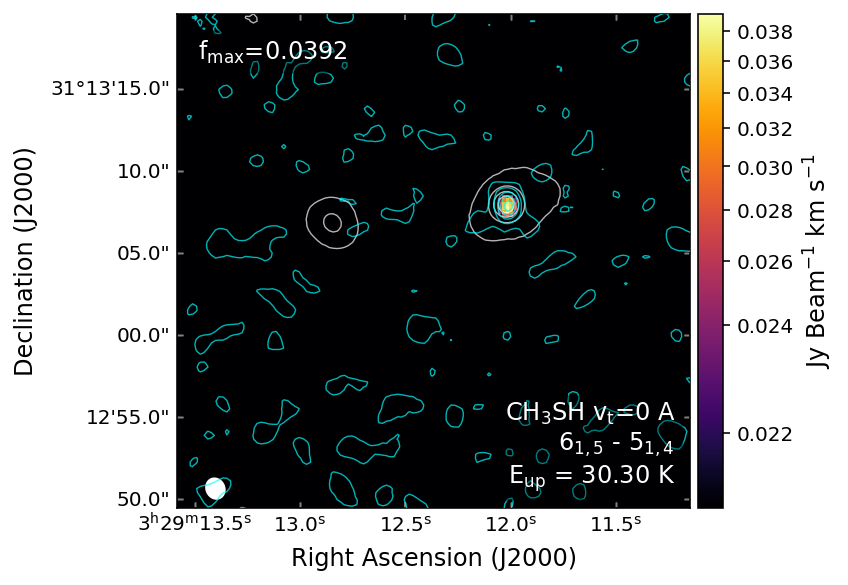}
    \caption{Integrated intensity map of molecules with only a single transition within the high-resolution spectral windows.}
    \label{fig:moment-0-single}
\end{figure}

\begin{longrotatetable}
\begin{deluxetable}{ccccccc}
\tablecaption{Molecules that have multiple transitions covered in the high-resolution windows. \label{tab:hr-lines-mult}}
\tablehead{
\colhead{\textbf{Molecule}} & \colhead{\textbf{Transition}} & \colhead{\textbf{Rest Frequency}} & \colhead{\textbf{E$_u$}} & \colhead{\textbf{SPW}} & \colhead{\textbf{Weighted $f_{max}$}} & \colhead{\textbf{Notes}}\\
\colhead{}& \colhead{QN} & \colhead{(MHz)} & \colhead{(K)} & \colhead{}  & \colhead{(Jy beam$^{-1}$)} & \colhead{} \\
}
\startdata
\ch{CH3OH} v$_t$=0-2 &&&&&&  \\
 Extended Emission &&&&& 1.6365& \\
 &6$_{-1,6}$--5$_{-0,5}$ E v$_t$=0 & 132890.8 & 54.31 & 8 && Absorption both cores\\
&3$_{-0,3}$--2$_{-0,2}$ E v$_t$=0 & 145093.8 & 27.05 & 20 &&Blended \ch{CH3OH} transitions (Set 1)\\
&3$_{1,3}$--2$_{1,2}$ E v$_t$=0 & 145097.4 & 19.50 & 20 &&Blended \ch{CH3OH} transitions (Set 1), Absorption both cores\\
& 3$_{0,3}$--2$_{0,2}$ A v$_t$=0 & 145103.2 & 13.93 & 20 &&Blended \ch{CH3OH} transitions (Set 1), Absorption both cores\\
&  3$_{2,2}$--2$_{2,1}$ A v$_t$=0 & 145124.3 & 51.64 & 20 &&Blended \ch{CH3OH} transitions (Set 2)\\
&  3$_{-2,2}$--2$_{-2,1}$ E v$_t$=0 & 145126.2 & 36.17 & 20 &&Blended \ch{CH3OH} transitions (Set 2)\\
 & 3$_{2,1}$--2$_{2,0}$ E v$_t$=0 & 145126.4 & 39.83 & 20 &&Blended \ch{CH3OH} transitions (Set 2)\\ 
 &3$_{-1,2}$--2$_{-1,1}$ E v$_t$=0 & 145131.9 & 34.98 & 20 &&Blended \ch{CH3OH} transitions (Set 2), Blended with \ch{CH2DOH}\\
 &3$_{1,2}$--2$_{1,1}$ A v$_t$=0 & 146368.3 & 28.59 & 23 && Absorption in 4B$'$\\
 &9$_{0,9}$--8$_{1,8}$ A v$_t$=0 & 146618.7 & 104.41 & 24&&\\
\hline
 \ch{CH3OH} v$_t=0-2$ &&&&&&  \\
 Compact Emission &&&&& 0.1060 &\\
&6$_{2,5}$--7$_{1,6}$ A v$_t$=0 & 132621.8 & 86.46 & 8 && Faint extended emission under 3$\times$rms\\
&20$_{-4,16}$--19$_{-5,14}$ E v$_t$=0 & 133260.9 & 583.10 & 9 &&\\
&5$_{2,3}$--6$_{1,6}$ E v$_t$=0 & 133605.4 & 60.72 & 10&& Faint extended emission under 3$\times$rms\\
 &12$_{3,9}$--13$_{2,11}$ E v$_t$=0 & 134231.0 & 243.74 & 11&&\\
 &8$_{2,7}$--7$_{3,4}$ A v$_t$=0 & 134896.9 & 121.27 & 12&&\\
 &7$_{-3,5}$--8$_{-2,7}$ E v$_t$=0 & 143169.5 & 112.71 & 15&&\\
 & 3$_{-0,3}$--2$_{-0,2}$ E v$_t$=2 & 144530.6 & 748.11 & 18&&\\
 &  3$_{2,1}$--2$_{2,0}$ E v$_t$=2 & 144571.3 & 658.82 & 18&&\\
 &3$_{0,3}$--2$_{0,2}$ A v$_t$=2 & 144572.0 & 522.08 & 18&&\\
 & 3$_{-2,2}$--2$_{-2,1}$ E v$_t$=2 & 144579.9 & 600.11 & 18&&\\
 &3$_{1,3}$--2$_{1,2}$ E v$_t$=2 & 144584.0 & 545.90 & 18&&\\  
 &3$_{1,3}$--2$_{1,2}$ A v$_t$=1 & 144589.9 & 339.14 & 18&&\\
 &3$_{2,2}$--2$_{2,1}$ E v$_t$=1 & 144728.8 & 378.50 & 19&&\\
  & 3$_{-2,1}$--2$_{-2,0}$ E v$_t$=1 & 144733.3 & 413.79 & 19&&Blended \ch{CH3OH} transitions (Set 3)\\
 &3$_{-1,3}$--2$_{-1,2}$ E v$_t$=1 & 144734.4 & 305.37 & 19&&Blended \ch{CH3OH} transitions (Set 3), Blended with \ch{C2H5OH}\\
 & 3$_{-0,3}$--2$_{-0,2}$ E v$_t$=1 & 144736.3 & 314.47 & 19 &&\\
 &3$_{1,2}$--2$_{1,1}$ E v$_t$=1 & 144750.3 & 427.27 & 19&&\\
 & 3$_{0,3}$--2$_{0,2}$ A v$_t$=1 & 144768.2 & 437.54 & 19&&\\
 &3$_{1,2}$--2$_{1,1}$ A v$_t$=1 & 144878.6 & 339.16 & 19&&\\
 &15$_{-0,15}$--15$_{1,15}$ E v$_t$=0 & 148112.0 & 290.74 & 26&&\\
 &14$_{-2,13}$--13$_{-3,11}$ E v$_t$=0 & 149532.5 & 266.13 & 27&&\\
 &14$_{-0,14}$--14$_{1,14}$ E v$_t$=0 & 150141.7 & 256.14 & 28&&\\
\hline
 \ch{CH2DOH} &&&&& 0.0692&\\
 &1$_{1,1}$--0$_{0,0}$ e$_0$ & 128656.0 & 6.17 & 3&&\\
 &18$_{0,18}$--18$_{1,18}$ o$_1$ & 128771.8 & 384.12 & 3&&\\
 &3$_{1,3}$--2$_{1,2}$ o$_1$ & 132777.8 & 35.62 & 8 &&\\
 &3$_{2,1}$--2$_{2,0}$ e$_0$ & 134185.4 & 29.01 & 11&&\\
 &4$_{1,3}$--3$_{0,3}$ e$_1$ & 144762.4 & 38.15 & 19&&\\
 &5$_{0,5}$--4$_{1,4}$ e$_0$ & 145132.4 & 32.16 & 20&&Blended with \ch{CH3OH}\\
 &10$_{1,10}$--9$_{2,7}$ e$_0$ & 146960.8 & 120.16 & 25&&\\
 &10$_{0,10}$--10$_{1,9}$ e$_1$ & 148125.2 & 130.81 & 26&&\\
 &17$_{1,16}$--17$_{0,17}$ o$_1$ & 148167.4 & 352.77 & 26&&\\
 &2$_{1,2}$--1$_{0,1}$ e$_1$ & 150651.4 & 22.84 & 28&&\\
\hline
\ch{CH3OD} &&&&& 0.0445\\
 & 9$_{-1,9}$--8$_{-2,7}$ E & 128049.4 & 102.15 & 2 && \\
 &5$_{-3,3}$--6$_{-2,5}$ A & 128675.1 & 73.67 & 3&&\\
 &3$_{-2,2}$--4$_{-1,2}$ A & 132619.5 & 34.85 & 8&&\\
 &2$_{-1,2}$--2$_{0,2}$ A & 135295.2 & 13.02 & 14&&\\
 &$7_{-3,5}$--$8_{-2,7}$ E & 143472.4 & 102.89 & 16&&\\
 &$5_{-1,5}$--$5_{0,5}$ A & 143741.7 & 39.53 & 17&&\\
\hline
\ch{CH3OCH3} &&&&& 0.0396\\
&17$_{2,16}$--16$_{3,13}$ AA & 132726.8 & 143.70 & 8 && Blended with \ch{C2H5OH}\\
&11$_{3,8}$--11$_{2,9}$  & 133265.3 & 72.87 & 9 && Either AE and/or EA\\
&11$_{3,8}$--11$_{2,9}$ EE & 133268.3 & 72.87 & 9&\\
&11$_{3,8}$--11$_{2,9}$ AA & 133271.3 & 72.87 & 9&\\    
&24$_{3,21}$--24$_{2,22}$ & 133314.7 & 290.14 & 9 && Either AA and/or E\\
&12$_{2,11}$--12$_{1,12}$ EE & 135266.6 & 76.28 & 14&&\\ 
&13$_{2,12}$--13$_{1,13}$ EE & 143163.0 & 88.00 & 15&&\\
& 6$_{3,3}$--6$_{2,4}$ EA & 144856.8 & 31.77 & 19&&\\
& 6$_{3,3}$--6$_{2,4}$ AA & 144862.0 & 31.77 & 19&&\\
&  7$_{1,7}$--6$_{0,6}$ & 147024.2 & 26.03 & 25 && Either AE and/or EA\\
&  7$_{1,7}$--6$_{0,6}$ EE & 147024.9 & 26.03 & 25&&\\
&  7$_{1,7}$--6$_{0,6}$ AA & 147025.6 & 26.03 & 25&& \\
&6$_{3,4}$--6$_{2,5}$ EE & 147206.8 & 31.77 & 25&&\\
& 6$_{3,4}$--6$_{2,5}$ AA & 147210.7 & 31.77 & 25&&\\
& 26$_{4,22}$--26$_{1,25}$ & 148119.4 & 344.84 & 26 && AA, EE, AE, and/or EA\\
& 9$_{3,7}$--9$_{2,8}$ AE & 149566.5 & 53.64 & 27&&\\
& 9$_{3,7}$--9$_{2,8}$ EE & 149569.8 & 53.64 & 27&&\\
\hline
\ch{HCOOCH3} &&&&& 0.0663\\
& 19$_{6,13}$--19$_{5,14}$ E $v_t$=0 & 132244.2 & 136.79 & 7&&\\
& 12$_{0,12}$--11$_{0,11}$ E $v_t$=0 & 132245.1 & 42.43 & 7&&\\
& 12$_{0,12}$--11$_{0,11}$ A $v_t$=0 & 132246.7 & 42.42 & 7&&\\
& 11$_{1,10}$--10$_{1,9}$ E $v_t$=0 & 132921.9 & 40.39 & 8&&\\
& 11$_{1,10}$--10$_{1,9}$ A $v_t$=0 & 132928.7 & 40.38 & 8&&\\
& 11$_{6,6}$--10$_{6,5}$ A $v_t$=1 & 134223.4 & 250.45 & 11&&\\
& 11$_{10,2}$--10$_{10,1}$ E $v_t$=0 & 134995.1 & 105.29 & 13 && Blended with \ch{CH3CHO}\\
& 11$_{7,4}$--10$_{7,3}$ E $v_t$=0 & 135290.5 & 71.48 & 14&& \\
&  11$_{7,5}$--10$_{7,4}$ A $v_t$=0 & 135302.3 & 71.46 & 14&&\\
&  11$_{7,5}$--10$_{7,4}$ E $v_t$=0 & 135303.0 & 71.46 & 14&&\\
&12$_{3,10}$--11$_{3,9}$ E $v_t$=0 & 146977.7 & 52.02 & 25&&\\
& 12$_{5,8}$--11$_{5,7}$ A $v_t$=1 & 146981.1 & 250.21 & 25&&\\
&  12$_{3,10}$--11$_{3,9}$ A $v_t$=0 & 146988.0 & 52.01 & 25&&\\
&  12$_{5,7}$--11$_{5,6}$ A $v_t$=1 & 147100.3 & 250.22 & 25 && Blended with \ch{CH3CN}\\
& 12$_{6,7}$--11$_{6,6}$ E $v_t$=1 & 147201.8 & 256.78 & 25&&\\
& 12$_{6,6}$--11$_{6,5}$ A $v_t$=0 & 148045.8 & 69.96 & 26 &&\\
& 12$_{4,8}$--11$_{4,7}$ E $v_t$=0 & 150600.8 & 57.04 & 28&&\\
& 12$_{4,8}$--11$_{4,7}$ A $v_t$=0 & 150618.3 & 57.02 & 28&&\\
\hline
\ch{CH3CHO} &&&&& 0.0595\\
&  12$_{1,11}$--11$_{2,10}$ A $v_t$=0 & 128051.6 & 76.13 & 2&&\\
& 7$_{2,5}$--7$_{1,6}$ E $v_t$=0 & 128627.4 & 35.09 & 3&&\\
&7$_{1,7}$--6$_{1,6}$ E $v_t$=1 & 132817.0 & 231.53  & 8&&\\
&  7$_{6,2}$--6$_{6,1}$ A $v_t$=0 & 134873.3 & 107.11 & 12&& Either transition (Set 1)\\
& 7$_{6,1}$--6$_{6,0}$ A $v_t$=0 & 134873.3 & 107.11 & 12 &&Either transition (Set 1)\\
&  7$_{6,1}$--6$_{6,0}$ E $v_t$=0 & 134877.3 & 107.10 & 12&& \\
&   15$_{1,14}$--15$_{0,15}$ E $v_t$=0 & 134880.0 & 115.72 & 12&&\\
& 7$_{5,3}$--6$_{5,2}$ A $v_t$=0 & 134881.7 & 82.36 & 12&&Either transition (Set 2)\\
& 7$_{5,2}$--6$_{5,1}$ A $v_t$=0 & 134881.7 & 82.36 & 12 &&Either transition (Set 2)\\
&  7$_{2,6}$--6$_{2,5}$ E $v_t$=0 & 134895.6 & 34.94 & 12&&\\
&   7$_{5,2}$--6$_{5,1}$ E $v_t$=0 & 134900.2 & 82.28 & 12&&\\
&   7$_{5,3}$--6$_{5,2}$ E $v_t$=0 & 134905.4 & 82.25 & 12&&\\
& 7$_{4,4}$--6$_{4,3}$ A $v_t$=0 & 134908.5 & 62.08 & 12&&\\   
& 7$_{4,3}$--6$_{4,2}$ E $v_t$=0 & 134922.2 & 62.02 & 12&&\\
&  7$_{3,5}$--6$_{3,4}$ E $v_t$=0 & 134996.1 & 46.19 & 13 &&Blended with \ch{HCOOCH3}\\
& 7$_{2,5}$--6$_{2,4}$ E $v_t$=2 & 135013.0 & 423.48 & 13&&\\
& 19$_{2,17}$--19$_{1,18}$ A $v_t$=0 & 135273.0 & 187.62 & 14&&\\
& 5$_{1,5}$--4$_{0,4}$ E $v_t$=0 & 135282.9 & 15.82 & 14&&\\
& 4$_{2,3}$--4$_{1,4}$ E $v_t$=0 & 144896.3 & 18.28 & 19&&\\
& 4$_{2,3}$--4$_{1,4}$ A $v_t$=0 & 147065.7 & 18.31 & 25&& \\
& 21$_{2,19}$--21$_{1,20}$ A $v_t$=0 & 150276.2 & 226.42 & 28&&\\
\hline
\ch{C2H5OH} &&&&& 0.0251& \\
&6$_{3,3}$--6$_{2,4}$ anti & 128689.7 & 28.94 & 3 &&Blended with \ch{CH3CN}\\
&15$_{1,14}$--15$_{0,15}$ anti & 132727.7 & 103.90 & 8 && Blended with \ch{CH3OCH3}\\
&3$_{2,1}$--2$_{1,2}$ anti & 132935.2 & 10.06 & 8&&\\
&6$_{3,4}$--5$_{2,4}$ gauche & 133282.6 & 85.31 & 9 &&Blended with \ch{CH2OHCHO}\\
&8$_{1,8}$--7$_{1,7}$ gauche & 133316.8 & 86.96 & 9&&\\
&7$_{1,7}$--6$_{0,6}$ anti & 133323.4 & 23.88 & 9&&\\
&23$_{4,19}$--23$_{3,20}$ anti & 134219.6 & 254.10 & 11&&\\
&27$_{4,23}$--27$_{3,24}$ anti & 135051.4 & 342.08 & 13&&\\
& 4$_{2,3}$--3$_{1,2}$ anti & 144734.1 & 13.41 & 19 && Blended with \ch{CH3OH}\\
\hline
cis-\ch{CH2OHCHO} &&&&& 0.0191& \\
& $20_{5,16}$--$20_{4,17}$ $v_t$=0 & 128021.5 & 132.83 & 2 &&\\
&$19_{4,16}$--$19_{3,17}$ $v_t$=0& 129120.2 & 115.67 &4&& \\
& $4_{4,0}$--$3_{3,1}$ $v_t$=0 & 134892.0 & 15.27 & 12&&\\
&$19_{2,17}-19_{1,18}$ $v_t$=0 & 143447.5 & 109.38 & 16&& \\
\hline
\ch{CH3CN} $v_t=0$ &&&&& 0.1236 &\\
& 7$_{0}$--6$_{0}$ & 128779.4 & 24.72 & 3&&\\
& 7$_{1}$--6$_{1}$ & 128776.9 & 31.87 & 3&&\\
& 7$_{2}$--6$_{2}$ & 128769.4 & 53.30 & 3&&\\
& 7$_{3}$--6$_{3}$ & 128757.0 & 89.02 & 3&&\\
& 7$_{4}$--6$_{4}$ & 128739.7 & 139.02 & 3&&\\
& 7$_{5}$--6$_{5}$ & 128716.9 & 203.28 & 3&&\\
& 7$_{6}$--6$_{-6}$ & 128690.1 & 281.79 & 3 && Blended with \ch{C2H5OH}\\
& 8$_{5}$--7$_{5}$ & 147103.7 & 210.34 & 25&&Blended with \ch{CH3CHO}\\
& 8$_{6}$--7$_{6}$ & 147072.6 & 288.84 & 25&&\\
& 8$_{0}$--7$_{0}$ & 147174.6 & 31.79 & 25&&\\
& 8$_{1}$--7$_{1}$ & 147171.8 & 38.93 & 25&&\\
& 8$_{2}$--7$_{2}$ & 147163.2 & 60.36 & 25&&\\
& 8$_{3}$--7$_{3}$ & 147149.0 & 96.08 & 25&&\\
& 8$_{4}$--7$_{4}$ & 147129.2 & 146.08 & 25&&\\
\hline
\ch{SO2} &&&&& 0.0475& \\
& 12$_{2,10}$--12$_{1,11}$ & 128605.1 & 82.58 & 3 &&\\
& 14$_{2,12}$--14$_{1,13}$ & 132744.9 & 108.12 & 8 &&\\
\hline
\ch{H2CO} &&&&& 2.3366&\\
& 2$_{0,2}$--1$_{0,1}$ & 145602.9 & 10.43 & 21&&Absorption in both cores\\
& 2$_{1,1}$--1$_{1,0}$ & 150498.3 & 22.62 & 28&&Absorption in both cores\\
\hline
\ch{$trans$-HCOOH} &&&&& 0.0226&\\
& 6$_{5,1}$--5$_{5,0}$ & 134920.2 & 102.12 & 12&&Either transition\\
& 6$_{5,2}$--5$_{5,1}$ & 134920.2 & 102.12 & 12&&Either transition\\
& 6$_{3,3}$--5$_{3,2}$ & 135005.0 & 51.29 & 13 & & \\
\enddata
\tablecomments{Multiple transitions have a weighted $f_{max}$ from stacked moment 0 maps listed in Jy beam$^{-1}$.}
\end{deluxetable}
\end{longrotatetable}

\begin{longrotatetable}
\begin{deluxetable}{ccccccc}
\tablecaption{Molecules that had a single transition within the high-resolution windows.\label{tab:hr-lines-single}}
\tablehead{
\colhead{\textbf{Molecule}} & \colhead{\textbf{Transition}} & \colhead{\textbf{Rest Frequency}} & \colhead{\textbf{E$_u$}} & \colhead{\textbf{SPW}} & \colhead{\textbf{$f_{max}$}} & \colhead{\textbf{Notes}}\\
\colhead{}& \colhead{QN} & \colhead{(MHz)} & \colhead{(K)} & \colhead{}  & \colhead{(Jy beam$^{-1}$)} & \colhead{} \\
}
\startdata
\ch{HDCO} & 2$_{0,2}$--1$_{0,1}$ & 128812.9 & 9.28 & 3 & 0.1498 &\\
& 2$_{1,1}$--1$_{1,0}$ & 134284.8 & 17.63 & 11&&Transition cut off by HR window\\
\hline
\ch{SO} $v_t$=0 & 3$_{3}$--2$_{2}$ & 129138.9 & 25.51 & 4 & 0.3494 &\\
\hline
\ch{SiO} & J=3-2 & 130268.7 & 12.50 & 6 & 1.2564 &\\
\hline
\ch{HCCCN} $v_t$=0 & J=16-15 & 145561.0 & 59.38 & 21 & 0.1829 &\\
\hline
\ch{OCS} $v_t$=0 & J=12-11 & 145946.8 & 45.53 & 22 & 0.4134 &\\
\hline
\ch{^{13}CH3OH} $v_t=0$ & 19$_{-2,18}$--18$_{-3,16}$ & 144618.2 & 457.04 & 18 & 0.2882 & Blended with \ch{C^{34}S} at 2.7 rms, but mostly \ch{C^{34}S}\\
\hline
\ch{C^{34}S} $v_t$=0 & J=3-2 & 144617.1 & 13.88 & 18 & 0.2882 & Blended with \ch{^{13}CH3OH}, but mostly \ch{C^{34}S}\\
\hline
\ch{CS} $v_t$=0 & J=3-2 & 146969.0 & 14.11 & 25 & 4.8750 & Absorption in both cores\\
\hline
\ch{DCN} & J=2-1 & 144828.0 & 10.43 & 19 & 0.1505 & Absorption in 4B$'$\\
\hline
\ch{H2C^{34}S} & 4$_{3,2}$--3$_{3,1}$ & 135028.1 & 134.79 & 13 & 0.2566 & Either transition\\
& 4$_{3,1}$--3$_{3,0}$ & 135028.1 & 134.79 & 13 && Either transition\\
\hline
\ch{DCCCN} $v_t$=0 & J=17-16 & 143524.9 & 61.99 & 16 & 0.0308 &\\
\hline
\ch{HDO} & 4$_{2,2}$--4$_{2,3}$ & 143727.2 & 319.17 & 17 & 0.0868 &\\
\hline
\ch{CNCN} & J=14-13, F$_1$=14-14, F=14-14 & 144866.4 & 52.15 & 19 & 0.0950 & Or F=13-13 or F=15-15\\
\hline
\ch{H2^{13}CO} & 2$_{1,1}$--1$_{1,0}$ & 146635.7 & 22.38 & 24 & 0.1422 &\\
\hline
\ch{CH3SH} & $6_{1,5} - 5_{1,4}$ A $v_t=0$ & 150146.9 & 30.30 & 28 & 0.0392 &\\
\hline
\enddata
\end{deluxetable}
\end{longrotatetable}

\begin{deluxetable}{cc}[ht]
    \tablecaption{Unidentified transitions covered in the high-resolution spectral windows.\label{tab:hr-lines-UI}}
    \tablehead{
    \colhead{\textbf{Frequency (MHz)}}&\colhead{\textbf{Spectral Window}}
    }
    \startdata
    128630.8 & 3 \\
    128773.5 & 3 \\
    132740.5 & 8 \\
    133242.7 & 9 \\
    133160.6 & 9 \\
    134175.1 & 11\\
    134914.6 & 12\\
    143547.1 & 16\\
    144637.7 & 18\\
    147107.6 & 25\\
    149549.5 & 27\\
    150476.4 & 28\\
    150470.0 & 28\\
    150218.9 & 28 \\
    \enddata  
\end{deluxetable}
\end{appendix}
\vspace{-40pt}

\bibliography{references}{}
\bibliographystyle{aasjournalv7}

\end{document}